%% file: ecethesis.tex
\documentclass[tocnosub,noragright,centerchapter,12pt]{uiucecethesis09}

\usepackage{amsmath}

\usepackage{amssymb}

\let\oldemptyset\emptyset
\let\emptyset\varnothing

\makeatletter
\usepackage{fixmath}
\usepackage[]{appendix}
\usepackage{setspace}
\usepackage{epsfig}  % for figures
\usepackage{graphicx}  % another package that works for figures
\usepackage{amsmath}  % for math spacing
\usepackage{amssymb}  % for math spacing
\usepackage{lscape}  % Useful for wide tables or figures.
\usepackage[justification=raggedright]{caption}	% makes captions ragged right - thanks to Bryce Lobdell

\usepackage{latexsym,amsfonts,epsf,cite,bbm,float}

\newtheorem{theorem}{Theorem}

\newtheorem{lemma}{Lemma}

\msthesis
\title{Near-Data Scheduling for Data Centers with Multiple Levels of Data Locality}
\author{Ali Yekkehkhany}
\department{Electrical and Computer Engineering}
\degreeyear{2016}

\advisor{Professor Yi Lu}

\begin{document}

%%%%%%%%%%%%%%%%%%%%%%%%%%%%%%%%%%%%%%%%%%%%%%%%%%%%%%%%%%%%%%%%%%%%%%%%%%%%%%%
% COPYRIGHT
%
%\copyrightpage
%\blankpage

%%%%%%%%%%%%%%%%%%%%%%%%%%%%%%%%%%%%%%%%%%%%%%%%%%%%%%%%%%%%%%%%%%%%%%%%%%%%%%%
% TITLE
%
\maketitle

%\raggedright
\parindent 1em%

\frontmatter

%%%%%%%%%%%%%%%%%%%%%%%%%%%%%%%%%%%%%%%%%%%%%%%%%%%%%%%%%%%%%%%%%%%%%%%%%%%%%%%
% ABSTRACT
%
\begin{abstract}
% Put the abstract in a file called "abs.tex" and it'll be inputted here.
\input{abs}
\end{abstract}

%%%%%%%%%%%%%%%%%%%%%%%%%%%%%%%%%%%%%%%%%%%%%%%%%%%%%%%%%%%%%%%%%%%%%%%%%%%%%%%
% DEDICATION
%
%\begin{dedication}
% Whatever dedication you want.

%\end{dedication}

%%%%%%%%%%%%%%%%%%%%%%%%%%%%%%%%%%%%%%%%%%%%%%%%%%%%%%%%%%%%%%%%%%%%%%%%%%%%%%%
% ACKNOWLEDGMENTS
%
% Put acknowledgments in a file called "ack.tex" and it'll be inputted here.
\begin{acknowledgments}
\input{ack}
\end{acknowledgments}

%%%%%%%%%%%%%%%%%%%%%%%%%%%%%%%%%%%%%%%%%%%%%%%%%%%%%%%%%%%%%%%%%%%%%%%%%%%%%%%
% TABLE OF CONTENTS
%
\tableofcontents

%%%%%%%%%%%%%%%%%%%%%%%%%%%%%%%%%%%%%%%%%%%%%%%%%%%%%%%%%%%%%%%%%%%%%%%%%%%%%%%
% LIST OF TABLES
%
% The List of Tables is not strictly necessary. Omitting the List of Tables will
% simplify the thesis check and reduce the number of corrections.
%%%%%%%%%%%%%%%%%%%%%%%%%%%%%%%\listoftables

%%%%%%%%%%%%%%%%%%%%%%%%%%%%%%%%%%%%%%%%%%%%%%%%%%%%%%%%%%%%%%%%%%%%%%%%%%%%%%%
% LIST OF FIGURES
%
% The List of Figures is not strictly necessary. Omitting the List of Figures will
% simplify the thesis check and reduce the number of corrections.
\listoffigures

%%%%%%%%%%%%%%%%%%%%%%%%%%%%%%%%%%%%%%%%%%%%%%%%%%%%%%%%%%%%%%%%%%%%%%%%%%%%%%%
% LIST OF ABBREVIATIONS
%
% The List of Abbreviations is not strictly necessary.
\chapter{LIST OF ABBREVIATIONS}
Pandas: Priority Algorithm for Near-Data Scheduling \\
JSQ-MW: Joint the Shortest Queue-MaxWeight \\
FCFS: First-Come-First-Served

%\begin{symbollist*}
%\item[EPIC] Explicitly Parallel Instruction Computing
%\item[GPU] Graphics Processing Unit
%\item[VLIW] Very Long Instruction Word
%\end{symbollist*}

%%%%%%%%%%%%%%%%%%%%%%%%%%%%%%%%%%%%%%%%%%%%%%%%%%%%%%%%%%%%%%%%%%%%%%%%%%%%%%%
% LIST OF SYMBOLS
%
%\begin{symbollist}[0.7in]
%\item[$\tau$] Time taken to drink one cup of coffee.
%\end{symbollist}

\mainmatter

%%%%%%%%%%%%%%%%%%%%%%%%%%%%%%%%%%%%%%%%%%%%%%%%%%%%%%%%%%%%%%%%%%%%%%%%%%%%%%%
% INSERT REAL CONTENT HERE
%

\include{intro}	% for INTRODUCTION in "intro.tex"

\include{litreview}

\include{JSQMW}

\include{simulationresults}

\include{AppendixA}

%\include{exper}
%\include{concl}

%%%%%%%%%%%%%%%%%%%%%%%%%%%%%%%%%%%%%%%%%%%%%%%%%%%%%%%%%%%%%%%%%%%%%%%%%%%%%%%
% APPENDIX
%
\newpage
\appendix

\backmatter

%%%%%%%%%%%%%%%%%%%%%%%%%%%%%%%%%%%%%%%%%%%%%%%%%%%%%%%%%%%%%%%%%%%%%%%%%%%%%%%
% BIBLIOGRAPHY
%
\bibliographystyle{IEEE_ECE}
% Put references in BibTeX format in thesisrefs.bib.
\bibliography{thesisrefs}

%%%%%%%%%%%%%%%%%%%%%%%%%%%%%%%%%%%%%%%%%%%%%%%%%%%%%%%%%%%%%%%%%%%%%%%%%%%%%%%
% AUTHOR'S BIOGRAPHY
% As of 10/03/2011, Author's Biography or Vita no longer accepted by Grad College

\end{document}

%% file: abs.tex
Data locality is a fundamental issue for data-parallel applications. Considering MapReduce in Hadoop, the map task scheduling part requires an efficient algorithm which takes data locality into consideration; otherwise, the system may become unstable under loads inside the system's capacity region and jobs may experience longer completion times which are not of interest. The data chunk needed for any map task can be in memory, on a local disk, in a local rack, in the same cluster or even in another data center. Hence, unless there has been much work on improving the speed of data center networks, different levels of service rates still exist for a task depending on where its data chunk is saved and from which server it receives service. Most of the theoretical work on load balancing is for systems with two levels of data locality including the Pandas algorithm by Xie et al. and the JSQ-MW algorithm by Wang et al., where the former is both throughput and heavy-traffic optimal, while the latter is only throughput optimal, but heavy-traffic optimal in only a special traffic load. We show that an extension of the JSQ-MW algorithm for a system with thee levels of data locality is throughput optimal, but not heavy-traffic optimal for all loads, only for a special traffic scenario. Furthermore, we show that the Pandas algorithm is not even throughput optimal for a system with three levels of data locality. We then propose a novel algorithm, Balanced-Pandas, which is both throughput and heavy-traffic optimal. To the best of our knowledge, this is the first theoretical work on load balancing for a system with more than two levels of data locality. This is more challenging than two levels of data locality as a dilemma between performance and throughput emerges.

%% file: ack.tex
%Firstly, I would like to thank Professor Yi Lu for her unsparing support, motivation, immense knowledge, and her dedication to research. She is not only a great researcher but also a great advisor and always acts professionally. I can hardly think of a better advisor than her and I'm so pleased to be her student. I have mostly been lucky in most of my life and working with Professor Lu was one of those big lucks in my life. She is indeed the best advisor I have ever seen.
Firstly, I would like to thank Professor Yi Lu for her invaluable support, motivation, immense knowledge, and her dedication to research. She is not only a great professor, but also a great advisor, and I feel so pleased to be her student. Working under the supervision of Professor Lu was one of the most fortunate happenings in my life. She is just the right advisor for me.

I would also like to thank my mother who has always been a great support to me. I am also grateful to my sister.

%% file: intro.tex
\chapter{Introduction}
\label{introduction}

\begin{figure}[h]
\centering
\includegraphics[scale=0.5]{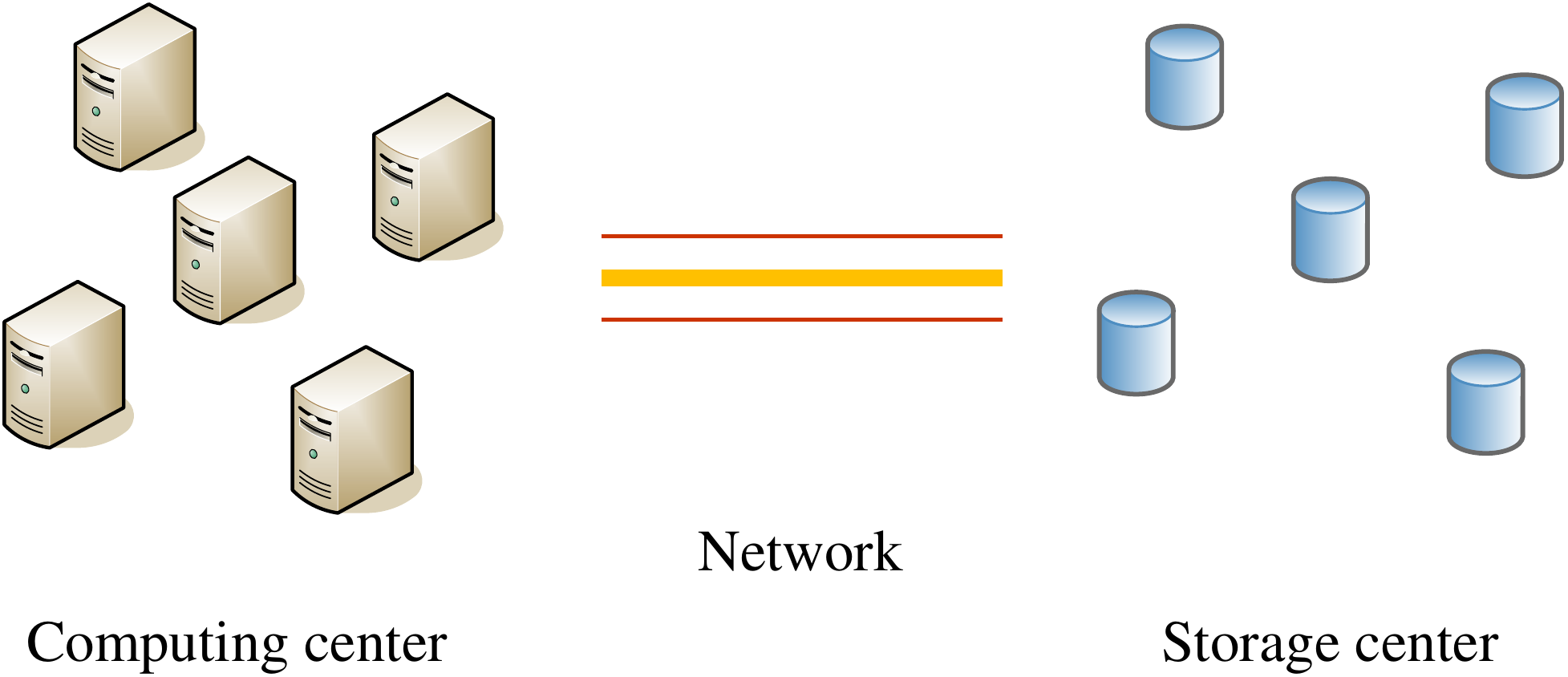}
\caption{Data center architecture when the data communication cost between the storage and computing centers is affordable.}
\label{DS1}
\end{figure}

Today's data centers need to keep pace with the explosion of data and processing the data \cite{xie2016scheduling}.
The emergence of large data sets by social networks such as Facebook \cite{facebook}, Twitter \cite{twitter}, LinkedIn \cite{linkedin}, health-care industry, search engines, and scientific research has pushed researchers to change the architecture of data centers in order to adapt them with the new needs of fast processing for large data sets.
The architecture of a data center is mainly determined by the storage and processor units and the way these two units are connected to each other.
The structure of a data center was once depicted as in Figure \ref{DS1}.
The data was stored in a large storage unit, and whenever the data was needed for a job, it was fetched by the computing unit.
Hence, every time that a chunk of data is needed for a process, it has to go through the network between storage and computing units.
Without the existence of large data sets in the past, this structure worked well.
However, with the appearance of large data sets, this structure lost its utility as the network between the two units was not capable of responding to the real-time applications.

\begin{figure}[h]
\centering
\includegraphics[scale=0.5]{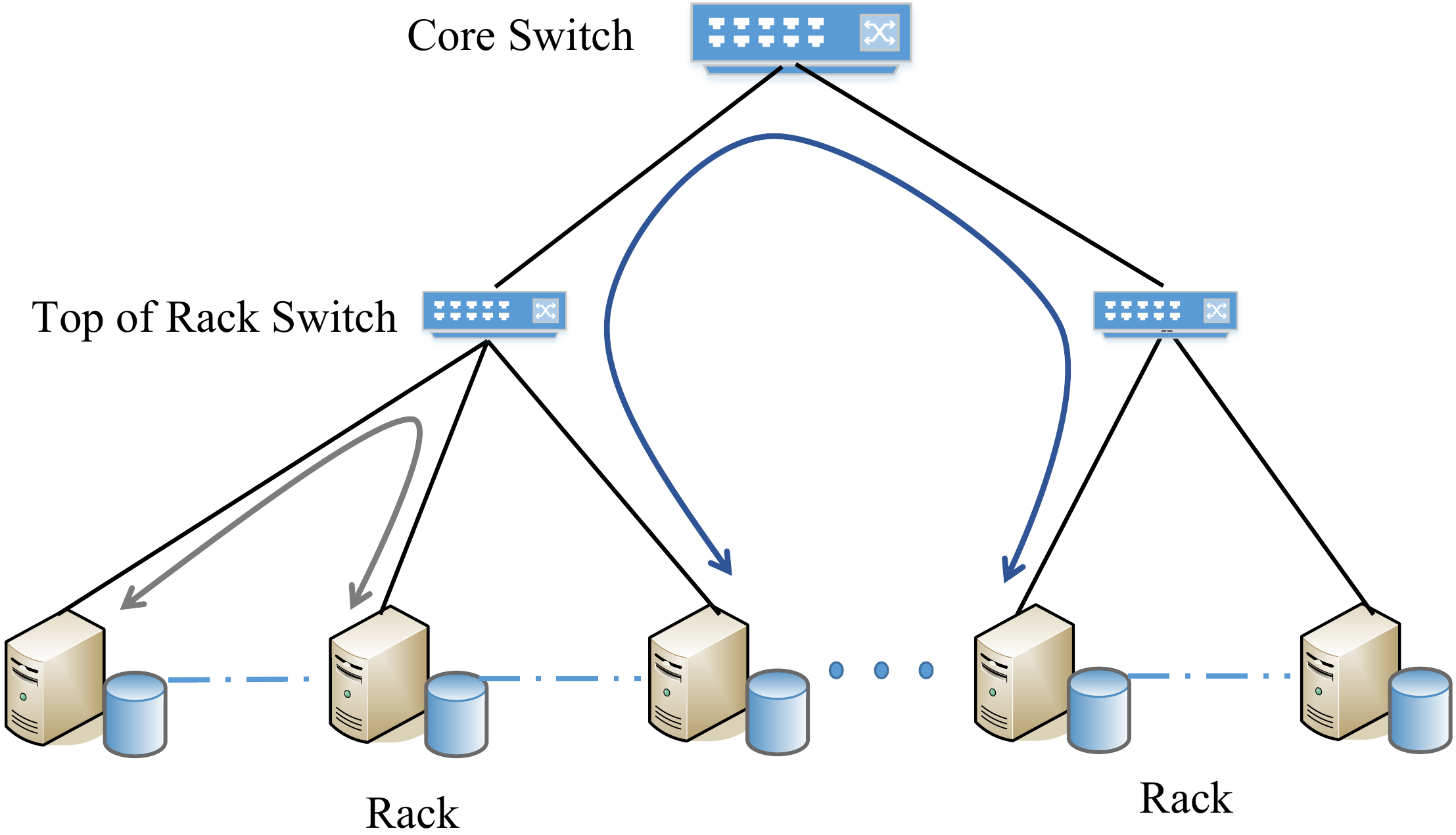}
\caption{The state-of-the-art data center architecture.}
\label{DS2}
\end{figure}

The objection to the data center architecture described above is that it is not consistent with the large data set processing applications as all the data needed for the process should be transmitted through the network.
Unless there has been a large body of research on increasing the speed of the network used in data centers, there is still a significantly large delay in data transmission compared to the service time \cite{ananthanarayanan2011scarlett, ananthanarayanan2012pacman, xie2016pandas, zaharia2010delay}.
Therefore, scientists changed the data center structure as the one depicted in Figure \ref{DS2}.
Both the large computing and storage centers are split into smaller units, and each small computing and storage center is combined with the others that we name it a server.
This way, data is moved to the computing unit and if an appropriate scheduler is used to assign tasks to servers, then very few data communication through a network is needed.
Assume that there are $M$ parallel servers in the system. The set of servers is denoted by $\mathcal{M} = \{ 1, 2, 3, \cdots, M \}$.

A large data set is split into small chunks of ordinary sizes of $64$, $128$, or $256$ megabytes.
Each data chunk is stored on the storage of a number of servers for easier accessibility and fault resilience.
If the data set is going to be processed, different servers process the data chunks stored on them, and then the results of all the servers are reduced to the final result (MapReduce).
Using such an architecture for data centers, the requirement for data transmission decreases as we try to assign each server to process a task with the needed data chunk saved on the storage of the server.
This concept is called \textit{Near-Data Scheduling} as each server prefers processing a data chunk saved on itself.
On the other hand, as we will see in Sections \ref{JSQMW2}, \ref{PandasPandas}, \ref{JSQMW}, and \ref{BP}, there are cases where we still need a data chunk to be transmitted from one server to another.
In order to give the data center such a flexibility, the servers are not completely isolated from each other.
Instead, there are \textit{rack switches} on top of servers in the same rack (a \textit{rack} consists of servers that are directly connected with each other through a switch called rack switch).
Furthermore, there is a core switch which is connected to all rack switches.
This structure for data centers allows the data chunks to be transmitted from a server to another server inside or outside of the rack where the data chunk is stored. The system consists of $K$ racks, denoted with the set $\mathcal{K} = \{ 1, 2, 3, \cdots, K \}$. A server $m$ belongs to a rack denoted by $K(m) \in \mathcal{K}$.

From the new underlying network architecture of data centers, it is obvious that the transmission of a data chunk between two servers in different racks on average takes more time other than two servers in the same rack as the data chunk should pass through three switches rather than one switch.

A task generated by a user requires its own data chunk to be processed. The data chunk is saved on $d$ servers for security and availability reasons.
In real applications a data chunk is stored on three servers in order that if one or two of the servers fail to work or become disconnected from the network, the data chunk needed for the task will still be available in the other servers.
Because of limited storage, the data chunks are usually not replicated on more than three servers.
We define the type of a task by the location where its data chunk is stored, denoted by $\bar{L} = (m_1, m_2, m_3)$ where $m_1, m_2,$ and $m_3$ are the servers storing the corresponding data chunk.
Then, the set of all task types is $\mathcal{L} = \{ (m_1, m_2, m_3) \in \mathcal{M}^3 | m_1 < m_2 < m_3 \}$.

When a server is allocated to a task to process it, the server could have the data chunk available on its memory, local disk, or the server could not have the data.
In case that the data is not available on the server, the server requires the data from another server which can be in the same rack, in a different rack, or even in another data center.
Therefore, different levels of data locality exists in data centers \cite{xie2012degree}. Most of the theoretical works that have been done on load balancing (scheduling) for data centers have been done on two levels of data locality which will be illustrated in more detail later.
The focus of this thesis is scheduling for three levels of data locality, but algorithms for two levels of data locality will also be discussed in the previous work found in Sections \ref{JSQMW2} and \ref{PandasPandas}.

As a convention, for a task, the three servers which have the data chunk associated to it ($\{ m | m \in \bar{L} \}$) are called the \textit{local servers}, and if the task receives service from one of these local servers, we say that the task is \textit{local} to the server and receives service \textit{locally}.
A task receives service \textit{rack-locally} from one of the servers that does not have the data chunk stored on it, but is in the same rack as the required data is stored in one of its servers.
The set of \textit{rack-local} servers for a task of type $\bar{L}$ is $\bar{L}_k = \{ m \notin \bar{L} | \exists n \in \bar{L} \ s.t. \ K(m) = K(n) \}$.
Finally, a task receives service \textit{remotely} if the server giving service not only does not have the required data, but the data is also not stored in another server in the same rack of the server.
In summary, $m \in \bar{L}, m \in \bar{L}_k,$ and $m \in \bar{L}_r$ denote that the server $m$ is a local, rack-local, and remote server to the task of type $\bar{L}$, respectively.

We analyze the system in a discrete-time regime, where the time slots are numbered by $t, t \geq 0$, with the following service and task arrival processes: \\
\textbf{Service process:} Assume that a local service, rack-local service, and remote service follows geometric distribution with mean $\frac{1}{\alpha}$, $\frac{1}{\beta}$, and $\frac{1}{\gamma}$, respectively.
It is clear from the structure of the data centers that because of the fetching time, it takes on average shortest time for a task to receive service from a local server, other than a rack-local server, and longest time on a remote server.
Hence, $\alpha > \beta > \gamma$.
A server can process at most one task at a time slot, and the task processing is assumed to be non-preemptive.
A task departs the system at the end of the time slot that the service is completed.
Note that the completion time of a task is not only the service time, but also the waiting time of the task to be assigned to a server for service. \\
\textbf{Arrival process:} Task arrival occurs in the beginning of a time slot. The number of incoming tasks of type $\bar{L}$ at the beginning of time slot $t$ is denoted by $A_{\bar{L}}(t)$.
The arrival process of different task types are independent of each other, with $\mathbb{E}[A_{\bar{L}}(t)] = \lambda_{\bar{L}}$.
The arrival rate vector of all task types is denoted by $\mathbold{\lambda} = (\lambda_{\bar{L}}: \bar{L} \in \mathcal{L})$.
We further assume a bounded total number of task arrival in each time slot.

The load balancing policy (consisting of routing and scheduling defined later in this section) for a data center decides which task should be assigned to an idle server for service.
The main two optimality criteria for a load balancing policy (scheduler) are \textit{throughput optimality} and \textit{heavy-traffic optimality} defined as follows:

\begin{itemize}
  \item \textbf{Throughput Optimality:} A load balancing algorithm is called to be throughput optimal if it stabilizes the data center for any arrival rate vector strictly within the capacity region.
  \item \textbf{Heavy-Traffic Optimality:} A load balancing algorithm is said to be heavy-traffic optimal if it asymptotically minimizes the mean task completion time as the arrival rate approaches the boundary of the capacity region.
\end{itemize}

The load on data centers changes frequently, so as long as the arrival rate is within the capacity region of the data center, a throughput optimal scheduler is robust to the changes.
In pick loads where the arrival rate is close to the boundary of the capacity region, a heavy-traffic optimal scheduler assigns tasks to servers efficiently, hence tasks experience the minimum mean completion time.
Most of the heuristic load balancing algorithms for data centers have not been studied in theory \cite{apache, he2011matchmaking, ibrahim2012maestro, isard2009quincy, jin2011bar, zaharia2010delay}.
In this thesis, we will discuss the literature for scheduling algorithms with theoretical guarantee for their optimality for two levels of data locality and the affinity scheduling case.
We claim that the extension of algorithms for two levels of data locality are either not optimal for three levels of data locality or not practical to be implemented.
We will then propose a novel throughput and heavy-traffic optimal algorithm for three levels of data locality.

Note that the arriving tasks to the system do not get service immediately in case where all the servers are busy processing other tasks.
Therefore, in such cases that all servers are busy, an incoming task is routed to a queue waiting for receiving service.
Based on the scheduler used in the system, different queueing structures are needed.
For example, only one queue is needed in order to implement \textit{First-Come-First-Served} (\textit{FCFS}) scheduler for a data center. For other algorithms fewer, the same or a greater number of queues as the number of servers may be needed.
The queue structure for different algorithms will be mentioned when the algorithms are illustrated in Chapters \ref{LiteratureReview} and \ref{ThreeLevelsofDataLocality}.

There are two parts for any load balancing policy (scheduler), \textit{routing} and \textit{scheduling} policies which are described as follows:

\begin{itemize}
  \item \textbf{Routing:} When a new task arrives at the system, the routing policy determines which queue it should be routed to in order to wait until it receives service from a server. %The act of determining which queue a new arriving task should be waiting at until it starts being serverd in a server is called routing policy.
  \item \textbf{Scheduling:} When a server becomes idle, and is ready to process a task, the scheduling policy determines which task receives service from the idle server. %the act of determining which task should be given service by the idle server is called scheduling policy.
\end{itemize}

The capacity region realization of a data center with three levels of data locality is calculated as follows.
Recall the arrival rate vector $\mathbold{\lambda} = (\lambda_{\bar{L}} : \bar{L} \in \mathcal{L})$.
A decomposition of the arrival rate of a task type, $\lambda_{\bar{L}}$ is $(\lambda_{\bar{L}, m}, m \in \mathcal{M})$, where $\lambda_{\bar{L}, m}$ is the arrival rate of task type $\bar{L}$ that is processed in server $m$.
Assuming that a server can afford total local, rack-local and remote load of $1$, a necessary condition that an arrival rate vector $\mathbold{\lambda}$ is supportable is the following:

\begin{equation}
\label{necessarycondition}
\sum_{\bar{L} : m \in \bar{L}} \frac{\lambda_{\bar{L}, m}}{\alpha} + \sum_{\bar{L} : m \in \bar{L}_k} \frac{\lambda_{\bar{L}, m}}{\beta} + \sum_{\bar{L} : m \in \bar{L}_r} \frac{\lambda_{\bar{L}, m}}{\gamma} < 1.
\end{equation}

Then, an outer bound of the capacity region can be characterized as the set of arrival rates $\mathbold{\lambda}$ such that there exists a decomposition $(\lambda_{\bar{L}, m}, m \in \mathcal{M})$ satisfying the necessary condition \eqref{necessarycondition}.
The outer bound of the capacity region denoted by $\Lambda$, which will be shown in Section \ref{JSQMW} that is the same as the capacity region itself, is formalized as below:

\begin{equation}
\begin{aligned}
\label{capacityregion}
\Lambda = & \{  \boldsymbol{\lambda} = (\lambda_{\bar{L}} : \bar{L} \in \mathcal{L}) \ | \\
& \ \exists \lambda_{\bar{L}, m} \geq 0, \forall \bar{L} \in \mathcal{L}, \forall m \in \mathcal{M}, s.t. \\
& \lambda_{\bar{L}} = \sum_{m = 1}^{M} \lambda_{\bar{L}, m}, \ \forall \bar{L} \in \mathcal{L}, \\
& \sum_{\bar{L}: m \in \bar{L}} \frac{\lambda_{\bar{L}, m}}{\alpha} + \sum_{\bar{L}: m \in \bar{L}_k} \frac{\lambda_{\bar{L}, m}}{\beta} + \sum_{\bar{L}: m \in \bar{L}_r} \frac{\lambda_{\bar{L}, m}}{\gamma} < 1, \forall m \}.
\end{aligned}
\end{equation}

Therefore, a linear programming optimization problem should be solved in order to find $\Lambda$.

%% file: litreview.tex
\chapter{Literature Review}
\label{LiteratureReview}
The near-data scheduling problem for the system illustrated in Chapter \ref{introduction} is a special case of \textit{affinity} scheduling \cite{squillante2001threshold, mandelbaum2004scheduling, harrison1999heavy, harrison1998heavy, bell2001dynamic}.
In an affinity scheduling problem, instead of having a number of locality levels, a task of type $\bar{L}$ can be processed by server $m$ with rate $\mu_{\bar{L}, m}$ (in our system model of Chapter \ref{introduction}, $\mu_{\bar{L}, m}$ can only be $\alpha, \beta,$ or $\gamma$ according to whether server $m$ is local, rack-local, or remote to the task of type $\bar{L}$, respectively, but in affinity scheduling problem $\mu_{\bar{L}, m}$ can take any non-negative value).
In the following, we briefly describe the \textit{Fluid Model Planning} and \textit{Generalized c$\mu$-rule} as the affinity scheduling algorithms and discuss their shortcomings.
Then we explain two algorithms for a system with two levels of data locality.

\section{Fluid Model Planning}
Harrison and Lopez \cite{harrison1999heavy, harrison1998heavy} proposed the fluid model planning algorithm for affinity scheduling problem.
The queueing structure needed to implement this algorithm is to have separate queues for different types of tasks (each queue is associated to a task type).
Then the routing and scheduling policies are as follows:
\begin{itemize}
\item \textbf{Routing:} An incoming task is routed to the queue associated to its type.
\item \textbf{Scheduling:} The arrival rate of each task type is needed to be known to solve a linear programming optimization and find the basic activities based on which the servers are assigned to process tasks.
\end{itemize}
Fluid model planning algorithm is both throughput and heavy-traffic optimal.
However, there are two main objections to this algorithm.
First, distinct queues are considered for different task types.
In the system model described in Chapter \ref{introduction} each task type has its data chunk stored on three servers out of a total of $M$ servers.
Therefore, there can be ${M}\choose{3}$ different number of task types.
This means that we should have in the order of $O(M^3)$ distinct queues for tasks, where each queue is associated to a task type.
However, in a data center there usually exists thousands of servers, so it is not logical to define $O(M^3)$ number of queues as it makes the underlying network, scheduling computation, and data base infrastructure complicated.
Second, the arrival rate of each task type is considered to be known to the system for the scheduling part, but in a real data center the load changes frequently and is not known as users can have random behavior.
As a result, fluid model planning cannot be used for practical issues, unless it has both optimality conditions.

\section{Generalized c$\mu$-Rule}
Stolyar and Mandelbaum \cite{stolyar2004maxweight,    mandelbaum2004scheduling} proposed the generalized c$\mu$-rule algorithm.
Similar to the fluid model planning queueing structure, this algorithms also requires the existence of one queue per task type.
In contrast to the fluid model planning algorithm, the arrival rate of task types are not required to be known to implement this algorithm.
Instead, the generalized c$\mu$-rule utilizes the MaxWeight procedure for the scheduling part which makes the algorithm needless of the prior knowledge of the task types' arrival rates.
Assume the cost rate incurred by the type $\bar{L}$ tasks is $C_{\bar{L}}(Q_{\bar{L}})$ where $Q_{\bar{L}}$ denotes the number of tasks of type $\bar{L}$ queued in the corresponding queue.
The cost function should have fairly normal conditions for which we can mention the following (for more detail refer to \cite{stolyar2004maxweight, mandelbaum2004scheduling}):
$C_{\bar{L}}(.)$ should be convex and continuous with $C_{\bar{L}}(0) = 0$.
The derivative of the cost function, $C_{\bar{L}}^{'}(.)$, should be strictly increasing and continuous with $C_{\bar{L}}^{'}(0) = 0$.
The routing and scheduling policies of the generalized c$\mu$-rule are as below:
\begin{itemize}
\item \textbf{Routing:} An incoming task is routed to the queue associated to its type.
\item \textbf{Scheduling:} An idle server $m \in \mathcal{M}$ is scheduled to a task of type $\bar{L}$ in the set below at time slot $t$:
\end{itemize}

\begin{equation}
\begin{aligned}
\underset{\bar{L}}{ArgMax} \ \bigg \{ C^{'}_{\bar{L}} (Q_{\bar{L}}(t)) \mu_{\bar{L}, m} \bigg \}.
\end{aligned}
\end{equation}

In the system model with three levels of data locality, $\mu_{\bar{L}, m}$ is $\alpha, \beta,$ or $\gamma$ if the task type $\bar{L}$ is local, rack-local, or remote to the idle server $m$, respectively.
Mandelbaum and Stolyar proved that the generalized c$\mu$-rule asymptotically minimizes both instantaneous and cumulative queueing costs in heavy traffic \cite{mandelbaum2004scheduling}.
Consider the same cost function for all task types as $C_{\bar{L}}(Q_{\bar{L}}) = Q_{\bar{L}}^{\beta + 1}, \ \forall \bar{L} \in \mathcal{L}$, where $\beta > 0$.
The function $Q^{\beta + 1}$ satisfies all the required conditions mentioned for a valid cost function.
Therefore, the generalized c$\mu$-rule asymptotically minimizes the holding cost $\sum_{\bar{L}} Q_{\bar{L}}^{\beta + 1}$, and as the constant $\beta$ should strictly be greater than zero, this algorithm cannot minimize $\sum_{\bar{L}} Q_{\bar{L}}$.
Hence, the generalized c$\mu$-rule is not heavy-traffic optimal.
Besides, we still need many queues in the order of cubic number of servers in order to implement this algorithm which makes the underlying system complicated and is not practical.

In Sections \ref{JSQMW2} and \ref{PandasPandas}, two algorithms for two levels of data locality will be discussed.
In both algorithms, no prior task types' arrival rate information is needed to be known.
Furthermore, only $M$ queues, one for each server, is needed for queueing the tasks that are waiting for service.
In a system with two levels of data locality there is no notion of rack structure and rack-local service, instead there is only a core switch connecting all servers to each other.
A task can only get service locally with rate $\alpha$ from one of the local servers ($m \in \bar{L}$), or it gets service remotely with rate $\gamma$ from any other servers ($m \notin \bar{L}$).
The capacity region of a system with two levels of data locality is given in equation (\ref{capacityregion2}) which can be driven with the same reasoning we had for a system with three levels of data locality.

\begin{equation}
\begin{aligned}
\label{capacityregion2}
\Lambda = & \{  \boldsymbol{\lambda} = (\lambda_{\bar{L}} : \bar{L} \in \mathcal{L}) \ | \\
& \ \exists \lambda_{\bar{L}, m} \geq 0, \forall \bar{L} \in \mathcal{L}, \forall m \in \mathcal{M}, s.t. \\
& \lambda_{\bar{L}} = \sum_{m = 1}^{M} \lambda_{\bar{L}, m}, \ \forall \bar{L} \in \mathcal{L}, \\
& \sum_{\bar{L}: m \in \bar{L}} \frac{\lambda_{\bar{L}, m}}{\alpha} + \sum_{\bar{L}: m \notin \bar{L}} \frac{\lambda_{\bar{L}, m}}{\gamma} < 1, \forall m \}.
\end{aligned}
\end{equation}

First, join the shortest queue-MaxWeight (\textit{JSQ-MW}) \cite{wang2016maptask} which is a throughput optimal, but not heavy-traffic optimal algorithm will be discussed.
Then the Priority Algorithm for Near-Data Scheduling (\textit{Pandas}) algorithm proposed by Xie and Lu \cite{xie2015priority} which is both throughput and heavy-traffic optimal will be presented.

\section{Join the Shortest Queue-MaxWeight (JSQ-MW)}
\label{JSQMW2}
Joining the shortest queue-MaxWeight algorithm proposed by Wang et al. \cite{wang2016maptask} requires the existence of one queue per server.
The length of the $m$-th queue at time slot $t$ is denoted by $Q_m(t)$.
The central scheduler that maintains all the queue lengths routes the new incoming tasks to a queue and schedules the idle servers to a task as follows:
\begin{itemize}
\item \textbf{Routing:} An arriving task of type $\bar{L}$ is routed to the shortest queue of the local servers in the set $\bar{L}$ (all ties are broken randomly throughout this paper). \\
\item \textbf{Scheduling:} At time slot $t$, the idle server $m$ is assigned to process a task from a queue in the set given in equation (\ref{felan}):
\begin{equation}
\label{felan}
\underset{n \in \mathcal{M}}{\arg \max} \ \{ \alpha Q_n(t)I_{\{ n = m \}}, \beta Q_n(t)I_{\{ n \neq m \}} \}.
\end{equation}
\end{itemize}

The JSQ-MW algorithm is proven to be throughput optimal for a system with two levels of data locality \cite{wang2016maptask}.
However, it is not heavy-traffic optimal.

As a definition, if the incoming load routed to a server exceeds the capacity of the server, the server is called to be a \textit{beneficiary} server.
On the other hand, if the incoming load routed to a server is less than what the server can process, the server is called to be a \textit{helper}.
Beneficiaries cannot process all the tasks routed to their queues, so they get help from helpers (helping servers).

Wang et al. \cite{wang2016maptask} proved that the JSQ-MW algorithm can minimize the mean task completion time in a specific traffic scenario as follows.
If all the incoming traffic is local to a set of servers where all of them are beneficiaries, and the rest of servers do not receive any traffic load, so to be helpers, the JSQ-MW algorithm minimizes the mean task completion time.

\section{Priority Algorithm for Near-Data Scheduling (Pandas)}
\label{PandasPandas}
To the best of our knowledge, the Pandas algorithm is the only throughput and heavy-traffic optimal algorithm for a system with two levels of data locality.
Assuming the existence of one queue per server, the routing and scheduling policies are as follows:
\begin{itemize}
\item \textbf{Routing:} An arriving task of type $\bar{L}$ is routed to the shortest queue of the local servers in the set $\bar{L}$.
\item \textbf{Scheduling:} As long as there exists a task available in the queue of the idle server $m$, it will be assigned to a local task at the $m$-th queue. If there is no local task at the $m$-th queue, the idle server is assigned to give service to a task in the longest queue of the system (the task in the longest queue can be local or remote to the idle server).
\end{itemize}

We can improve the Pandas algorithm by adding the following two features to the algorithm.
First, when an idle server $m$ does not have any tasks queued in front of it and is assigned to process a task from the longest queue, the scheduler can assign a local task to server $m$ queued at the longest queue if available.
Second, an idle server $m$ is assigned to serve a remote task from the longest queue in the system if $\underset{n \in \mathcal{M}}{\max} \ {Q_n} > \frac{\alpha}{\gamma}$ in order to make sure that the remote task will experience less service time if it gets service from the idle server, other than waiting at its current queue and receiving service locally.

In Chapter \ref{ThreeLevelsofDataLocality}, our theoretical analysis of a system with three levels of data locality will be discussed.

%% file: JSQMW.tex
\chapter{Three Levels of Data Locality}
\label{ThreeLevelsofDataLocality}
In this chapter we propose our two algorithms: the JSQ-MW algorithm and the Balanced-Pandas (Weighted-Workload Routing and Priority Scheduling) algorithm \cite{xie2016scheduling}.
Our proposed JSQ-MW algorithm is an extension of the algorithm used for two levels of data locality which is throughput optimal, but not heavy-traffic optimal.
We will show that the extension of the JSQ-MW algorithm also minimizes the mean task completion time in specific workloads, but not all loads.
On the other hand, the Balanced-Pandas algorithm is a novel algorithm proposed by us which is throughput optimal for three levels of data locality.
It is heavy-traffic optimal in the case that $\beta^2 > \alpha \gamma$, which means that the rack-local service is much faster than the remote service.
This condition usually holds in real systems.
To the best of our knowledge the Balanced-Pandas algorithm is the only throughput and heavy-traffic optimal algorithm proposed for a system with three levels of data locality.

The difficulty of designing an algorithm for a system with three levels of data locality is illustrated in Section \ref{felanfelan}.

\section{The Performance versus Throughput Dilemma}
\label{felanfelan}
\begin{figure}[h]
\centering
\includegraphics[scale=0.5]{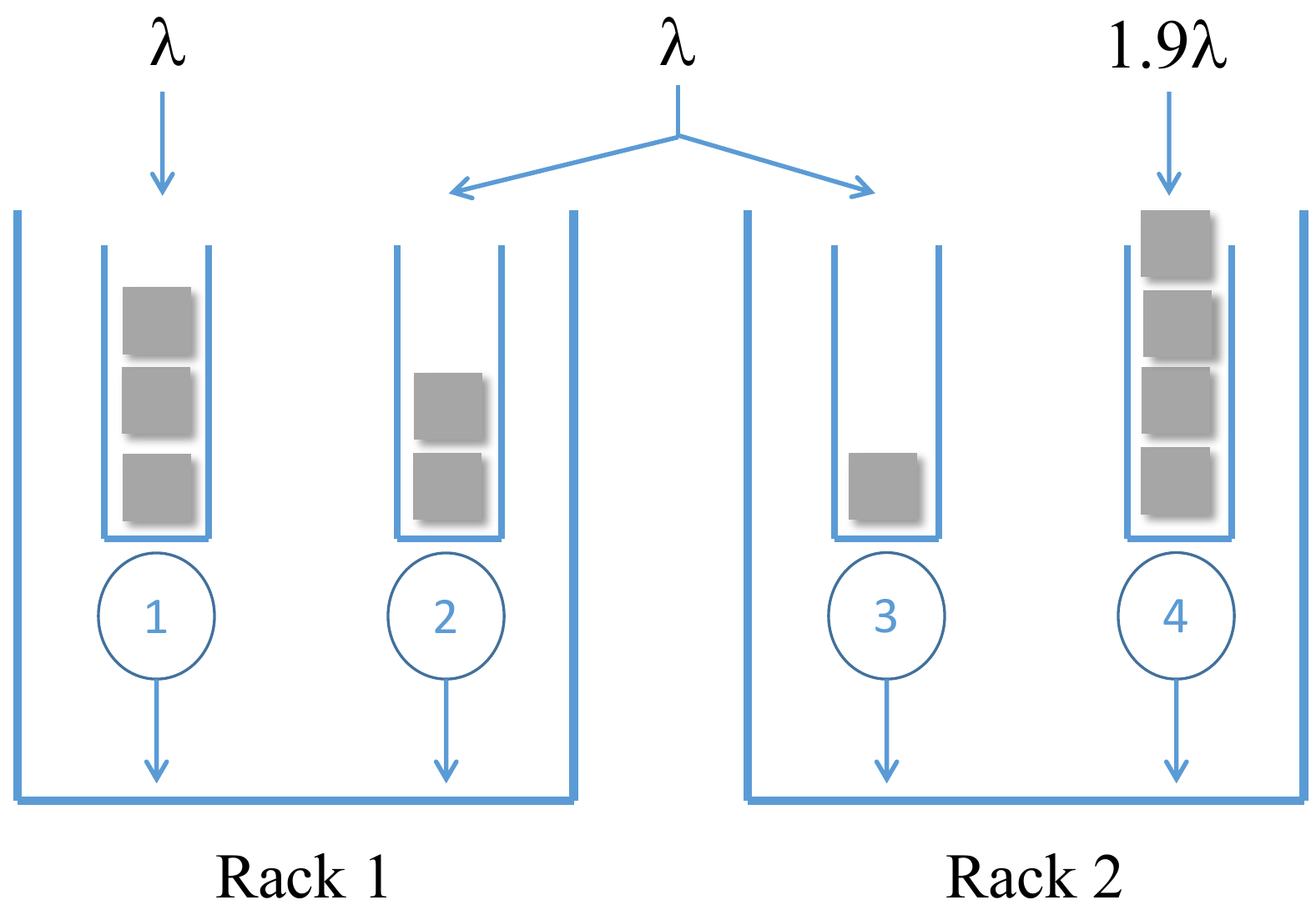}
\caption{A system with two racks showing how performance should be sacrificed in order to achieve throughput optimality.}
\label{PTD}
\end{figure}

Under the Pandas algorithm, each server has a queue maintaining tasks local to it.
The Pandas routing policy balances tasks across their local servers.
An idle server processes local tasks as long as there exists one in its queue; otherwise, it processes a remote task from the longest queue of the system.
In other words, the Pandas algorithm forces servers to process as many local tasks as possible, then process remote tasks if they do not have any local tasks available.
Therefore, in a system with three levels of data locality, the Pandas algorithm has good performance in low and medium loads by maximizing the number of tasks served locally.
However, the Pandas algorithm sacrifices throughput optimality at high loads.
The example depicted in Figure \ref{PTD} and the explanation afterward makes the performance versus throughput dilemma clear.

Assume each of the two racks has two servers as depicted in Figure \ref{PTD}.
Three types of tasks receive service from the system as follows: one type of task with arrival rate $\lambda$ is only local to the first server, another type with arrival rate $\lambda$ is local to both the second and third servers, and the other type of task with arrival rate $1.9 \lambda$ is only local to the fourth server.
Using the Pandas algorithm to maximize the number of tasks served locally, the second type of tasks is split between the second and third servers evenly.
Assuming that local, rack-local, and remote service rates are $\alpha = 1$, $\beta = 0.9$, and $\gamma = 0.5$, respectively, the Pandas algorithm can stabilize the system in the following region:

\[
1.9 \lambda < \alpha + \beta (1 - \frac{0.5 \lambda}{\alpha}) + \gamma (1 - \frac{0.5 \lambda}{\alpha}) + \gamma (1 - \frac{\gamma}{\alpha}),
\]
which gives $\lambda < 0.9355$.
However, if the second type of tasks is only routed to the second server, so the third server does not process any local tasks, but only processes rack-local tasks, then the system is stable as long as $\lambda < 1$ which is a bigger capacity region other than the one for the Pandas algorithm.
Our proposed two algorithms described in Sections \ref{JSQMW} and \ref{BP} stabilize the system in the capacity region given in equation \eqref{capacityregion} without the knowledge of the tasks' arrival rates.
In the next section, we draw an equivalent capacity region for a system with rack structure which will be used in the optimality proofs of our proposed algorithms.

\section{Equivalent Capacity Region}
\label{ECR}
In this section we will show that the outer bound of capacity region proposed in equation \eqref{capacityregion} is actually the capacity region of a system with three levels of data locality.
In the following lemma, we propose an equivalent capacity region with the one in equation \eqref{capacityregion} which will be used in our proofs.

\begin{lemma}
\label{Ecapacityregion}
The following set $\bar{\Lambda}$ is equivalent to $\Lambda$ defined in equation \eqref{capacityregion}:

\begin{equation}
\begin{aligned}
\label{ECRR}
\bar{\Lambda} = & \{  \boldsymbol{\lambda} = (\lambda_{\bar{L}} : \bar{L} \in \mathcal{L}) \ | \\
& \ \exists \lambda_{\bar{L}, n, m} \geq 0, \forall \bar{L} \in \mathcal{L}, \forall n \in \bar{L}, \forall m \in \mathcal{M}, s.t. \\
& \lambda_{\bar{L}} = \sum_{n: n \in \bar{L}} \sum_{m = 1}^{M} \lambda_{\bar{L}, n, m}, \ \forall \bar{L} \in \mathcal{L}, \\
& \sum_{\bar{L}: m \in \bar{L}} \sum_{n: n \in \bar{L}} \frac{\lambda_{\bar{L}, n, m}}{\alpha} + \sum_{\bar{L}: m \in \bar{L}_k} \sum_{n: n \in \bar{L}} \frac{\lambda_{\bar{L}, n, m}}{\beta} + \sum_{\bar{L}: m \in \bar{L}_r} \sum_{n: n \in \bar{L}} \frac{\lambda_{\bar{L}, n, m}}{\gamma} < 1, \forall m \},
\end{aligned}
\end{equation}
where $\lambda_{\bar{L}, n, m}$ is the arrival rate of type $\bar{L}$ tasks that are local to server $n$, but are scheduled to be processed at server $m$.
$\lambda_{\bar{L}, n, m}$ is actually the decomposition of $\lambda_{\bar{L}, m}$ and $\lambda_{\bar{L}, m} = \sum_{n \in \mathcal{M}} \lambda_{\bar{L}, n, m}$.

\end{lemma}

%\begin*{\textbf{proof:}}
\textbf{proof:}
In order to prove that $\bar{\Lambda} = \Lambda$, we show that $\bar{\Lambda} \subset \Lambda$ and $\Lambda \subset \bar{\Lambda}$. \\
$\bar{\Lambda} \subset \Lambda$: If $\mathbold{\lambda} \in \bar{\Lambda}$, there exists a load decomposition $\{ \lambda_{\bar{L}, n, m} \}$ such that it satisfies all the conditions in \eqref{ECRR}.
By defining $\lambda_{\bar{L}, m} \equiv \sum_{n: n \in \bar{L}} \lambda_{\bar{L}, n, m}$, it is clear that this decomposition of $\mathbold{\lambda}$, that is $\{ \lambda_{\bar{L}, m} \}$ satisfies the conditions in equation \eqref{capacityregion}, so $\mathbold{\lambda} \in \Lambda$. Hence $\bar{\Lambda} \subset \Lambda$. \\
$\Lambda  \subset \bar{\Lambda}$: If $\mathbold{\lambda} \in \Lambda$, there exists a load decomposition $\{ \lambda_{\bar{L}, m} \}$ such that it satisfies all the conditions in equation \eqref{capacityregion}.
By defining $\lambda_{\bar{L}, n, m} \equiv \frac{\lambda_{\bar{L}, m}}{|\bar{L}|}$, it is clear that this decomposition of $\mathbold{\lambda}$, that is $\{ \lambda_{\bar{L}, n, m} \}$ satisfies the conditions in \eqref{ECRR}, so $\mathbold{\lambda} \in \bar{\Lambda}$. Hence $\bar{\Lambda} \subset \bar{\Lambda}$. \\
%\end*{proof}

\section{Join the Shortest Queue-MaxWeight (JSQ-MW)}
\label{JSQMW}
In order to implement the JSQ-MW algorithm, the central scheduler keeps one queue per server, where the length of $m$-th queue associated to the $m$-th server at time slot $t$ is denoted by $Q_m(t)$.
The $m$-th queue only keeps tasks that are local to the $m$-th server.
Then, the JSQ-MW routing and scheduling policies are as follows:
\begin{itemize}
\item \textbf{JSQ-MW Routing:} An arriving task of type $\bar{L}$ is routed to its shortest local queue. That is, the central scheduler inserts the new task to the shortest queue in the set $\{ Q_m | m \in \bar{L} \}$, where ties are broken randomly.
\item \textbf{JSQ-MW Scheduling:} The scheduling decision $\eta_m(t)$ of an idle server $m$ at time slot $t$ is chosen from the following set where the ties are broken randomly. That is, idle server $m$ is scheduled to give service to a task queued in a queue in the following set:
\[
\underset{n \in \mathcal{M}}{\arg \max} \ \{ \alpha Q_n(t) I_{\{ n = m \}}, \ \beta Q_n(t) I_{\{ K(n) = K(m) \}}, \ \gamma Q_n(t) I_{\{ K(n) \neq K(m) \}} \}.
\]
\end{itemize}

In order to describe the queue evolution of a system with three levels of data locality using the JSQ-MW algorithm, we define the following terminologies.
Let the number of type $\bar{L}$ tasks that are routed to the $m$-th queue at time slot $t$ be denoted by $A_{\bar{L}, m}(t)$. Then the total number of task arrivals to $Q_m$ at time slot $t$ is as follows:

\[
A_m(t) = \sum_{\bar{L}: m \in \bar{L}} A_{\bar{L}, m} (t).
\]
Server $m$ provides local, rack-local, and remote services denoted by $S_m^l(t)$, $R_m^k(t)$, and $R_m^r(t)$, respectively.
Service times are assumed to follow geometric distribution, so $S_m^l(t)$, $R_m^k(t)$, and $R_m^r(t)$ are Bernoulli random variables in each time slot as follows:
$S_m^l(t) \sim Bern(\alpha I_{\{ \eta_m(t) = m \}})$, $R_m^k(t) \sim Bern(\beta I_{\{ K(\eta_m(t)) = K(m), \eta_m(t) \neq m \}})$, and $R_m^r(t) \sim Bern(\gamma I_{\{ K(\eta_m(t)) \neq K(m) \}})$.
Queue $m$ can receive local, rack-local, and remote services as follows.
The local service is received from server $m$ which is $S_m^l(t)$, while rack-local service can be received from any other servers in the same rack of $Q_m$ which we denote by $S_m^k(t) = \sum_{n: K(n) = K(m), n \neq m} R_n^k(t) I_{\{ \eta_n(t) = m \}}$, and $Q_m$ receives remote services from all servers out of its rack which is denoted by $S_m^r(t) = \sum_{n: K(n) \neq K(m)} R_n^r(t) I_{\{ \eta_n(t) = m \}}$.
The total number of task departures for $Q_m$ at time slot $t$ is equal to the summation of local, rack-local, and remote services given to $Q_m$ at time slot $t$ which we denote by $S_m(t) \equiv S_m^l(t) + S_m^k(t) + S_m^r(t)$.
Defining $U_m(t) = \max \{ 0, \ S_m(t) - A_m(t) - Q_m(t) \}$ as the unused service allocated to the $m$-th queue, the queues evolve from one time slot to the next one as follows:

\[
Q_m(t + 1) = Q_m(t) + A_m(t) - S_m(t) + U_m(t).
\]

Note that the queue length vector $\mathbold{Q}(t) = (Q_1(t), Q_2(t), \cdots, Q_M(t))$ is not a Markov chain since given the queue lengths at a time slot, the future of the queue lengths is not independent from the past.
The reason is that given the queue lengths, we cannot figure out the status of each server in the system.
Therefore, we define the working status of server $m$ at time slot $t$, $f_m(t)$ as follows:

\[
f_m(t) = \begin{cases}
               -1 \ \ \ \ \text{if server $m$ is idle} \\
               \ n \ \ \ \ \ \text{if server $m$ processes a task from $Q_n$}
         \end{cases}.
\]

If the $m$-th server is in idle mode, not processing any tasks, its working status is equal to $-1$.
Otherwise, if server $m$ processes a local task from $Q_m$, then $f_m(t) = m$.
If server $m$ processes a task from $Q_n$, where $n \neq m$, but $K(n) = K(m)$ ($K(n) \neq K(m)$), it means that it is serving a rack-local (remote) task and $f_m(t) = n$.

Defining the working status vector $\mathbold{f}(t) = (f_1(t), f_2(t), \cdots, f_M(t))$, as the service times follow geometric distributions, both queue length vector $\mathbold{Q}(t)$ and $\mathbold{f}(t)$ together, $\{ \mathbold{Z}(t) = (\mathbold{Q}(t), \mathbold{f}(t)), t \geq 0 \}$, form an irreducible and aperiodic Markov chain.
The following theorem indicates the capacity region and throughput optimality of the JSQ-MW algorithm.
What we mean by the system being stabilized is that the queue lengths are bounded in steady state.

\begin{theorem}
\label{JSQMWTOPT}
The JSQ-MW algorithm can stabilize the system with three levels of data locality, as long as the arrival rate vector of the task types is strictly within the outer bound of the capacity region, $\Lambda$.
This means that $\Lambda$ is the capacity region of the system and the JSQ-MW algorithm is throughput optimal.
\end{theorem}

%\begin*{\textbf{proof:}}
\textbf{proof:}
An extension of the Foster-Lyapunov theorem, where the $T$-time slot drift of the Lyapunov function is studied, is used to prove the throughput optimality of the JSQ-MW algorithm.
We choose the function $V_1(t) = || \mathbold{Q}(t) ||^2 = \sum_{m = 1}^{M} Q_m^2(t)$ as the Lyapunov function. Note that this choice of the Lyapunov function satisfies the requirements of non-negativity, being equal to zero only at $\mathbold{Q}(t) = \bar{0}$, and going to infinity as any elements of $\mathbold{Q}(t)$ goes to infinity.
In Appendix \ref{PROOFJSQMWTOPT}, we show that as long as the arrival rate vector of task types is strictly within $\Lambda$, under the JSQ-MW algorithm, there exists an integer $T > 0$ where the expected $T$-time slot drift of $V_1(t)$ is negative outside of a bounded region of the state space, and is finite inside this bounded region.
Therefore, the fact that $\Lambda$ is the capacity region of the system and the throughput optimality of the JSQ-MW algorithm are followed by the extension of the Foster-Lyapunov theorem.
%\end*{proof}

A corollary of Theorem \ref{JSQMWTOPT} is that the outer bound of capacity region proposed in equation \eqref{capacityregion} is actually the capacity region of a system with three levels of data locality.

It was mentioned in Section \ref{JSQMW2} that the JSQ-MW algorithm is not heavy-traffic optimal for two levels of data locality, but minimizes the mean task completion time at high loads under a specific traffic scenario.
We should also mention that it is very rare that such a traffic load occurs in real-world applications.
In the simulation results in Chapter \ref{simresults}, it would be clear from the simulation results that the JSQ-MW algorithm is not heavy-traffic optimal for a system with three levels of data locality.
In the following, we will show the traffic scenario in which the JSQ-MW algorithm can minimize the mean task completion time at high loads for a system with three levels of data locality.

Under the following three conditions, the JSQ-MW algorithm minimizes the mean task completion time at high loads:
\begin{enumerate}
\item All the incoming traffic concentrates on a subset of racks, so the complement set of racks does not have any incoming local tasks.
\item The racks that receive nonzero incoming local tasks cannot process their incoming local tasks without getting help from other servers in the racks with no incoming local task.
\item The servers in the racks that receive local incoming tasks either have zero incoming local tasks or are overloaded.
\end{enumerate}
Here we formalize the traffic scenario in which the JSQ-MW algorithm is heavy-traffic optimal after setting some notations.
The set of racks that receive nonzero local tasks is shown by $\mathcal{O}$, and the racks belonging to this set are called overloaded racks.
The set of all servers that receive nonzero local tasks is denoted by $\mathcal{M}_l$.
It is clear that all the servers in the set $\mathcal{M}_l$ belong to the racks in the set $\mathcal{O}$.
The other set of servers in the overloaded racks that receive zero local tasks is shown by $\mathcal{M}_k = \{ m \in \mathcal{M} | K(m) \in \mathcal{O}, \text{and} \ m \notin \mathcal{M}_l \}$.
The set of all other servers not belonging to the overloaded racks that do not receive any local tasks is denoted by $\mathcal{M}_r = \{ m \in \mathcal{M} | K(m) \notin \mathcal{O} \}$.
Assuming a set of servers $\mathcal{S} \subset \mathcal{M}_l$, the set of all task types having a local server in the set $\mathcal{S}$ is denoted by $\mathcal{N}(\mathcal{S}) = \{ \bar{L} \in \mathcal{L} | \exists m \in \mathcal{S}, \text{s.t.} \ m \in \bar{L} \}$.
Likewise, for a set of racks $\mathcal{R} \subset \mathcal{O}$, the set of task types that have a local server in the set $\mathcal{R}$ is denoted by $\mathcal{N}(\mathcal{R}) = \{ \bar{L} \in \mathcal{L} | \exists m, \text{s.t.} \ K(m) \in \mathcal{R}, \text{and} \ m \in \bar{L} \}$.
Furthermore, the set of servers belonging to a rack in the set $\mathcal{R}$ that receive local incoming tasks (receive zero local task) is denoted by $\mathcal{M}_l^{\mathcal{R}} = \{ m \in \mathcal{M}_l | K(m) \in \mathcal{R} \}$ ($\mathcal{M}_k^{\mathcal{R}} = \{ m \in \mathcal{M}_k | K(m) \in \mathcal{R} \}$).

The heavy-traffic regime is characterized as follows.
Any subset of servers receiving local tasks are overloaded, and any subset of racks receiving local tasks are also overloaded.
However, the arrival rate vector of task types should be in the capacity region in order to be supportable, that is, $\sum_{\bar{L} \in \mathcal{L}} \lambda_{\bar{L}} < |\mathcal{M}_l| \alpha + |\mathcal{M}_k| \beta + |\mathcal{M}_r| \gamma$.
The following three conditions characterize a heavy-traffic regime with a parameter $\epsilon > 0$ which is the $L^1$-norm of the difference of the arrival rate vector and the nearest point on the boundary of the capacity region to the arrival rate vector.

\begin{equation}
\begin{aligned}
\label{HTC}
& \forall \mathcal{S} \subset \mathcal{M}_l, \ \ \sum_{\bar{L} \in \mathcal{N}(\mathcal{S})} \lambda_{\bar{L}} > |\mathcal{S}| \alpha, \\
& \forall \mathcal{R} \subset \mathcal{O}, \ \ \sum_{\bar{L} \in \mathcal{R}(\mathcal{S})} \lambda_{\bar{L}} > |\mathcal{M}_l^{\mathcal{R}}| \alpha + |\mathcal{M}_k^{\mathcal{R}}| \beta, \\
& \sum_{\bar{L} \in \mathcal{L}} \lambda_{\bar{L}} = |\mathcal{M}_l| \alpha + |\mathcal{M}_k| \beta + |\mathcal{M}_r| \gamma - \epsilon. \\
\end{aligned}
\end{equation}
Theorem \ref{JSQMWOPT} clarifies the heavy-traffic optimality of the JSQ-MW algorithm.

\begin{theorem}
\label{JSQMWOPT}
The JSQ-MW algorithm minimizes the mean task completion time in the steady state as long as the task arrival process $\{ A_{\bar{L}}^{\epsilon} (t), \ t \geq 0 \}_{\bar{L} \in \mathcal{L}}$ with arrival rate vector $\mathbold{\lambda}^{\epsilon}$ satisfies the conditions in \eqref{HTC}, where servers are either overloaded or do not receive any local tasks, and the racks are also either overloaded or do not receive any local tasks.
\end{theorem}

%\begin*{\textbf{proof:}}
\textbf{proof:}
The complete proof of Theorem \ref{JSQMWOPT} appears in Appendix \ref{PROOFJSQMWOPT}.
%\end*{proof}

As mentioned, the JSQ-MW algorithm is not heavy-traffic optimal in all loads.
The reason is that if any task is local to helper servers in overloaded or under-loaded racks, their local queues grow and local tasks receive service by unnecessary delay.
In the next section, we propose our novel algorithm called Balanced-Pandas, which is both throughput and heavy-traffic optimal.

\section{Balanced-Pandas}
\label{BP}
In order to implement the Balanced-Pandas algorithm, the central scheduler should keep three queues per server since using this algorithm the incoming tasks are not necessarily routed to the queue of their local server, so we want to keep track of the local, rack-local and remote tasks routed to a server by considering different queues for each of them.
The tasks routed to server $m$ which are local (rack-local or remote) to it are queued in the first (second or third) queue of the server, which is denoted by $Q_m^l$ ($Q_m^k$ or $Q_m^r$).
The three queue lengths of the $m$-th server at time slot $t$ are denoted by the vector notation $\bar{\mathbold{Q}}_m(t) = (Q_m^l(t), Q_m^k(t), Q_m^r(t))$, and the central scheduler maintains the vector of queue lengths $\bar{\mathbold{Q}}(t) = (\bar{\mathbold{Q}}_1(t), \bar{\mathbold{Q}}_2(t), \cdots, \bar{\mathbold{Q}}_M(t))$.
As the service time of local, rack-local, and remote tasks follow geometric distributions with means $\frac{1}{\alpha}, \frac{1}{\beta},$ and $\frac{1}{\gamma}$, respectively, the mean time needed for server $m$ to process all the tasks queued at $Q_m^l, Q_m^k,$ and $Q_m^r$ at time slot $t$ is as follows:
\[
W_m(t) = \frac{Q_m^l(t)}{\alpha} + \frac{Q_m^k(t)}{\beta} + \frac{Q_m^r(t)}{\gamma}.
\]
We call $W_m(t)$ the \textit{workload} on the $m$-th server.

\begin{figure}[h]
\centering
\includegraphics[scale=0.4]{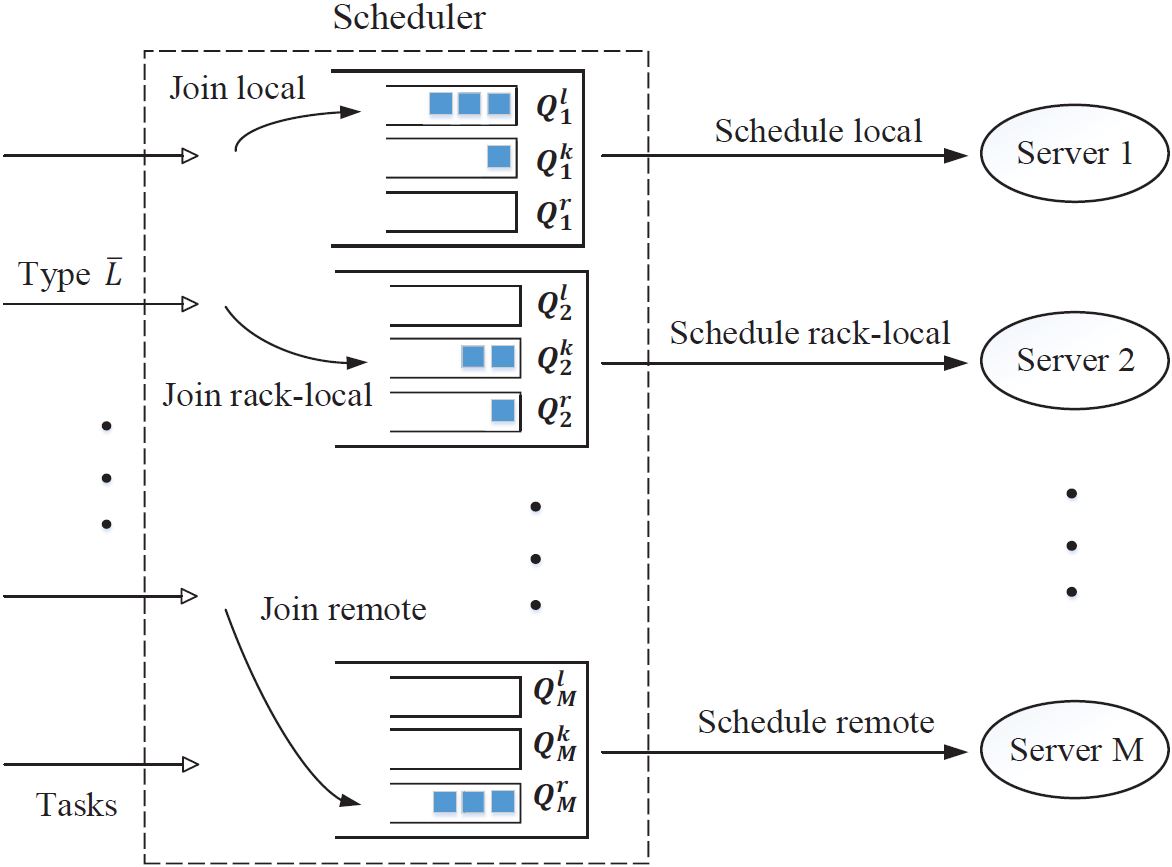}
\caption{The queueing structure needed for the Balanced-Pandas algorithm.}
\label{BPS}
\end{figure}

Figure \ref{BPS} along with the Balanced-Pandas routing and scheduling policies presented in the following make the queueing structure and the Balanced-Pandas algorithm clear.
\begin{itemize}
\item \textbf{Balanced-Pandas Routing (Weighted-Workload Routing):}
The routing decision for an incoming task of type $\bar{L}$ is based on both data locality and the workload on the servers. In order to decide routing for an incoming task, the workloads of servers local (rack-local or remote) to the incoming task are each divided by $\alpha$ ($\beta$ or $\gamma$).
The task is routed to the corresponding sub-queue ($Q_{m^*}^l, Q_{m^*}^k,$ or $Q_{m^*}^r$) of the server $m^*$ with the minimum weighted workload.
That is, if the incoming task is local, rack-local, or remote to the server with the minimum weighted workload, it is routed to the first, second, or third queue, respectively.
Formally speaking, an arriving task of type $\bar{L}$ is routed to the corresponding sub-queue of a server in the set below:
\[
\underset{m \in \mathcal{M}}{\arg \min} \bigg \{ \frac{W_m(t)}{\alpha} I_{\{ m \in \bar{L} \}}, \frac{W_m(t)}{\beta} I_{\{ m \in \bar{L}_k \}}, \frac{W_m(t)}{\gamma} I_{\{ m \in \bar{L}_r \}} \bigg \}.
\]
\item \textbf{Balanced-Pandas Scheduling (Prioritized Scheduling):} As server $m$ becomes idle at time $t^-$, the central scheduler assigns it to a local task queued at $Q_m^l$ at time slot $t$, if available.
However, if $Q_m^l(t) = 0$, server $m$ will be assigned to a rack-local task queued at $Q_m^k$, if available.
If both local and rack-local sub-queues of server $m$ are empty, the server is assigned to process a remote task queued at $Q_m^r$.
In other words, the highest priority to process a task for an idle server is given to local, then rack-local, and finally remote tasks queued in front of it.
If all sub-queues of the idle server are empty, the idle server remains idle until a new task joins any of the sub-queues.
\end{itemize}
In the following, we will first propose two optimality theorems of the Balanced-Pandas algorithm, then we will define the notations which will be used in the proof of these theorems in Appendices \ref{PROOFTTOBP} and \ref{PROOFBPHO}.

\begin{theorem}
\label{TTOBP}
The Balanced-Pandas algorithm stabilizes a system with three levels of data locality as long as the arrival rate is strictly inside the capacity region, which means that the Balanced-Pandas algorithm is throughput optimal.
\end{theorem}

%\begin{proof}
\textbf{proof:}
We use the Foster-Lyapunov theorem to prove the throughput optimality.
We use the $l^2$-norm of the workload vector of servers as the Lyapunov function:
\[
V_3(Z(t)) = || \mathbold{W}(t) ||^2.
\] \\
This choice of Lyapunov function is non-negative, is equal to zero just at $\mathbold{W}(t) = \bar{0}$, and goes to infinity as any elements of $\mathbold{W}(t)$ goes to infinity.
We show that there exists a finite integer $T > 0$ where the expectation of the $T$-time slot drift of the Lyapunov function is negative outside of a bounded region of the state space, and is positive and finite inside this bounded region.
We should note that in the proof of throughput optimality, we do not use the fact of using prioritized scheduling.
Therefore, to the purpose of throughput optimality, an idle server can serve any task in its three sub-queues as local, rack-local and remote tasks decrease the expected workload at the same rate.
The prioritized scheduling is to minimize the mean task completion time experienced by tasks which will be of interest in heavy-traffic optimality.
For the complete proof refer to Appendix \ref{PROOFTTOBP}
%\end{proof}

\begin{theorem}
\label{THOBP}
As long as $\beta^2 > \alpha \gamma$, the Balanced-Pandas algorithm is heavy-traffic optimal, i.e., minimizes the mean task completion time as the arrival rate vector of task types approaches the boundary of the capacity region.
\end{theorem}

%\begin{proof}
\textbf{proof:}
For the complete proof look at Appendix \ref{PROOFBPHO}.
%\end{proof}

In the next three subsections, the queue dynamics when the Balanced-Pandas algorithm is used, overloaded servers and racks, and ideal load decomposition will be discussed which will be used in the proofs of Theorems \ref{TTOBP} and \ref{THOBP}.

\subsection{Queue Dynamics}
Recall that $A_{\bar{L}, m}(t)$ denotes the number of type $\bar{L}$ tasks that are routed to the $m$-th queue, and $\bar{L}, \bar{L}_k,$ and $\bar{L}_r$ denote the set of local, rack-local and remote servers to a task of type $\bar{L}$.
Using these definitions, we can formalize the local, rack-local, and remote tasks routed to the three sub-queues of the $m$-th server at time slot $t$ denoted by $A_m^l(t), A_m^k(t),$ and $A_m^r(t)$, respectively as follows: $A_m^l(t) = \sum_{\bar{L}: m \in \bar{L}} A_{\bar{L}, m}(t), A_m^k(t) = \sum_{\bar{L}: m \in \bar{L}_k} A_{\bar{L}, m}(t),$ and $A_m^r(t) = \sum_{\bar{L}: m \in \bar{L}_r} A_{\bar{L}, m}(t)$.

Remembering the Markov chain defined in Section \ref{JSQMW}, the queue dynamics themselves cannot form a Markov chain alone.
Therefore, we define the working status of server $m$ at time slot $t$ as follows:

\[
f_m(t) = \begin{cases}
               -1 \ \ \ \ \text{if server $m$ is idle} \\
               \ 0 \ \ \ \ \ \text{if server $m$ processes a local task from $Q_m^l$} \\
			   \ 1 \ \ \ \ \ \text{if server $m$ processes a rack-local task from $Q_m^k$} \\
			   \ 2 \ \ \ \ \ \text{if server $m$ processes a remote task from $Q_m^r$}
         \end{cases}.
\] \\
When server $m$ is done processing a task at time slot $t-1$, so its working status is $f_m(t^-) = -1$, the scheduling decision $\eta_m(t)$ for this server is made based on both working status of servers $\mathbold{f}(t) = (f_1(t), f_2(t), \cdots, f_M(t))$ and the queue length vector $\bar{\mathbold{Q}}(t)$.
Note that $\eta_m(t) = f_m(t)$ as long as server $m$ is busy processing a task, and when the server becomes idle at the end of a time slot, $\eta_m(t)$ is determined by the scheduling policy.

As described in Chapter \ref{introduction}, in a system with three levels of data locality the service distributions are as follows.
If server $m$ is working on a local (rack-local, or remote) task at time slot $t$, the service provided by the server at time slot $t$ follows a Bernoulli random variable with mean $\alpha$ ($\beta,$ or $\gamma$) which is denoted by $S_m^l(t)$ ($S_m^k(t),$ or $S_m^r(t)$).
In other words, the local (rack-local, or remote) service provided by the $m$-th server at time slot $t$ is $S_m^l(t) \sim Bern(\alpha I_{\{ \eta_m(t) = 0 \}})$ ($S_m^k(t) \sim Bern(\beta I_{\{ \eta_m(t) = 1 \}}),$ or $S_m^r(t) \sim Bern(\gamma I_{\{ \eta_m(t) = 2 \}})$).

Defining the unused service of server $m$ as $U_m(t) = \max \{ 0, S_m^r(t) - A_m^r(t) - Q_m^r(t) \}$, the three sub-queues of server $m$ evolve as follows:

\begin{equation}
\begin{aligned}
\label{queueevolution}
& Q_m^l(t + 1) = Q_m^l(t) + A_m^l(t) - S_m^l(t), \\
& Q_m^k(t + 1) = Q_m^k(t) + A_m^k(t) - S_m^k(t), \\
& Q_m^r(t + 1) = Q_m^r(t) + A_m^r(t) - S_m^r(t) + U_m(t).
\end{aligned}
\end{equation} \\
The service times are all geometrically distributed, so the queue length vector $\bar{\mathbold{Q}}(t)$ together with the working status vector of servers $\mathbold{f}(t)$ form the irreducible and aperiodic Markov chain ($\{ \mathbold{Z}(t) = (\bar{\mathbold{Q}}(t), \mathbold{f}(t)), t \geq 0 \}$).

\subsection{Overloaded Servers and Racks}
Server $m$ is overloaded if it cannot process the local tasks that are routed to it without the help of other servers, and its local load cannot be distributed with under-loaded servers by load balancing.
In order to describe the overloaded servers formally, we define the following notation:

\[
\psi_n = \sum_{\bar{L}: n \in \bar{L}} \sum_{m = 1}^{M} \lambda_{\bar{L}, n, m} \ \ \ \forall n \in \mathcal{M}.
\] \\
$\psi_n$ is the pseudo-arrival rate of type $\bar{L}$ tasks routed to server $n$ under the task types' arrival rates $\{ \lambda_{\bar{L}, n, m} \}$.
Server $n$ is overloaded under the load decomposition $\{ \lambda_{\bar{L}, n, m} \}$ if $\psi_n > \alpha$.
For a subset of servers $\mathcal{S} \subset \mathcal{M}$, we define $\mathcal{L}_{\mathcal{S}}$ as the set of task types only local to servers of the set $\mathcal{S}$, that is, $\mathcal{L}_{\mathcal{S}} = \{ \bar{L} \in \mathcal{L} | \bar{L} \subset \mathcal{S} \}$.
On the other hand, for the same set $\mathcal{S}$, we define $\mathcal{L}_{\mathcal{S}}^*$ as the set of task types that are local to at least one server of the set $\mathcal{S}$, that is, $\mathcal{L}_{\mathcal{S}}^* = \{ \bar{L} \in \mathcal{L} | \bar{L} \cap \mathcal{S} \neq \varnothing \}$.
Lemma \ref{lemmaOLS} shows that there exists a load decomposition under which the truly overloaded set of servers, denoted by $\mathcal{D}$, do not receive any task type that has at least one local task out of this set $\mathcal{D}$, that is tasks that are local to a server out of the set $\mathcal{D}$ are routed to under-loaded servers which are out of set $\mathcal{D}$.

\begin{lemma}
\label{lemmaOLS}
There exists a load decomposition $\{ \widetilde{\lambda}_{\bar{L}, n, m} \}$ for any arrival rate vector $\mathbold{\lambda} \in \bar{\Lambda}$ that satisfies the following two conditions \cite{xie2015priority}:
\begin{equation}
\begin{aligned}
\label{condition1}
\sum_{\bar{L}: m \in \bar{L}} \sum_{n: n \in \bar{L}} \frac{\widetilde{\lambda}_{\bar{L}, n, m}}{\alpha} + \sum_{\bar{L}: m \in \bar{L}_k} \sum_{n: n \in \bar{L}} \frac{\widetilde{\lambda}_{\bar{L}, n, m}}{\beta} + \sum_{\bar{L}: m \in \bar{L}_r} \sum_{n: n \in \bar{L}} \frac{\widetilde{\lambda}_{\bar{L}, n, m}}{\gamma} < 1, \forall m \in \mathcal{M}, \\
\end{aligned}
\end{equation}
\begin{equation}
\begin{aligned}
\label{condition2}
\forall n \in \mathcal{D} = \{ n \in \mathcal{M} : \psi_n \geq \alpha \}, \ \widetilde{\lambda}_{\bar{L}, n, m} = 0, \ \forall \bar{L} \notin \mathcal{L}_{\mathcal{D}}, \ \forall m \in \mathcal{M}.
\end{aligned}
\end{equation}
\end{lemma}

%\begin{proof}
\textbf{proof:}
This lemma is lent from Lemma $2$ in \cite{xie2015priority}.
For any arrival rate vector $\mathbold{\lambda}$ in the capacity region, there exists a load decomposition $\{ \lambda_{\bar{L}, n, m} \}$ that satisfies \eqref{condition1}.
The proof iteratively refines the load decomposition $\{ \lambda_{\bar{L}, n, m} \}$ as follows.
In each iteration, an appropriate amount of the local load to both temporary overloaded and under-loaded servers that are routed to overloaded servers are moved to under-loaded servers.
What we mean by moving an appropriate amount of local load from an overloaded server to a local under-loaded server is that we move the shared load until either both servers become overloaded or under-loaded, or there is no more shared local load between them to move.
By such load movements, the load on the whole system reduces, so \eqref{condition1} still holds for the new load decomposition in each iteration.
We continue the load movements from overloaded servers to under-loaded ones until there is no more load local to both kinds of servers that are routed to overloaded servers.
We name the ultimate local decomposition $\{ \widetilde{\lambda}_{\bar{L}, n, m} \}$.
For more details refer to the proof provided in the Appendix $A$, Section $7.1$ in \cite{xie2015priority}.
%\end{proof}

A rack is overloaded if the under-loaded servers in the rack cannot process the whole extra load on the overloaded servers in the same rack, and the local load on this rack cannot be distributed on other servers in under-loaded racks by load balancing.
For a subset of racks, $\mathcal{R} \subset \mathcal{K}$, we define $\mathcal{L}_{\mathcal{R}}$ as the set of task types that are only local to the servers in this set of racks, that is, $\mathcal{L}_{\mathcal{R}} = \{ \bar{L} \in \mathcal{L} | \forall m \in \bar{L}, K(m) \in \mathcal{R} \}$.
Formally, rack $k$ is overloaded under a load decomposition $\{ \lambda_{\bar{L}, n, m} \}$ if the following inequality holds:

\begin{equation}
\label{OLRC}
\sum_{m: K(m) = k, \psi_m \geq \alpha} (\psi_m - \alpha) \geq \beta \sum_{n: K(n) = k, \psi_n < \alpha} (1 - \frac{\psi_n}{\alpha}).
\end{equation}

The left-hand side of the above inequality is the extra load on overloaded servers of the overloaded rack $k$ that cannot be served locally, and the right-hand side is the maximum rack-local service that can be afforded by the under-loaded servers in the rack.
Therefore, if \eqref{OLRC} holds for rack $k$, the rack is overloaded and needs to receive remote service in order not to overflow.
The following lemma for overloaded racks is an equivalent to Lemma \ref{lemmaOLS} for overloaded servers.

\begin{lemma}
\label{lemmaOLR}
Assuming $\beta^2 > \alpha \gamma$, there exists a load decomposition $\{ \hat{\lambda}_{\bar{L}, n, m} \}$ for any arrival rate vector $\mathbold{\lambda} \in \bar{\Lambda}$ that satisfies not only conditions \eqref{condition1} and \eqref{condition2}, but also the following condition. Note that $\mathcal{O}$ is the set of overloaded racks satisfying equation \eqref{OLRC} under the load decomposition $\{ \hat{\lambda}_{\bar{L}, n, m} \}$.

\begin{equation}
\begin{aligned}
\label{condition3}
\forall n \ \text{s.t.} \ K(n) \in \mathcal{O}, \ \hat{\lambda}_{\bar{L}, n, m} = 0, \ \forall \bar{L} \notin \mathcal{L}_{\mathcal{O}}, \ \forall m \in \mathcal{M}.
\end{aligned}
\end{equation}
\end{lemma}

Analogous to overloaded servers, an overloaded rack only receives task types that are only local to the servers in the overloaded set of racks.

%\begin{proof}
\textbf{proof:}
The proof is similar to proof of Lemma \ref{lemmaOLS}.
Starting from the load decomposition $\{ \widetilde{\lambda}_{\bar{L}, n, m} \}$ that satisfies both conditions \eqref{condition1} and \eqref{condition2}, the load local to both overloaded and under-loaded racks that are routed to overloaded racks are iteratively moved to under-loaded ones.
Note that this load can be moved from beneficiary servers of overloaded racks to beneficiary servers of under-loaded racks, or from helper servers of overloaded racks to helper servers of under-loaded racks.
By moving the load from under-loaded servers in overloaded racks to under-loaded servers in under-loaded racks, there is a possibility that the under-loaded servers in under-loaded racks become overloaded.
By moving $\Delta$ amount of traffic from $\mathcal{H}_o$ to $\mathcal{H}_u$ (where $\mathcal{H}_u$ is about to become $\mathcal{B}_u$), the added rack-local load on the under-loaded rack is $\frac{\Delta}{\beta}$ which means that the reduced amount of remote load on the under-loaded rack is $\gamma \frac{\Delta}{\beta}$.
On the other hand, be removing $\Delta$ amount of traffic from $\mathcal{H}_o$, $\mathcal{H}_o$ can process additional rack-local traffic of $\beta \frac{\Delta}{\alpha}$.
This load movement reduces traffic on the whole system in case that $\beta \frac{\Delta}{\alpha} > \gamma \frac{\Delta}{\beta}$, which is equivalent to $\beta^2 > \alpha \gamma$.
In summary, the condition $\beta^2 > \alpha \gamma$ dictates that any load local to both overloaded and under-loaded racks should be routed to the under-loaded racks regardless of the load on servers.
The load movement from an overloaded rack to an under-loaded one continues until both racks become overloaded or under-loaded, or no task local to both ones is routed to the overloaded rack.
After load movements, some overloaded racks may become under-loaded or some under-loaded racks may become overloaded.
By such load movements, the overall load on system reduces and both conditions \eqref{condition1} and \eqref{condition2} hold for the obtained load decomposition.
%The load movement continues until no task local to both overloaded and under-loaded racks that is routed to overloaded racks is remained.
We call the ultimate load decomposition $\{ \hat{\lambda}_{\bar{L}, n, m} \}$ that satisfies all the conditions in Lemma \ref{lemmaOLR}.
%\end{proof}

\subsection{Ideal Load Decomposition}
\begin{figure}[h]
\centering
\includegraphics[scale=0.27]{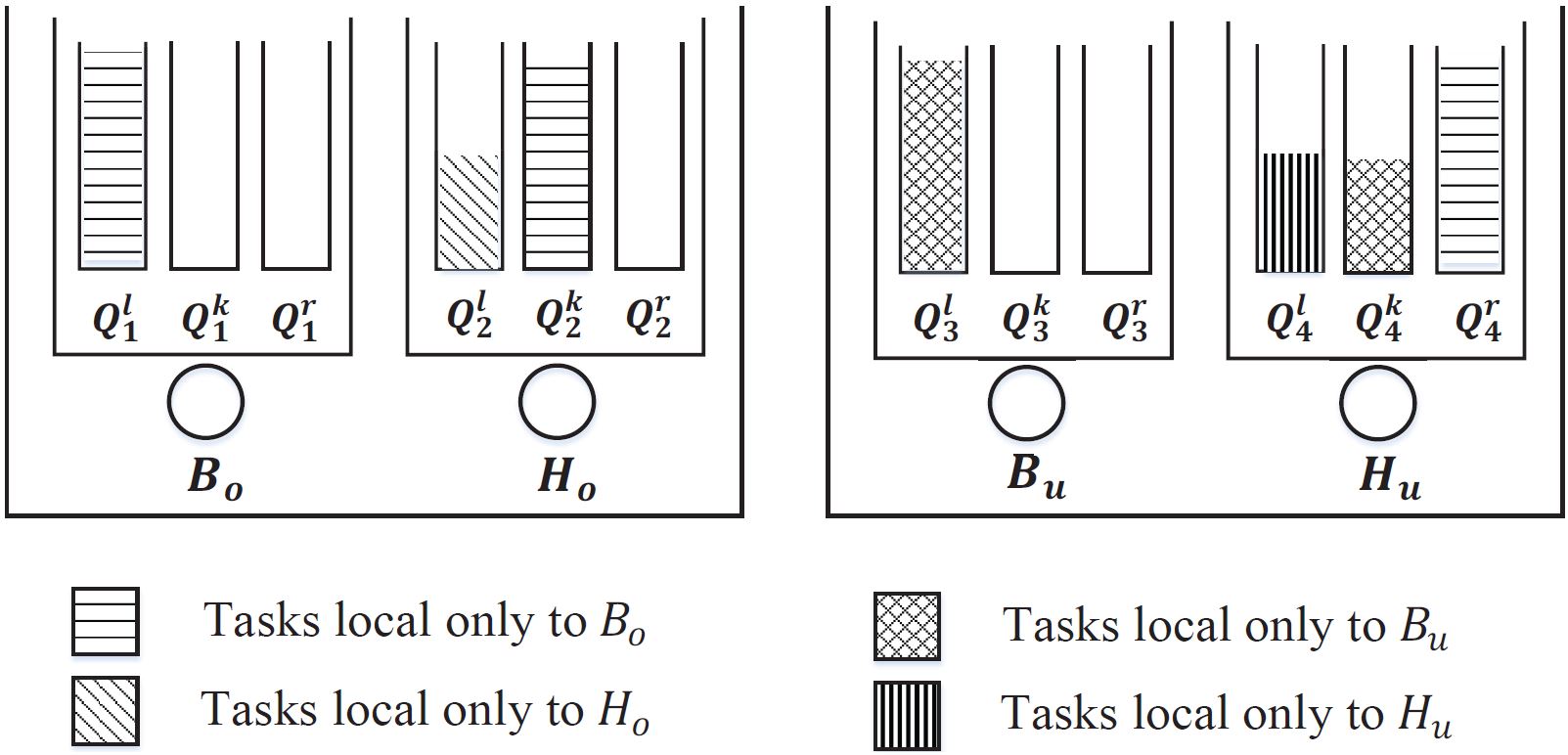}
\caption{The four types of servers and their load under the ideal load decomposition.}
\label{FTS}
\end{figure}

Under an arrival rate vector, servers can be classified into four types: helper servers in under-loaded racks, beneficiary servers in under-loaded racks, helper servers in overloaded racks, and beneficiary servers in overloaded racks.
The definition of these four types of servers is as follows:
\begin{itemize}
\item Helpers in under-loaded racks ($\mathcal{H}_u$): The set of under-loaded servers that are in under-loaded racks form the set $\mathcal{H}_u$.
The tasks local to this set of servers all receive service locally. The remaining capacity of this set of servers is scheduled for processing rack-local and remote tasks.
\item Beneficiaries in under-loaded racks ($\mathcal{B}_u$): The set of overloaded servers that are in under-loaded racks form the set $\mathcal{B}_u$.
The tasks local to this set of servers all receive service locally or rack-locally, but not remotely. The servers in this set only process local tasks, not rack-local or remote tasks.
\item Helpers in overloaded racks ($\mathcal{H}_o$): The set of under-loaded servers that are in overloaded racks form the set $\mathcal{H}_o$.
The tasks local to this set of servers all receive service locally. The remaining capacity of this set of servers is scheduled for processing only rack-local tasks, not remote tasks.
\item Beneficiaries in overloaded racks ($\mathcal{B}_o$): The set of overloaded servers that are in overloaded racks form the set $\mathcal{B}_o$.
The tasks local to this set of servers receive service locally, rack-locally, or remotely. The servers in this set only process local tasks, not rack-local or remote tasks.
\end{itemize}
Figure \ref{FTS} depicts the four types of servers and their load under ideal load decomposition.

Unless no real helper or beneficiary servers, under-loaded or overloaded racks exist in a real system, we will use this concept in the heavy-traffic optimality proof.
The following lemma formalizes the definition of four types of servers.

\begin{lemma}
\label{lemmaILD}
Assuming $\beta^2 > \alpha \gamma$, there exists a load decomposition $\{ \lambda_{\bar{L}, n, m}^* \}$ for any arrival rate vector $\mathbold{\lambda} \in \Lambda$ that satisfies conditions \eqref{condition1}, \eqref{condition2}, and \eqref{condition3} in Lemmas \ref{lemmaOLS} and \ref{lemmaOLR}, and under this load decomposition any server belongs to one of the four types described below.
Note that $\mathcal{O}$ and $\mathcal{U}$ stand for the set of overloaded and under-loaded racks, respectively.

\[
\begin{aligned}
\mathcal{H}_u & = \{ n: K(n) \in \mathcal{U} | \psi_n < \alpha, \text{and} \ \forall \bar{L} \in \mathcal{L}, \forall m \neq n, \lambda_{\bar{L}, n, m}^* = 0 \}, \\
\mathcal{B}_u & = \{ n: K(n) \in \mathcal{U} | \psi_n \geq \alpha, \text{and} \ \forall \bar{L} \in \mathcal{L}, \forall m \neq n, \lambda_{\bar{L}, m, n}^* = 0, \\
& \ \ \ \ \ \ \ \ \ \ \ \ \ \ \text{and} \ \forall \bar{L} \in \mathcal{L}, \forall m \text{ s.t.} \ K(m) \neq K(n), \lambda_{\bar{L}, n, m}^* = 0 \}, \\
\mathcal{H}_o & = \{ n: K(n) \in \mathcal{O} | \psi_n < \alpha, \text{and} \ \forall \bar{L} \in \mathcal{L}, \forall m \neq n, \lambda_{\bar{L}, n, m}^* = 0, \\
& \ \ \ \ \ \ \ \ \ \ \ \ \ \ \text{and} \ \forall \bar{L} \in \mathcal{L}, \forall m \text{ s.t.} \ K(m) \neq K(n), \lambda_{\bar{L}, m, n}^* = 0 \}, \\
\mathcal{B}_o & = \{ n: K(n) \in \mathcal{O} | \psi_n \geq \alpha, \text{and} \ \forall \bar{L} \in \mathcal{L}, \forall m \neq n, \lambda_{\bar{L}, m, n}^* = 0 \}.
\end{aligned}
\]
\end{lemma}

%\begin{proof}
\textbf{proof:}
In order to achieve the ideal load decomposition $\{ \lambda_{\bar{L}, n, m}^* \}$ that satisfies the conditions in Lemma \ref{lemmaILD}, we start from the load decomposition $\{ \hat{\lambda}_{\bar{L}, n, m} \}$ which satisfies conditions \eqref{condition1}, \eqref{condition2}, and \eqref{condition3}.
The following four steps should be taken to achieve the ideal load decomposition:
\begin{enumerate}
\item If server $n$ is an under-loaded server in an under-loaded rack, and $\hat{\lambda}_{\bar{L}, n, m} \neq 0$, where $m \neq n$, we move this load to be scheduled locally at server $n$.
This way, the local load to the under-loaded servers in under-loaded racks which were scheduled to be served rack-locally or remotely will be served locally, so the load on the whole system decreases (the rack-local or remote load on server $n$ may be required to be rescheduled to other servers with removed load).
\item Under the updated load decomposition in the previous step, we offload any rack-local or remote load on any overloaded server $n$ in an under-loaded rack. Hence, server $n$ is only scheduled to process its local load.
This way, there would be empty capacity on the servers that used to serve the overloaded load of server $n$. This empty capacity can be used for the previous rack-local and remote load that were being processed by server $n$.
On the other hand, if local load to server $n$ receive service remotely, it can be scheduled to under-loaded servers in the same rack of server $n$ to receive service rack-locally.
All these load movements reduce the load on the whole system.
\item Under the updated load decomposition in step 2, if the load local to an under-loaded server $n$ in an overloaded rack receive rack-local or remote service, we reschedule it to be processed in its local server $n$ (the rack-local or remote load on server $n$ may be required to be rescheduled to other servers with removed load).
Furthermore, we remove any remote load on server $n$ to make more space for rack-local load of overloaded servers in the same rack of server $n$ which used to receive service remotely.
By these load adjustments, the overall load decreases on the whole system.
\item Under the updated load decomposition of step 3, the rack-local or remote load scheduled to overloaded servers in overloaded racks should be removed. Instead, local loads to these servers should be assigned to them.
This way, the required remote service of these servers decreases more than the remote load that was removed from them.
Hence, the overall load on the whole system decreases under this load movement.
\end{enumerate}

%\end{proof}

%% file: simulationresults.tex
\chapter{Simulation Results}
\label{simresults}
The performances of FCFS scheduler which is the Hadoop's default scheduler, and Hadoop Fair Scheduler (HFS) are studied against the JSQ-MW algorithm in a system with two levels of data locality in \cite{wang2016maptask}.
As FCFS scheduler does not take the data locality into account, it performs worst than other algorithms like the JSQ-MW algorithm, specially at high loads. Hence, performance of FCFS is not given in the analysis.
In this chapter, we compare three algorithms, the Balanced-Pandas algorithm, the JSQ-MaxWeight algorithm, and the Pandas algorithm implemented on a system with three levels of data locality through simulation.
The configuration of the simulated system is as follows: we assume a continuous time system consisting of $10$ racks ($K = 10$), each of which consists of $50$ servers, that is $M = 500$.
The task arrival follows Poisson process, and the service time for a local, rack-local or remote task follows exponential distribution with rate $\alpha = 1, \beta = 0.9,$ or $\gamma = 0.5$, respectively.
The two times slowdown service for a remote task is consistent to the measurements in \cite{zaharia2010delay}.
In our simulation environment, the three local servers to a task (the task type) is determined at the task's arrival among a set of servers uniformly at random.
The set of servers among which the local servers are chosen determines the load on the system.
We investigate two traffic scenarios as follows:
\begin{enumerate}
\item In this traffic scenario, all the incoming task have their data chunks stored in three servers that are uniformly selected among the first five racks.
This means that, the incoming load is uniformly distributed over all the $250$ servers in the first five racks.
If the mean arrival rate $\lambda \equiv \sum_{\bar{L}} \frac{\lambda_{\bar{L}}}{M}$ is larger than or equal to $0.5$, the first $250$ servers in the first five racks are beneficiaries, and the first five racks are overloaded.
The rest of the servers are helpers, and their five corresponding racks are under-loaded.
The JSQ-MW algorithm achieves heavy-traffic optimality under this specific load.
The Balanced-Pandas algorithm is also heavy-traffic optimal in all loads.
Therefore, both algorithms achieve the minimum mean task completion time in this traffic scenario.
Figure \ref{HTOF} affirms the above statement.
\begin{figure}[t]
\centering
\includegraphics[scale=0.301]{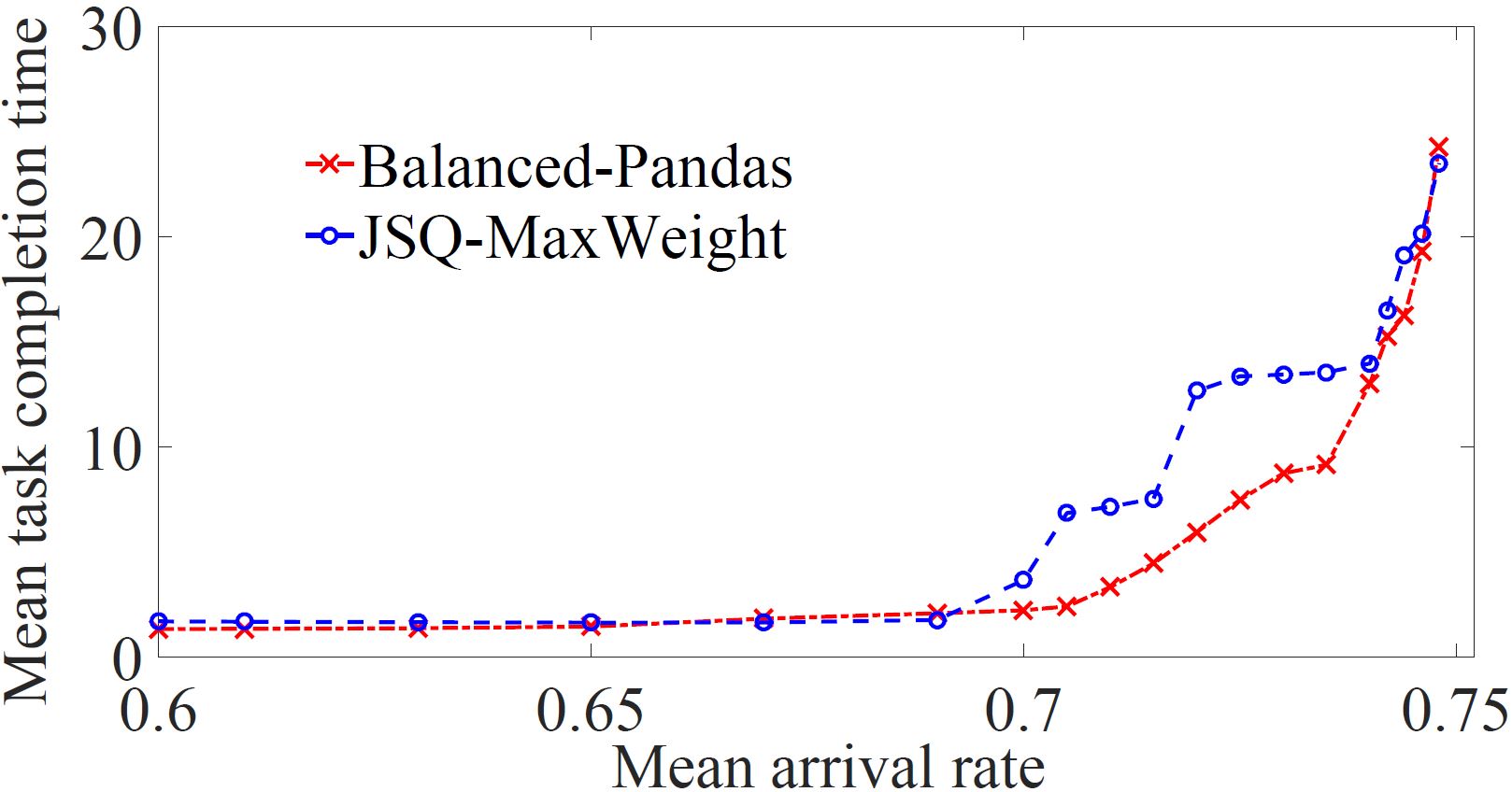}
\caption{The mean task completion time versus the mean arrival rate for two algorithms, the Balanced-Pandas algorithm and the JSQ-MW algorithm, under a load that both algorithms minimize the mean task completion time at high loads.}
\label{HTOF}
\end{figure}
\item Under this load, $20$ percent of the arriving tasks have their three local servers chosen uniformly at random from the first $10$ servers of the first rack, and six percent of the incoming tasks have their three local servers chosen uniformly at random from the first $25$ servers in the second rack.
All the other $74$ percent of the incoming tasks have their three local servers chosen uniformly at random from the rest of $465$ servers in the system.
This way, at high loads, the first 10 servers in the first rack and the 25 first servers in the second rack are beneficiaries, and the rest of servers are helpers.
The first rack is overloaded and the rest of racks are under-loaded at high loads.
Therefore, all four kinds of servers exist in the system under this traffic scenario at high loads.
The mean task completion time of three algorithms is shown in Figure \ref{TOF}.
\begin{figure}[h]
\centering
\includegraphics[scale=0.301]{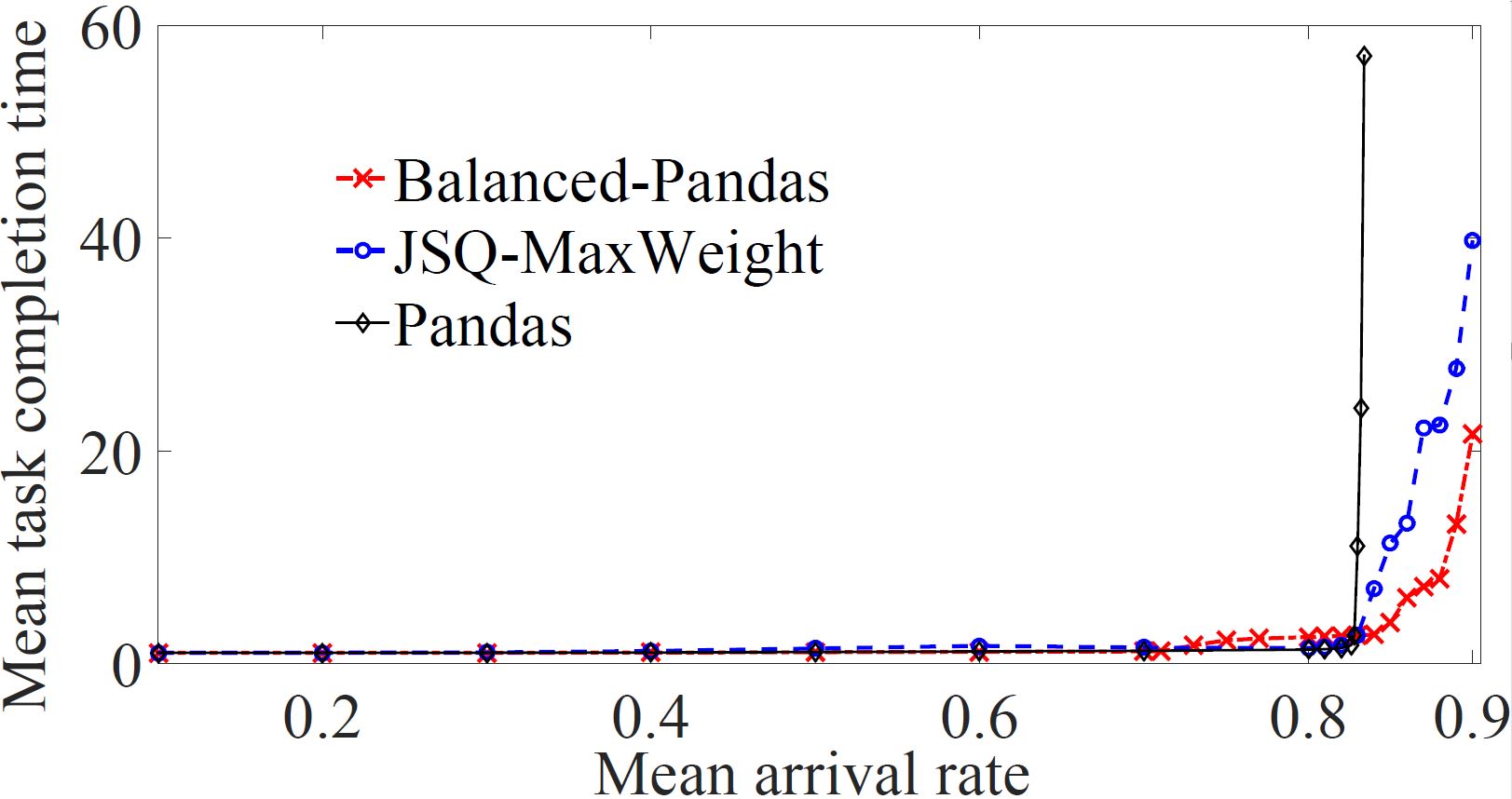}
\caption{The mean task completion time versus the mean arrival rate for three algorithms the Balanced-Pandas, JSQ-MW, and Pandas algorithms under a general load that all four kinds of servers exist in the system.}
\label{TOF}
\end{figure}
As Figure \ref{TOF} affirms, the Pandas algorithm is not throughput optimal as there exists other algorithms that can stabilize the system at higher loads.
Calculating the capacity region, the system is stabilizable as long as $\lambda < 0.9027$.
Both the Balanced-Pandas and JSQ-MW  algorithms stabilize the system in this capacity region, but the Pandas algorithm makes the system unstable at load $\lambda \approx 0.83$, so the Pandas algorithm is not throughput optimal.
Taking a more careful look at high loads, Figure \ref{HLHTOF} shows a significant up to fourfold outperformance of the Balanced-Pandas algorithm compared to the JSQ-MW algorithm.
This fact affirms that the JSQ-MW algorithm is not a heavy-traffic optimal algorithm.

\begin{figure}[h]
\centering
\includegraphics[scale=0.301]{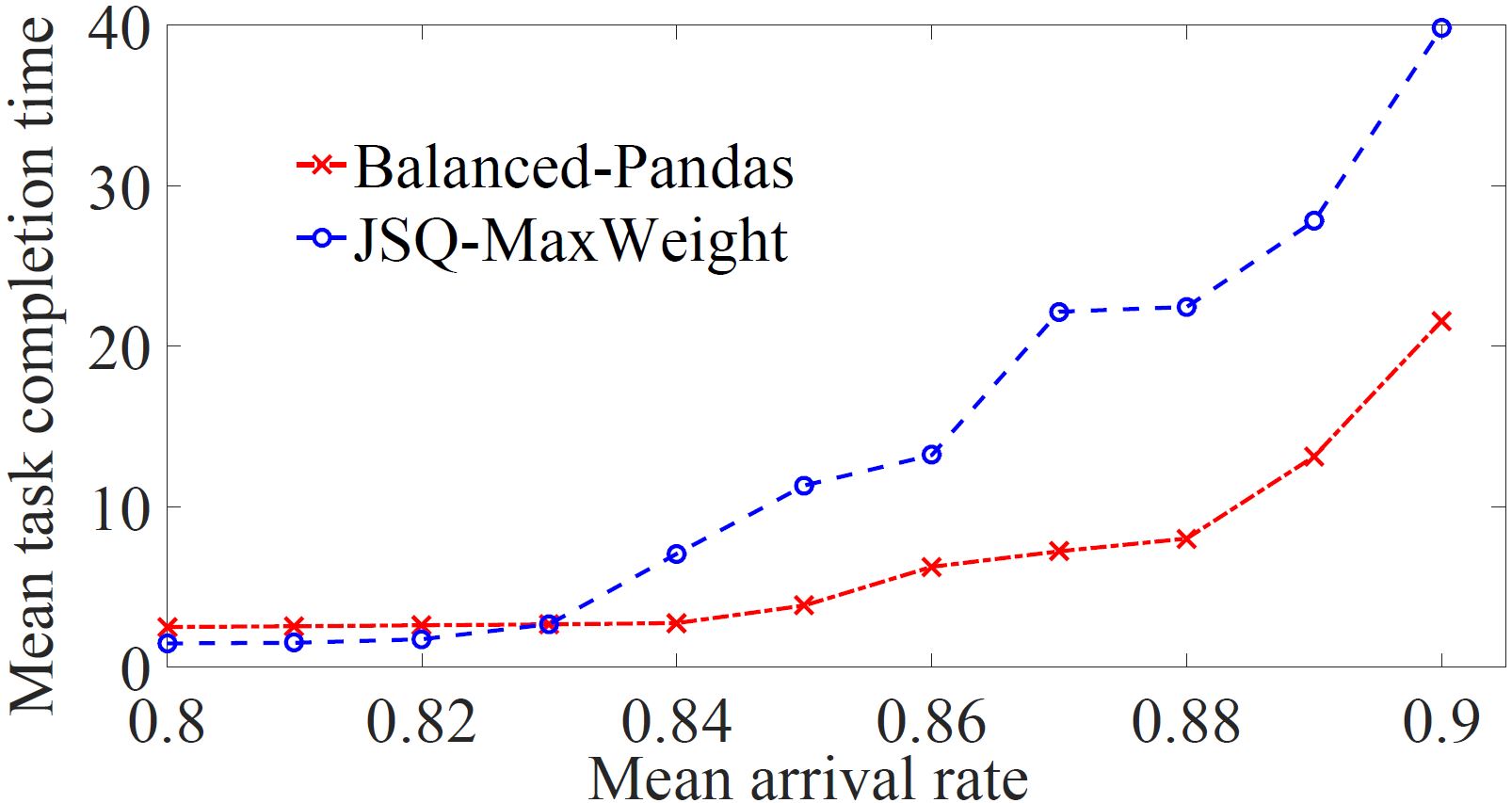}
\caption{The performance of the JSQ-MW  algorithm versus the Balanced-Pandas algorithm at high loads.}
\label{HLHTOF}
\end{figure}
\end{enumerate}

%% file: AppendixA.tex
\begin{appendices}
\chapter{Theorem Proofs}
\section{Proof of Theorem \ref{JSQMWTOPT}}
\label{PROOFJSQMWTOPT}
We prove that the JSQ-MW algorithm stabilizes the system as long as the arrival rate vector is strictly inside the outer bound of the capacity region.
This means that the outer bound $\bar{\Lambda}$ is the capacity region and the JSQ-MW algorithm is a throughput optimal algorithm.
Assume that $\mathbold{\lambda} \in \bar{\Lambda}$, then there exists $\delta > 0$ such that $\mathbold{\lambda}(1 + \delta) = \mathbold{\lambda}^{'} \in \bar{\Lambda}$.
As $\mathbold{\lambda}^{'} \in \bar{\Lambda}$ there exists a load decomposition for $\mathbold{\lambda}^{'}$, $\{ \lambda_{\bar{L}, n, m}^{'} \}$, such that it satisfies the conditions in \eqref{ECRR} specifically the following:
\[
\sum_{\bar{L}: m \in \bar{L}} \sum_{n: n \in \bar{L}} \frac{\lambda^{'}_{\bar{L}, n, m}}{\alpha} + \sum_{\bar{L}: m \in \bar{L}_k} \sum_{n: n \in \bar{L}} \frac{\lambda^{'}_{\bar{L}, n, m}}{\beta} + \sum_{\bar{L}: m \in \bar{L}_r} \sum_{n: n \in \bar{L}} \frac{\lambda^{'}_{\bar{L}, n, m}}{\gamma} < 1, \ \forall m.
\] \\
By our choice of arrival rate vector, we have the following:
\[
\{ \lambda_{\bar{L}, n, m}, \ \forall \bar{L} \in \mathcal{L}, \ \forall  n, m \in \mathcal{M} \} = \{ \frac{\lambda^{'}_{\bar{L}, n, m}}{1 + \delta}, \ \forall \bar{L} \in \mathcal{L}, \ \forall n, m \in \mathcal{M} \}.
\] \\
Hence, we conclude the following:
\[
\sum_{\bar{L}: m \in \bar{L}} \sum_{n: n \in \bar{L}} \frac{\lambda_{\bar{L}, n, m}}{\alpha} + \sum_{\bar{L}: m \in \bar{L}_k} \sum_{n: n \in \bar{L}} \frac{\lambda_{\bar{L}, n, m}}{\beta} + \sum_{\bar{L}: m \in \bar{L}_r} \sum_{n: n \in \bar{L}} \frac{\lambda_{\bar{L}, n, m}}{\gamma} < \frac{1}{1 + \delta}, \ \forall m.
\] \\
We define the pseudo arrival rate vector of servers, $\mathbold{\psi} = (\psi_1, \psi_2, \cdots, \psi_M)$, as follows:
\begin{equation}
\label{psi}
\psi_n = \sum_{\bar{L}: n \in \bar{L}} \sum_{m = 1}^{M} \lambda_{\bar{L}, n, m}, \ \forall n.
\end{equation} \\
We use $\mathbold{\psi}$ as an intermediary to prove this theorem.
In the proof, we will use the following three lemmas where the first two lemmas are analogous to Lemmas $2$ and $3$ in \cite{wang2016maptask}.
We eliminate the proofs of the first two lemmas as they mostly do not change for a system with three levels of data locality other than for a system with two levels of data locality.

\begin{lemma}
\label{b00}
For any arrival rate vector strictly inside the capacity region, $\mathbold{\lambda} \in \bar{\Lambda}$, and its corresponding pseudo arrival rate vector of servers $\mathbold{\psi}$ defined in \eqref{psi}, under the Joining the Shortest Queue routing policy we have the following inequality:
\[
\mathbb{E}[\langle \mathbold{Q}(t), \mathbold{A}(t) \rangle - \langle \mathbold{Q}(t), \mathbold{\psi} \rangle | Z(t_0)] \leq 0, \ \forall t_0, \text{and} \ \forall t \geq t_0.
\]
\end{lemma}

\begin{lemma}
\label{b0}
For any arrival rate vector strictly inside the capacity region, $\mathbold{\lambda} \in \bar{\Lambda}$, and its corresponding pseudo arrival rate vector of servers $\mathbold{\psi}$ defined in \eqref{psi}, under MaxWeight scheduling policy we have the following inequality:
\[
\begin{aligned}
& \forall T > T_1, \text{and} \ \forall t_0, \exists T_1 > 0 \text{ such that}, \\
& \mathbb{E} \bigg [ \sum_{t = t_0}^{t_0 + T - 1} \bigg ( \langle \mathbold{Q}(t), \mathbold{\psi} \rangle - \langle \mathbold{Q}(t), \mathbold{S}(t) \rangle \bigg ) \bigg | Z(t_0) \bigg ] \leq - \theta_1 || \mathbold{Q}(t_0)||_1 + c_1,
\end{aligned}
\] \\
where the constants $\theta_1 > 0$ and $c_1$ are independent from $Z(t_0)$.
\end{lemma}

\begin{lemma}
\label{b1}
\[
\langle \mathbold{Q}(t), \mathbold{U}(t) \rangle < M^2, \ \forall t.
\]
\end{lemma}

%\begin{proof}
\textbf{proof:}
If $U_m(t) > 0$, then it implies that $Q_m(t) < M$.
The reason is that queue $m$ can receive at most $M$ services at a time slot, so if there exists any unused services, the queue length should have been less than the whole services which is $M$.
If $U_m(t) = 0$, then $Q_m(t) \times U_m(t) = 0$.
Therefore, $Q_m(t) \times U_m(t) < M \times U_m(t)$.
On the other hand it is clear that the summation of all the unused services in a time slot is less than or equal to the number of servers, that is $\sum_{m \in \mathcal{M}}U_m(t) < M$.
Hence, $\langle \mathbold{Q}(t), \mathbold{U}(t) \rangle < \sum_{m \in \mathcal{M}} M \times U_m(t) \leq M^2 $ for any time slot $t$.
%\end{proof}

Proof of Theorem \ref{JSQMWTOPT} mainly starts here. Choosing the Lyapunov function $V_1(Z(t)) = \sum_{m \in \mathcal{M}} Q_m^2(t) = || \mathbold{Q}(t) ||^2$, it satisfies the conditions in the Foster-Lyapunov theorem to be non-negative, to be equal to zero only at $\mathbold{Q}(t) = 0$, and to go to infinity as any elements of $\mathbold{Q}(t)$ goes to infinity. Then the expected $T$-time slot drift of the Lyapunov function is as follows:
\[
\begin{aligned}
& \mathbb{E} [\Delta V_1 (Z(t_0))] \\
& = \mathbb{E} [V_1(t_0 + T) - V_1(t_0) | Z(t_0)] \\
& = \mathbb{E} \bigg [ \sum_{t = t_0}^{t_0 + T - 1} \bigg ( V_1(t + 1) - V_1(t) \bigg ) \bigg | Z(t_0) \bigg ] \\
& = \mathbb{E} \bigg [ \sum_{t = t_0}^{t_0 + T -1} \bigg ( || \mathbold{Q}(t + 1) ||^2 - || \mathbold{Q}(t) ||^2 \bigg ) \bigg | Z(t_0) \bigg ] \\
& = \mathbb{E} \bigg [ \sum_{t = t_0}^{t_0 + T -1} \bigg ( || \mathbold{Q}(t) + (\mathbold{A}(t) - \mathbold{S}(t) + \mathbold{U}(t)) ||^2 - || \mathbold{Q}(t) ||^2 \bigg ) \bigg | Z(t_0) \bigg ] \\
& = \mathbb{E} \bigg [ \sum_{t = t_0}^{t_0 + T -1} \bigg ( 2 \langle \mathbold{Q}(t), \mathbold{A}(t) - \mathbold{S}(t) \rangle + 2 \langle \mathbold{Q}(t), \mathbold{U}(t) \rangle \\
& \ \ \ \ \ \ \ \ \ \ \ \ \ \ \ \ \ \ \ \ \ \ \ \ \ \ \ \ \ \ \ \ \ \ \ \ \ \ \ \ \ \ + || \mathbold{A}(t) - \mathbold{S}(t) + \mathbold{U}(t) ||^2 \bigg ) \bigg | Z(t_0) \bigg ].
\end{aligned}
\] \\
We assumed that the task arrival process at a time slot is bounded with probability one and it is clear that the provided services and the unused services are also bounded.
Hence, $|| \mathbold{A}(t) - \mathbold{S}(t) + \mathbold{U}(t) ||^2$ is bounded.
Also using Lemma \ref{b1}, we have $2 \langle \mathbold{Q}(t), \mathbold{U}(t) \rangle + || \mathbold{A}(t) - \mathbold{S}(t) + \mathbold{U}(t) ||^2 = c_2$, where $c_2 > 0$ is a constant independent of $Z(t_0)$.
Then for any arrival rate vector $\mathbold{\lambda} \in \bar{\Lambda}$ we can use the corresponding $\mathbold{\psi}$ defined in \eqref{psi} as an intermidiary to write the expected Lyapunov function drift as follows:

\[
\begin{aligned}
\mathbb{E} [\Delta V_1 & (Z(t_0))] \\
& = \mathbb{E} \bigg [ \sum_{t = t_0}^{t_0 + T -1} \bigg ( 2 \langle \mathbold{Q}(t), \mathbold{A}(t) - \mathbold{S}(t) \rangle \bigg ) \bigg | Z(t_0) \bigg ] + c_2 \\
& = 2 \mathbb{E} \bigg [ \sum_{t = t_0}^{t_0 + T -1} \bigg ( \langle \mathbold{Q}(t), \mathbold{A}(t) \rangle - \langle \mathbold{Q}(t), \mathbold{\psi} \rangle \bigg ) \bigg | Z(t_0) \bigg ] \\
& \ \ \ + 2 \mathbb{E} \bigg [ \sum_{t = t_0}^{t_0 + T -1} \bigg ( \langle \mathbold{Q}(t), \mathbold{\psi} \rangle - \langle \mathbold{Q}(t), \mathbold{S}(t) \rangle \bigg ) \bigg | Z(t_0) \bigg ] + c_2 \\
& \overset{(a)}{\leq} -2 \theta_1 || \mathbold{Q}(t_0) ||_1 + c_2,
\end{aligned}
\] \\
where $(a)$ in the last inequality follows from Lemmas \ref{b00} and \ref{b0}.
Hence, for any $\epsilon > 0$, there exists $T \geq T_0$ such that for any $Z(t_0) \in \mathcal{P}^c$, we have $\mathbb{E}[V_1(Z(t_0 + T)) - V_1(Z(t_0))] \leq -\epsilon$, where $\mathcal{P}$ is a finite subset of state spaces and it is defined as $\mathcal{P} = \bigg \{ Z = (\mathbold{Q}, \mathbold{f}) \bigg | || \mathbold{Q} ||_1 \leq \frac{c_2 + \epsilon}{2 \theta_1} \bigg \}$.
It is also obvious that the expected $T$-period drift of the Lyapunov function is bounded as long as $Z(t_0) \in \mathcal{P}$.
Therefore, from the Foster-Lyapunov theorem we conclude that $\{ Z(t), t \geq 0 \}$ is positive recurrent, which means that the JSQ-MaxWeight algorithm stabilizes the system under any arrival rate vector $\mathbold{\lambda} \in \bar{\Lambda}$.
This means that $\bar{\Lambda}$ and $\Lambda$ are both the capacity region of the system.

%%%%%%%%%%%%%%%%%%%%%%%%%%%%%%%%%%%%%%%%%%%%%%%%%%%%%%%%%%%%%%%%%%%%%%%%%

\section{Proof of Theorem \ref{JSQMWOPT}}
\label{PROOFJSQMWOPT}
The proof consists of three parts.
First we obtain a lower bound for the expected queue length.
Then, we prove the state space collapse of the queue lengths.
Finally, we use a Lyapunov drift based approach presented in \cite{eryilmaz2012asymptotically} which uses the state space collapse result to find an upper bound for the expected queue length.
If the lower and upper bounds in the first and last steps match each other, the algorithm is heavy-traffic optimal in the traffic load that we considered.
We summarize the proof as follows as it is similar to the proof in \cite{wang2016maptask} for a system with two levels of data locality.

The lower bound on the expected sum of all queue lengths is obtained as follows. Assume we have a single server with the following arrival and service processes, respectively:
\[
\begin{aligned}
& \sum_{\bar{L}} A_{\bar{L}}^{\epsilon}(t), \\
& b_1(t) = \sum_{i \in \mathcal{B}_o} X_i(t) + \sum_{j \in \mathcal{H}_o} Y_j(t) + \sum_{n \in \mathcal{H}_u} V_n(t),
\end{aligned}
\]
where $\{ X_i(t) \sim Bern(\alpha) \}_{i \in \mathcal{B}_o}$, $\{ Y_j(t) \sim Bern(\beta) \}_{j \in \mathcal{H}_o}$, and $\ \{ V_n(t) \sim Bern(\gamma) \}_{n \in \mathcal{H}_u}$ .
$\{ X_i \}_{i \in \mathcal{B}_o}$, $\{ Y_j \}_{j \in \mathcal{B}_u}$, and $\{ V_n \}_{n \in \mathcal{H}_u}$ are independent from each other and each of them are i.i.d. processes.
We define the variances of the arrival and service processes of this single server model as $\sigma_1^{\epsilon} = \mathrm{var}(\sum_{\bar{L}} A_{\bar{L}}^{\epsilon}(t))$ and $\nu_1^2 = \mathrm{var}(b_1(t))$.
It is obvious that the queue length of this single server/single queue model is stochastically smaller than the sum of the queue lengths in the original system model with three levels of data locality.
Hence, we have the following lower bound on the expected sum of queue lengths:
\[
\mathbb{E} \bigg [ \sum_{m = 1}^{M} Q_m^{\epsilon}(t) \bigg ] \geq \frac{(\sigma_1^{\epsilon})^2 + \nu_1^2 + \epsilon^2}{2 \epsilon} - \frac{M}{2}.
\] \\
Therefore, if we let $\epsilon$ go to zero to create the heavy-traffic regime, we have the following lower bound:
\begin{equation}
\liminf_{\epsilon \rightarrow 0^+} \epsilon \mathbb{E} \bigg [ \sum_{m = 1}^{M} Q_m^{\epsilon}(t) \bigg ] \geq \frac{\sigma_1^2 + \nu_1^2}{2}.
\end{equation}

As $\epsilon$ goes to zero, we expect the queue lengths of beneficiary servers to grow to infinity and have somehow equal lengths.
We define the $M$-dimensional vector $\mathbold{c}_1 \in \mathbb{R}^M$ and define the parallel and perpendicular components of $\mathbold{Q}$ with respect to $\mathbold{c}_1$ as follows:

\[
\mathbold{c}_1(m) = \begin{cases}
               \frac{1}{\sqrt{M_{B_o}}} \ \ \ \ \forall m \in \mathcal{B}_o \\
               \ \ \ \ 0 \ \ \ \ \ \ \ \text{else} \\
         \end{cases},
\]

\[
\mathbold{Q}_{||} = \langle \mathbold{c}_1, \mathbold{Q} \rangle \mathbold{c}_1,
\]
\[
\mathbold{Q}_{\perp} = \mathbold{Q} - \mathbold{Q}_{||}.
\]
By taking $V_2(\mathbold{Z}) = ||\mathbold{Q}_{\perp}||$ as the Lyapunov function, we can show that using the JSQ-MW scheduling algorithm, the expected drift of this Lyapunov function is bounded, and becomes negative for sufficiently large $\mathbold{Q}_{\perp}$.
Therefore, we have the following theorem for state space collapse (the proof for the following theorem is eliminated as it is similar to the corresponding theorem in \cite{wang2016maptask}).

\begin{theorem}
There exists finite sequence of numbers $\{ C_r : r \in \mathbb{N} \}$ such that
\[
\mathbb{E} [|| \mathbold{Q}_{\perp} ||^r] \leq C_r, \ \forall r \in \mathbb{N}.
\]
\end{theorem}

We can then use the state space collapse result to prove the following upper bound for the mean sum of queue lengths:

\[
\mathbb{E} \bigg [ \sum_{m = 1}^{M} Q_m^{\epsilon} (t) \bigg ] \leq \frac{(\sigma_1^{\epsilon})^2 + \nu_1^2 }{2 \epsilon} + B^{\epsilon},
\]
where $B^{\epsilon} = o(\frac{1}{\epsilon})$. Hence, letting $\epsilon$ to go to zero, we have the following upper bound in the heavy-traffic regime:

\[
\limsup_{\epsilon \rightarrow 0^+} \epsilon \mathbb{E} \bigg [ \sum_{m = 1}^{M} Q_m^{\epsilon} (t) \bigg ] \leq \frac{\sigma^2 + \nu_1^2}{2}.
\]
As the upper bound of the mean sum of queue lengths coincides with the lower bound under using the JSQ-MW algorithm, this algorithm is heavy-traffic optimal under the load we specified in Theorem \ref{JSQMWOPT} (but it is not heavy-traffic optimal in all traffic scenarios).

\section{Proof of Theorem \ref{TTOBP}}
\label{PROOFTTOBP}
A corollary of Theorem \ref{JSQMWTOPT} is that $\Lambda$ is the capacity region of a system with three levels of data locality.
Hence, to prove the throughput optimality of the Balanced-Pandas algorithm, it is enough to show that this scheduling algorithm can stabilize the system as long as the arrival rate vector is strictly inside the capacity region, $\mathbold{\lambda} \in \Lambda$.
For any $\mathbold{\lambda} \in \Lambda$, there exists $\delta > 0$ such that $\mathbold{\lambda}^{'} = \mathbold{\lambda} (1 + \delta) \in \Lambda$.
As $\mathbold{\lambda}^{'} \in \Lambda$, there exists a load decomposition $\{ \lambda_{\bar{L}, m}^{'} \}$ such that it satisfies the following:
\[
\sum_{\bar{L}: m \in \bar{L}} \frac{\lambda^{'}_{\bar{L}, m}}{\alpha} + \sum_{\bar{L}: m \in \bar{L}_k} \frac{\lambda^{'}_{\bar{L}, m}}{\beta} + \sum_{\bar{L}: m \in \bar{L}_r} \frac{\lambda^{'}_{\bar{L}, m}}{\gamma} < 1, \ \forall m \in \mathcal{M},
\] \\
then by our choice of $\mathbold{\lambda}^{'}$ to be $\mathbold{\lambda}(1 + \delta)$, we have  the following:
\begin{equation}
\label{delta}
\sum_{\bar{L}: m \in \bar{L}} \frac{\lambda_{\bar{L}, m}}{\alpha} + \sum_{\bar{L}: m \in \bar{L}_k} \frac{\lambda_{\bar{L}, m}}{\beta} + \sum_{\bar{L}: m \in \bar{L}_r} \frac{\lambda_{\bar{L}, m}}{\gamma} < \frac{1}{1 + \delta}, \ \forall m \in \mathcal{M}.
\end{equation}

Define the workload vector of servers, $\mathbold{w} = (w_1, w_2, \cdots, w_M)$, under the load decomposition $\{ \lambda_{\bar{L}, m} \}$ as follows:

\begin{equation}
\label{workloadm}
w_m = \sum_{\bar{L}: m \in \bar{L}} \frac{\lambda_{\bar{L}, m}}{\alpha} + \sum_{\bar{L}: m \in \bar{L}_k} \frac{\lambda_{\bar{L}, m}}{\beta} + \sum_{\bar{L}: m \in \bar{L}_r} \frac{\lambda_{\bar{L}, m}}{\gamma}, \ \forall m \in \mathcal{M}.
\end{equation}

The workload on a server evolves as follows:

\[
\begin{aligned}
W_m(t+1) & = \frac{Q_m^l(t+1)}{\alpha} + \frac{Q_m^k(t+1)}{\beta} + \frac{Q_m^r(t+1)}{\gamma} \\
& \overset{(a)}{=} \frac{Q_m^l(t) + A_m^l(t) - S_m^l(t)}{\alpha} + \frac{Q_m^k(t) + A_m^k(t) - S_m^k(t)}{\beta} \\
& \ \ \ \ \ \ \ \ \ \ \ \ \ \ \ \ \ \ \ \ \ \ \ \ \ \ \ \ \ \ \ \ \ \ \ \ + \frac{Q_m^r(t) + A_m^r(t) - S_m^r(t) + U_m(t)}{\gamma} \\
& = W_m(t) + \bigg (\frac{A_m^l(t)}{\alpha} + \frac{A_m^k(t)}{\beta} + \frac{A_m^r(t)}{\gamma} \bigg ) \\
& \ \ \ \ \ \ \ \ \ \ \ \ \ \ \ \ \ \ \ \ \ \ \ \ \ \ \ \ \ \ - \bigg (\frac{S_m^l(t)}{\alpha} + \frac{S_m^k(t)}{\beta} + \frac{S_m^r(t)}{\gamma} \bigg ) + \frac{U_m(t)}{\gamma},
\end{aligned}
\]
where $(a)$ follows from the queue evolution in \eqref{queueevolution}.
Define the pseudo arrival, service and unused service processes as $\mathbold{A} = (A_1, A_2, \cdots, A_M)$, $\mathbold{S} = (S_1, S_2, \cdots, S_M)$, and $\widetilde{\mathbold{U}} = (\widetilde{U}_1, \widetilde{U}_2, \cdots, \widetilde{U}_M)$, respectively, where

\[
A_m(t) = \frac{A_m^l(t)}{\alpha} + \frac{A_m^k(t)}{\beta} + \frac{A_m^r(t)}{\gamma}, \ \forall m \in \mathcal{M},
\]

\[
S_m(t) = \frac{S_m^l(t)}{\alpha} + \frac{S_m^k(t)}{\beta} + \frac{S_m^r(t)}{\gamma}, \ \forall m \in \mathcal{M},
\]
\[
\widetilde{U}_m(t) = \frac{U_m(t)}{\gamma}, \ \forall m \in \mathcal{M}.
\]

By the above definitions, we can write the dynamics of the queue workloads, $\mathbold{W} = (W_1, W_2, \cdots, W_M)$, as follows:

\begin{equation}
\label{evolW}
\mathbold{W}(t + 1) = \mathbold{W}(t) + \mathbold{A}(t) - \mathbold{S}(t) + \widetilde{\mathbold{U}}(t).
\end{equation}

The following three lemmas will be used in the proof of Theorem \ref{TTOBP}.

\begin{lemma}
\label{lemma123}
\begin{equation}
\langle \mathbold{W}(t) , \widetilde{\mathbold{U}}(t) \rangle = 0, \ \forall t.
\end{equation}
\end{lemma}

%\begin{proof}
\textbf{proof:}
The expression simplifies as follows:
\[
\langle \mathbold{W}(t) , \widetilde{\mathbold{U}}(t) \rangle = \sum_m \bigg ( \frac{Q_m^l(t)}{\alpha} + \frac{Q_m^k(t)}{\beta} + \frac{Q_m^r(t)}{\gamma} \bigg ) \frac{U_m(t)}{\gamma}.
\]
Note that for any server $m$, if $U_m(t) = 0$, then $\big ( \frac{Q_m^l(t)}{\alpha} + \frac{Q_m^k(t)}{\beta} + \frac{Q_m^r(t)}{\gamma} \big ) \frac{U_m(t)}{\gamma} = 0 $.
Otherwise, $U_m(t) > 0$ implies that all sub-queues of server $m$ are empty which again results in $U_m(t) = 0$, then $\big ( \frac{Q_m^l(t)}{\alpha} + \frac{Q_m^k(t)}{\beta} + \frac{Q_m^r(t)}{\gamma} \big ) \frac{U_m(t)}{\gamma} = 0 $.
Therefore, $\langle \mathbold{W}(t) , \widetilde{\mathbold{U}}(t) \rangle = 0$ for all time slots.
%\end{proof}

\begin{lemma}
\label{lemma1234}
\label{lemmma}
For any arrival rate vector strictly inside the capacity region, $\mathbold{\lambda} \in \Lambda$, and the corresponding workload vector of servers $\mathbold{w}$ defined in \eqref{workloadm}, we have the following inequality by using the Balanced-Pandas algorithm:
\begin{equation}
\mathbb{E}[ \langle \mathbold{W}(t), \mathbold{A}(t) \rangle - \langle \mathbold{W}(t), \mathbold{w} \rangle | Z(t) ] \leq 0, \ \forall t \geq 0.
\end{equation}
\end{lemma}

%\begin{proof}
\textbf{proof:}
We first define the minimum weighted workload for a task type, $\bar{L} \in \mathcal{L}$ as follows:
\begin{equation}
W_{\bar{L}}^*(t) = \min_{m \in \mathcal{M}} \bigg \{ \frac{W_m(t)}{\alpha} I_{\{ m \in \bar{L} \}}, \frac{W_m(t)}{\beta} I_{\{ m \in \bar{L}_k \}}, \frac{W_m(t)}{\gamma} I_{\{ m \in \bar{L}_r \}} \bigg \}.
\end{equation}
At the beginning of time slot $t$, an incoming task of type $\bar{L}$  is routed to queue $m^*$ with the minimum expected workload $W_{\bar{L}}^*(t)$.
Therefore, for any task type $\bar{L} \in \mathcal{L}$ we have the following:
\begin{equation}
\label{MWW}
\begin{aligned}
& \frac{W_m(t)}{\alpha} \geq W_{\bar{L}}^*(t), \ \forall m \in \bar{L}, \\
& \frac{W_m(t)}{\beta} \geq W_{\bar{L}}^*(t), \ \forall m \in \bar{L}_k, \\
& \frac{W_m(t)}{\gamma} \geq W_{\bar{L}}^*(t), \ \forall m \in \bar{L}_r.
\end{aligned}
\end{equation}
In other words, task of type $\bar{L}$ does not join a server $m$ with weighted workload greater than $W_{\bar{L}}^*$.
Then we have the following:

\begin{equation}
\begin{aligned}
\label{eq1}
& \mathbb{E} \big [ \langle \mathbold{W}(t), \mathbold{A}(t) \rangle | Z(t) \big ] \\
 & = \mathbb{E} \bigg [ \sum_m W_m(t) \bigg ( \frac{A_m^l(t)}{\alpha} + \frac{A_m^k(t)}{\beta} + \frac{A_m^r(t)}{\gamma} \bigg ) \bigg | Z(t) \bigg ] \\
& = \mathbb{E} \bigg [ \sum_m W_m(t) \bigg ( \frac{1}{\alpha} \sum_{\bar{L}: m \in \bar{L}} A_{\bar{L}, m}(t) + \frac{1}{\beta} \sum_{\bar{L}: m \in \bar{L}_k} A_{\bar{L}, m}(t) \\
& \ \ \ \ \ \ \ \ \ \ \ \ \ \ \ \ \ \  \ \ \ \ \ \ \ \ \ \ \ \ \ \ \ \ \ \ \ \ \ \ \ \ \ \ \ \ \ \ \ \ \ \ \ \ \ \ \ \ \ \ \ \ \ \ + \frac{1}{\gamma} \sum_{\bar{L}: m \in \bar{L}_r} A_{\bar{L}, m}(t) \bigg ) \bigg | Z(t) \bigg ] \\
& \overset{(a)}{=} \mathbb{E} \bigg [ \sum_{\bar{L} \in \mathcal{L}} \bigg ( \sum_{m: m \in \bar{L}} \frac{W_m(t)}{\alpha} A_{\bar{L}, m}(t) + \sum_{m: m \in \bar{L}_k} \frac{W_m(t)}{\beta} A_{\bar{L}, m}(t)  \\ 
& \ \ \ \ \ \ \ \ \ \ \ \ \ \ \ \ \ \ \ \ \ \ \ \ \ \ \ \ \ \ \ \ \ \ \ \ \ \ \ \ \ \ \ \ \ \ \ \ \ \ \ \ \ \ \ \ \ + \sum_{m: m \in \bar{L}_r} \frac{W_m(t)}{\gamma} A_{\bar{L}, m}(t) \bigg ) \bigg | Z(t) \bigg ] \\
& \overset{(b)}{=} \mathbb{E} \bigg [ \sum_{\bar{L} \in \mathcal{L}} W_{\bar{L}}^*(t) A_{\bar{L}}(t) \bigg | Z(t) \bigg ] \\
& = \sum_{\bar{L} \in \mathcal{L}} W_{\bar{L}}^*(t) \lambda_{\bar{L}},
\end{aligned}
\end{equation}
where $(a)$ is true by changing the order of the summations, and $(b)$ follows from the Balanced-Pandas routing policy that task of type $\bar{L}$ is routed to the queue with the minimum weighted workload, $W_{\bar{L}}^*$.
On the other hand,

\begin{equation}
\begin{aligned}
\label{eq2}
& \mathbb{E} \big [ \langle \mathbold{W}(t), \mathbold{w} \rangle | Z(t) \big ] \\
& = \sum_m W_m(t) w_m \\
& = \sum_m W_m(t) \bigg ( \sum_{\bar{L}: m \in \bar{L}} \frac{\lambda_{\bar{L}, m}}{\alpha} + \sum_{\bar{L}: m \in \bar{L}_k} \frac{\lambda_{\bar{L}, m}}{\beta} + \sum_{\bar{L}: m \in \bar{L}_r} \frac{\lambda_{\bar{L}, m}}{\gamma} \bigg ) \\
& \overset{(a)}{=} \sum_{\bar{L} \in \mathcal{L}} \bigg ( \sum_{m: m \in \bar{L}} \frac{W_m(t)}{\alpha} \lambda_{\bar{L}, m} + \sum_{m: m \in \bar{L}_k} \frac{W_m(t)}{\beta} \lambda_{\bar{L}, m} + \sum_{m: m \in \bar{L}_r} \frac{W_m(t)}{\gamma} \lambda_{\bar{L}, m} \bigg ) \\
& \overset{(b)}{\geq} \sum_{\bar{L} \in \mathcal{L}} \sum_{m \in \mathcal{M}} W_{\bar{L}}^*(t)\lambda_{\bar{L}, m}  \\
& = \sum_{\bar{L} \in \mathcal{L}} W_{\bar{L}}^*(t) \lambda_{\bar{L}},
\end{aligned}
\end{equation}
where $(a)$ is true by changing the order of summations, and $(b)$ follows from \eqref{MWW}.
Lemma \ref{lemmma} is concluded from expressions \eqref{eq1} and \eqref{eq2}.

%\end{proof}

\begin{lemma}
\label{lemma12345}
For any arrival rate vector strictly inside the capacity region, $\mathbold{\lambda} \in \Lambda$, and the corresponding workload vector of servers $\mathbold{w}$ defined in \eqref{workloadm}, we have the following inequality by using the Balanced-Pandas algorithm:
\begin{equation}
\mathbb{E}[ \langle \mathbold{W}(t), \mathbold{w} \rangle - \langle \mathbold{W}(t), \mathbold{S}(t) \rangle | Z(t) ] \leq - \theta_2 || \bar{\mathbold{Q}}(t) ||_1, \ \forall t \geq 0,
\end{equation} \\
where the constant $\theta_2 > 0$ is independent of $Z(t)$.
\end{lemma}

%\begin{proof}
\textbf{proof:}
Using \eqref{delta}, the mean workload vector on servers defined in \eqref{workloadm} can be bounded as follows:
\[
w_m \leq \frac{1}{1 + \delta}, \ \forall m \in \mathcal{M}.
\]
Hence,
\begin{equation}
\label{eq21}
\mathbb{E} [ \langle \mathbold{W}(t), \mathbold{w} \rangle | Z(t) ] = \sum_m W_m(t) w_m \leq \frac{1}{1 + \delta} \sum_m W_m(t).
\end{equation}
We also have the following:
\begin{equation}
\label{eq22}
\begin{aligned}
& \mathbb{E} [ \langle \mathbold{W}(t), \mathbold{S}(t) \rangle | Z(t) ] \\
& = \mathbb{E} \bigg [ \sum_m W_m(t) \bigg ( \frac{S_m^l(t)}{\alpha} + \frac{S_m^k(t)}{\beta} + \frac{S_m^r(t)}{\gamma} \bigg ) \bigg | Z(t) \bigg ] \\
& = \sum_m W_m(t) \mathbb{E} \bigg [ \frac{S_m^l(t)}{\alpha} + \frac{S_m^k(t)}{\beta} + \frac{S_m^r(t)}{\gamma} \bigg | Z(t) \bigg ] \\
& = \sum_m W_m(t) \mathbb{E} \bigg [ \sum_{i = 0}^{2} \mathbb{E} \bigg [ \frac{S_m^l(t)}{\alpha} + \frac{S_m^k(t)}{\beta} + \frac{S_m^r(t)}{\gamma} \bigg | Z(t), \eta_m(t) = i \bigg ] | Z(t) \bigg ] \\
& = \sum_m W_m(t) \bigg \{ \mathbb{E} \bigg [ \mathbb{E} \bigg [ \frac{S_m^l(t)}{\alpha} \bigg | Z(t), \eta_m(t) = 0 \bigg ] | Z(t) \bigg ] \\
& + \mathbb{E} \bigg [ \mathbb{E} \bigg [ \frac{S_m^k(t)}{\beta} \bigg | Z(t), \eta_m(t) = 1 \bigg ] | Z(t) \bigg ] + \mathbb{E} \bigg [ \mathbb{E} \bigg [ \frac{S_m^r(t)}{\gamma} \bigg | Z(t), \eta_m(t) = 2 \bigg ] | Z(t) \bigg ] \bigg \} \\
& = \sum_m W_m(t).
\end{aligned}
\end{equation}
Therefore,

\[
\begin{aligned}
& \mathbb{E} [ \langle \mathbold{W}(t), \mathbold{w} \rangle - \langle \mathbold{W}(t), \mathbold{S}(t) \rangle | Z(t) ] \\
& \leq \frac{1}{1 + \delta} \sum_m W_m(t) - \sum_m W_m(t) \\
& = - \frac{\delta}{1 + \delta} \sum_m W_m(t) \\
& = - \frac{\delta}{1 + \delta} \sum_m \bigg ( \frac{Q_m^l(t)}{\alpha} + \frac{Q_m^k(t)}{\beta} + \frac{Q_m^r(t)}{\gamma} \bigg ) \\
& \leq - \frac{\delta}{\alpha (1 + \delta)} \sum_m (Q_m^l(t) + Q_m^k(t) + Q_m^r(t)) \\
& = -\theta_2 || \bar{\mathbold{Q}}(t) ||_1.
\end{aligned}
\]

Using Lemmas \ref{lemma123}, \ref{lemma1234}, and \ref{lemma12345}, we prove Theorem \ref{TTOBP} as follows.
Assume the Lyapunov function is chosen as
\[
V_3(Z(t)) = || \mathbold{W}(t) ||^2,
\]
then its expected drift is as follows:

\[
\begin{aligned}
& \mathbb{E}[ \Delta(Z(t)) ] \\
& = \mathbb{E}[V_3(t + 1) - V_3(t) | Z(t)] \\
& = \mathbb{E} \bigg [|| \mathbold{W}(t + 1) ||^2 - || \mathbold{W}(t) ||^2 \bigg | Z(t) \bigg ] \\
& \overset{(a)}{=} \mathbb{E} \bigg [|| \mathbold{W}(t) + \mathbold{A}(t) - \mathbold{S}(t) + \widetilde{\mathbold{U}}(t) ||^2 - || \mathbold{W}(t) ||^2 \bigg | Z(t) \bigg ] \\
& = \mathbb{E} \bigg [ 2 \langle \mathbold{W}(t), \mathbold{A}(t) - \mathbold{S}(t) \rangle + 2 \langle \mathbold{W}(t), \widetilde{\mathbold{U}}(t) \rangle + || \mathbold{A}(t) - \mathbold{S}(t) + \widetilde{\mathbold{U}}(t) ||^2 \bigg | Z(t) \bigg ] \\
& \overset{(b)}{=} 2 \mathbb{E} \bigg [ \langle \mathbold{W}(t), \mathbold{A}(t) - \mathbold{S}(t) \rangle \bigg | Z(t) \bigg ] + c_3 \\
& = 2 \mathbb{E} \bigg [ \langle \mathbold{W}(t), \mathbold{A}(t) \rangle - \langle \mathbold{W}(t), \mathbold{w} \rangle \bigg | Z(t) \bigg ] \\
& \ \ \ \ \ \ \ \ \ \ \ \ \ \ \ \ \ \ \ \ \ \ \ \ \ \ \ \ \ \ \ \ \ \ \ \ \ \ \ \ \ \ + 2 \mathbb{E} \bigg [ \langle \mathbold{W}(t), \mathbold{w} \rangle - \langle \mathbold{W}(t), \mathbold{S}(t) \rangle \bigg | Z(t) \bigg ] + c_3 \\
& \overset{(c)}{\leq} - 2 \theta_2 || \bar{\mathbold{Q}}(t) ||_1 + c_3,
\end{aligned}
\]
where $(a)$ follows from \eqref{evolW}, $(b)$ follows from Lemma \ref{lemma123}, and the fact that $\mathbold{A}(t)$, $\mathbold{S}(t)$, and $\widetilde{\mathbold{U}}(t)$ are all bounded, and $(c)$ is true by Lemmas \ref{lemma1234} and \ref{lemma12345}.
By choosing any positive constant $\epsilon > 0$, let $\mathcal{P} = \bigg \{ Z = (\mathbold{Q}, \mathbold{f}) \bigg | || \mathbold{Q} ||_1 \leq \frac{c_3 + \epsilon}{2 \theta_2} \bigg \}$, where $\mathcal{P}$ is a bounded subset of the state space.
For any $Z \in \mathcal{P}$, $\Delta V_3(Z)$ is bounded and for any $Z \in \mathcal{P}^c$, $\Delta V_3(Z) \leq - \epsilon$.
Hence, for any $\mathbold{\lambda} \in \Lambda$, the Markov process $\{ Z(t), t \geq 0 \}$ is positive recurrent and the Balanced-Pandas algorithm makes the system stable, which means that this algorithm is throughput optimal.
%\end{proof}

\section{Proof of Theorem \ref{THOBP}}
\label{PROOFBPHO}
\begin{figure}[h]
\centering
\includegraphics[scale=0.27]{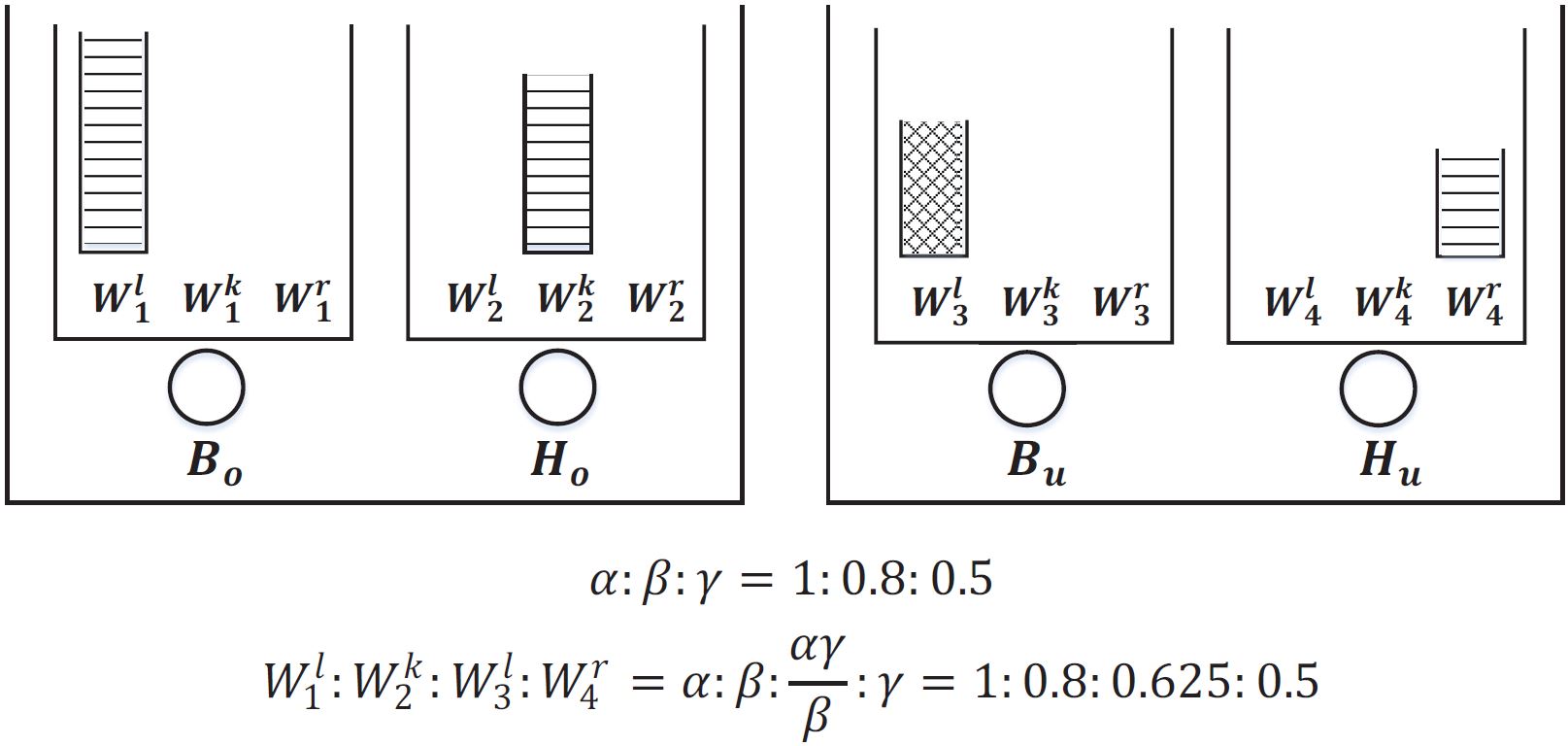}
\caption{The queue compositions of the four types of servers in the heavy-traffic regime with $\alpha = 1, \beta = 0.8, \gamma = 0.5.$ The workload at four types of servers maintain the ratio $\alpha : \beta : \frac{\alpha \gamma}{\beta} : \gamma = 1: 0.8 : 0.625 : 0.5$.}
\label{state-space}
\end{figure}

For simplicity, this proof is for the special case where $\mathcal{O} \neq  \oldemptyset$ and $\mathcal{B}_u = \oldemptyset$. For the general proof refer to \cite{xie2016schedulingPhD}. The heavy-traffic optimality is driven through the following three steps:
\begin{enumerate}
\item Establishing the state-space collapse in the heavy traffic regime.
\item Finding a lower bound on the expected sum of the queue lengths as $\epsilon \rightarrow 0$.
\item Finding an upper bound on the expected sum of the queue lengths as $\epsilon \rightarrow 0$, which matches the lower bound found in step 2.
\end{enumerate}
In heavy traffic regime, the system collapses to the one-dimensional state space vector shown in Figure \ref{state-space}.

Note that the prioritized service uniformly bounds the helper subsystem in heavy-traffic regime. This results in disappearance of local and rack-local queues of servers in the set $\mathcal{H}_u$ and local queues of servers in the set $\mathcal{H}_o$. On the other hand, the weighted-workload routing policy distributes tasks that are only local to beneficiary servers in overloaded racks across $\mathcal{B}_o, \mathcal{H}_o$, and $\mathcal{H}_u$ in the ratio of $\alpha : \beta : \gamma$ in terms of server workload. Furthermore, the tasks only local to servers in the set $\mathcal{B}_u$ are just helped rack-locally by servers in the set $\mathcal{H}_u$, and the weighted-workload scheduling policy maintains the ratio $\alpha : \beta$ in terms of workload on beneficiary and helper servers in under-loaded racks. Hence, the workload is distributed over servers in this proportion: $W_1^l : W_2^k : W_3^l : W_4^r = \alpha : \beta : \frac{\alpha \gamma}{\beta} : \gamma$.

Denote the local traffic on $\mathcal{H}_u$ and $\mathcal{H}_o$ by $\sum_{\bar{L} \in \mathcal{L}_{\mathcal{H}_u}^*} \lambda_{\bar{L}} \equiv \Phi_u \alpha$ and $\sum_{\bar{L} \in \mathcal{L}_{\mathcal{H}_o}} \lambda_{\bar{L}}$ $\equiv \Phi_o \alpha$, respectively, where $\mathcal{L}_{\mathcal{H}_u}^* = \{ \bar{L} : \exists m \in \mathcal{H}_u \ s.t. \ m \in \bar{L} \}$, and $\mathcal{L}_{\mathcal{H}_o} = \{ \bar{L} : \forall m \in \bar{L}, m \in \mathcal{H}_o \cup \mathcal{B}_o, \text{ and } \exists n \in \mathcal{H}_o \ s.t. \ n \in \bar{L} \}$. Then, the heavy-traffic regime parameterized by $\epsilon > 0$, where $\epsilon$ shows the distance of the arrival rate vector from the boundary of the capacity region, is defined as follows:

$$\sum_{\bar{L} \in \mathcal{L}_{\mathcal{B}_o}} \lambda_{\bar{L}} = \alpha |\mathcal{B}_o| + \beta (|\mathcal{H}_o| - \Phi_o) + \gamma(|\mathcal{H}_u| - \Phi_u) - \epsilon.$$

Consider the arrival process $\{A_{\bar{L}}^{(\epsilon)}(t)\}_{\bar{L} \in \mathcal{L}}$ with arrival rate vector ${\mathbold{\lambda}}^{(\epsilon)}$. An assumption is made that the total local load for helpers is fixed, that is $\{ \lambda_{\bar{L}} : \bar{L} \in \mathcal{L}_{\mathcal{H}_u}^* \cup \mathcal{L}_{\mathcal{H}_o} \}$ is independent of $\epsilon$. Hence, the variance of $\{A_{\bar{L}}^{(\epsilon)}(t)\}_{\bar{L} \in \mathcal{L}_{\mathcal{H}_u}^* \cup \mathcal{L}_{\mathcal{H}_o}}$ is independent of $\epsilon$. On the other hand, the variance of the number of tasks that are only local to beneficiary servers in overloaded racks is denoted by $(\sigma^{(\epsilon)})^2$ that converges to $\sigma^2$ as $\epsilon \downarrow 0$, that is $Var \left (  \sum_{\bar{L} \in \mathcal{L}_{\mathcal{B}_o}} A_{\bar{L}^{(\epsilon)}(t)} \right ) = (\sigma^{(\epsilon)})^2 \overset{\epsilon \rightarrow 0}{\longrightarrow} \sigma^2$. The system state under the Balanced-Pandas algorithm when the arrival rate is ${\mathbold{\lambda}}^{(\epsilon)}$ is denoted by $\Big \{ {\bf{Z}}^{( \epsilon )}(t) =$ $\Big ( \bar{{\bf{Q}}}^{(\epsilon)}(t),$ ${\bf{f}}^{(\epsilon)}(t) \Big ), t \geq 0 \Big \}$. Then, the Markov chain ${\bf{Z}}^{(\epsilon)}(t)$ is positive recurrent and has a steady state distribution as long as ${\mathbold{\lambda}}^{(\epsilon)} \in \Lambda$.
The following theorem states that local and rack-local queues of $\mathcal{H}_u$ and local queues of $\mathcal{H}_o$ are uniformly bounded and the bound is independent of $\epsilon$.

\begin{theorem} \textbf{(Helper Queues)}
$$\lim_{\epsilon \downarrow 0} \epsilon \mathbb{E} \left [ \sum_{m \in \mathcal{H}_u} \left ( Q_m^{l(\epsilon)}(t) + Q_m^{k(\epsilon)}(t) \right) \right ] = 0,$$
$$\lim_{\epsilon \downarrow 0} \epsilon \mathbb{E} \left [ \sum_{m \in \mathcal{H}_o} Q_m^{l(\epsilon)}(t) \right ] = 0.$$
\label{helperqueues}
\end{theorem}

The proof of Theorem \ref{helperqueues} is given in Section \ref{proofhelperqueues}.

As the arrival rate vector approaches the boundary of the capacity region, that is $\epsilon \downarrow 0$, the mean sum of queue lengths approaches infinity in steady state, that is $\mathbb{E} \left [ ||\bar{{\bf{Q}}}|| \right ] = \mathbb{E} \left[ \sum_m \left( Q_m^l + Q_m^k + Q_m^r \right) \right] \rightarrow \infty$.
By Theorem \ref{helperqueues}, it is enough to consider the following to characterize the scaling order of $\mathbb{E} \left [ \bar{{\bf{Q}}} \right ]$:

$$\Phi = \sum_{m \in \mathcal{H}_u} Q_m^r + \sum_{m \in \mathcal{H}_o} \left ( Q_m^k + Q_m^r \right ) + \sum_{m \in \mathcal{B}_o} \left ( Q_m^l + Q_m^k + Q_m^r \right ).$$
Define $\tilde{{\bf{c}}} \in \mathbb{R}_+^M$ as follows:
$$\tilde{c}_m=
\begin{cases}
	\gamma, &  \forall m \in \mathcal{H}_u \\
	\beta, &  \forall m \in \mathcal{H}_o \\
    \alpha, & \forall m \in \mathcal{B}_o
\end{cases}.$$
By defining ${\bf{c}} = \frac{\tilde{{\bf{c}}}}{||\tilde{{\bf{c}}}||}$, the parallel and perpendicular components of the steady-state weighted queue-length vector, ${\bf{W}}$, with respect to vector ${\bf{c}}$ are as follows:
$${\bf{W}}_{||} = \langle {\bf c}, {\bf W} \rangle {\bf c}, \ \ \ \ \ {\bf W}_{\perp} = {\bf W} - {\bf W}_{||}.$$

The following theorem states that the deviation of ${\bf W}$ from direction ${\bf c}$ is bounded and is independent of the heavy-traffic parameter, $\epsilon$.

\begin{theorem} \textbf{(State Space Collapse)}

There exists a sequence of finite numbers $\{ C_r : r \in \mathbb{N} \}$ such that for each positive integer $r$ we have the following:
$$\mathbb{E} \left [ || {\bf W}_{\perp} ||^r \right ] \leq C_r.$$

\label{state-space-collapse}
\end{theorem}

The proof of Theorem \ref{state-space-collapse} is given in Section \ref{proofstate-space-collapse}.

Define the service process as follows:

$$b^{(\epsilon)}(t) = \sum_{i \in \mathcal{B}_o} X_i(t) + \sum_{j \in \mathcal{H}_o} Y_j(t) + \sum_{n \in \mathcal{H}_u} V_n(t),$$
where $\{ X_i(t) \}_{i \in \mathcal{B}_o}$, $\{ Y_j(t) \}_{j \in \mathcal{H}_o}$, and $\{ V_n(t) \}_{n \in \mathcal{H}_u}$ are independent from each other and each process is i.i.d. and,
$$
\begin{cases}
	X_i(t) \sim Bern(\alpha) & \forall i \in \mathcal{B}_o \\
	Y_j(t) \sim Bern(\beta(1 - \rho_j^l)) & \forall j \in \mathcal{H}_o \\
	V_n(t) \sim Bern(\gamma(1 - \rho_n)) & \forall n \in \mathcal{H}_u
\end{cases}.
$$
where $\rho_j^l$ is the proportion of time that helper server $j$ gives service to local tasks in steady state, and $\rho_n$ is the proportion of time that helper server $n$ gives service to local and rack-local tasks in steady state. Let $Var \left ( b^{(\epsilon)}(t) \right ) = \left ( \nu^{(\epsilon)} \right )^2$ that converges to $\nu^2$ as $\epsilon \downarrow 0$. Then, we have the following two theorems.

\begin{theorem} \textbf{(Lower Bound)}

$$\mathbb{E} \left [ \Phi^{(\epsilon)}(t) \right ] \geq \frac{ \left ( \sigma^{(\epsilon)} \right )^2 + \left ( \nu^{(\epsilon)} \right )^2 + \epsilon^2}{2 \epsilon} - \frac{M}{2}.$$
Hence,

$$\liminf_{\epsilon \downarrow 0} \epsilon \mathbb{E} \left [ \Phi^{(\epsilon)}(t) \right ] \geq \frac{\sigma^2 + \nu^2}{2}.$$

\label{lower-bound}
\end{theorem}

The proof of Theorem \ref{lower-bound} is given in Section \ref{prooflower-bound}.

\begin{theorem} \textbf{(Upper Bound)}

$$\mathbb{E} \left [ \Phi^{(\epsilon)}(t) \right ] \leq \frac{ \left ( \sigma^{(\epsilon)} \right )^2 + \left ( \nu^{(\epsilon)} \right )^2 }{2 \epsilon} + B^{(\epsilon)},$$
where $B^{(\epsilon)} = o(\frac{1}{\epsilon})$, that is $\lim_{\epsilon \downarrow 0} \epsilon B^{(\epsilon)} = 0$; hence,

$$\limsup_{\epsilon \downarrow 0} \epsilon \mathbb{E} \left [ \Phi^{(\epsilon)}(t) \right ] \leq \frac{\sigma^2 + \nu^2}{2}.$$
This upper bound matches with the lower bound found in Theorem \ref{lower-bound}. %, so the proof of heavy-traffic optimality is complete.

\label{upper-bound}
\end{theorem}

The proof of Theorem \ref{upper-bound} is given in Section \ref{proofupper-bound}.

Note that,

\begin{equation*}
\begin{aligned}
& \mathbb{E} \left [ \sum_m \left ( Q_m^{l(\epsilon)}(t) + Q_m^{k(\epsilon)}(t) + Q_m^{r(\epsilon)}(t) \right ) \right ] \\
& = \mathbb{E} \left [ \sum_{m \in \mathcal{H}_u} \left ( Q_m^{l(\epsilon)}(t) + Q_m^{k(\epsilon)}(t) \right ) + \sum_{m \in \mathcal{H}_o} Q_m^{l(\epsilon)}(t) \right ] + \mathbb{E} \left [ \Phi^{(\epsilon)}(t) \right ],
\end{aligned}
\end{equation*}
where Theorems \ref{lower-bound} and \ref{upper-bound} give the coincidence of lower and upper bounds of the term $\epsilon \mathbb{E} \left [ \Phi^{(\epsilon)}(t) \right ]$ as $\epsilon \rightarrow 0.$ Using Theorem \ref{helperqueues}, the proof of heavy-traffic optimality is complete and Theorem \ref{PROOFBPHO} is proved.

%The proofs for Theorems \ref{helperqueues}, \ref{state-space-collapse}, \ref{lower-bound}, and \ref{upper-bound} are given in the following.

\subsection{Proof of Theorem \ref{helperqueues}}
\label{proofhelperqueues}

Considering the system in steady state, define the following for any $m \in \mathcal{H}_u$,

$$\hat{Q}_m(t) = Q_m^l(t) + Q_m^k(t),$$
$$\hat{A}_m(t) = A_m^l(t) + A_m^k(t),$$
$$\hat{S}_m(t) = S_m^l(t) + S_m^k(t),$$
where $\hat{{\bf Q}}$ evolves as below:
$$\hat{{\bf Q}}(t + 1) = \hat{{\bf Q}}(t) + \hat{{\bf A}}(t) - \hat{{\bf S}}(t).$$

Let $\hat{F}_m(t) = F_m^l(t) + F_m^k(t)$, where the ideal arrival process ${\bf F}(t)$ is defined in the proof of Theorem \ref{upper-bound}. Now we can rewrite the dynamics of $\hat{{\bf Q}}$ in the following way:
$$\hat{{\bf Q}}(t + 1) = \hat{{\bf Q}}(t) + \hat{{\bf F}}(t) - \hat{{\bf S}}(t) + \hat{{\bf A}}(t) - \hat{{\bf F}}(t).$$
Define the unit vector ${\bf c}_h \in \mathbb{R}_+^{M_{\mathcal{H}_o}}$ as follows:
$${\bf c}_h = \frac{1}{\sqrt{M_{\mathcal{H}_o}}} \underbrace{(1, 1, \cdots, 1)}_{M_{\mathcal{H}_o}}.$$
The drift of the function $ || \hat{{\bf Q}}_{||} ||^2 = || \langle {\bf c}_h, \hat{{\bf Q}}_{||} ||^2$ is zero in steady state, so

\begin{equation}
\begin{aligned}
& 2\mathbb{E} \left [ \langle {\bf c}_h, \hat{{\bf Q}}(t) \rangle \langle {\bf c}_h, \hat{{\bf S}}(t) - \hat{{\bf F}}(t) \rangle  \right ] \\
& \ \ \ \ \ = \mathbb{E} \left [ \langle {\bf c}_h , \hat{{\bf F}}(t) - \hat{{\bf S}}(t) \rangle^2 \right ] + \mathbb{E} \left [ \langle {\bf c}_h , \hat{{\bf A}}(t) - \hat{{\bf F}}(t) \rangle^2 \right ]
\end{aligned}
\label{38}
\end{equation}
\begin{equation}
\ \ \ \ \ \ \ \ \ \ \ \ \ \ \ \ \ \ \ \ \ + 2 \mathbb{E} \left [ \langle {\bf c}_h , \hat{{\bf Q}}(t) + \hat{{\bf F}}(t) - \hat{{\bf S}}(t) \rangle \langle {\bf c}_h , \hat{{\bf A}}(t) - \hat{{\bf F}}(t) \rangle \right ].
\label{39}
\end{equation}

The definition of the ideal arrival process yields that,

$$\langle {\bf c}_h , \hat{{\bf F}}(t) \rangle  = \frac{1}{\sqrt{M_{\mathcal{H}_o}}} \sum_{m \in \mathcal{H}_u} F_m^l(t) = \frac{1}{\sqrt{M_{\mathcal{H}_o}}} \sum_{\bar{L} \in \mathcal{L}_{\mathcal{H}_u}^*} A_{\bar{L}}(t).$$

Hence the sum of the ideal arrivals on $\mathcal{H}_u$ and the queue lengths are independent.

\begin{equation*}
\begin{aligned}
& \mathbb{E} \left [ \langle {\bf c}_h, \hat{{\bf Q}}(t) \rangle \langle {\bf c}_h, \hat{{\bf S}}(t) - \hat{{\bf F}}(t) \rangle  \right ] \\
& = \frac{1}{M_{\mathcal{H}_o}} \mathbb{E} \left [ \left ( \sum_{m \in \mathcal{H}_u} \hat{Q}_m(t) \right ) \left ( \sum_{m \in \mathcal{H}_u} \hat{S}_m(t) \right ) \right ] - \frac{1}{M_{\mathcal{H}_o}} \mathbb{E} \left [ \sum_{m \in \mathcal{H}_u} \hat{Q}_m(t) \right ] \left ( \sum_{\bar{L} \in \mathcal{L}_{\mathcal{H}_u}^*} \lambda_{\bar{L}} \right )
\end{aligned}
\end{equation*}

Note that $\hat{S}_m(t) = S_m^l(t) + S_m^k(t)$ only depends on the state of the $m$-th queue, so

\begin{equation}
\begin{aligned}
& \mathbb{E} \left [ \left ( \sum_{m \in \mathcal{H}_u} \hat{Q}_m(t) \right ) \left ( \sum_{m \in \mathcal{H}_u} \hat{S}_m(t) \right ) \right ] \\
& = \sum_{m \in \mathcal{H}_u} \mathbb{E} \left [ \hat{S}_m(t) \hat{Q}_m(t) \right ] + \sum_{m \in \mathcal{H}_u} \mathbb{E} \left [ \hat{S}_m(t) \right ] \mathbb{E} \left [ \sum_{n \in \mathcal{H}_u : n \neq m} \hat{Q}_n(t) \right ] \\
& = \sum_{m \in \mathcal{H}_u} \mathbb{E} \left [ \hat{S}_m(t) \hat{Q}_m(t) \right ] + \mathbb{E} \left [ \sum_{m \in \mathcal{H}_u} \hat{S}_m(t) \right ] \mathbb{E} \left [ \sum_{n \in \mathcal{H}_u} \hat{Q}_n(t) \right ] \\
& \ \ \ \ \ \ \ \ \ \ \ \ \ \ \ \ \ \ \ \ \ \ \ \ \ \ \ \ \ \ \ \ \ \ \ \ \ \ \ - \sum_{m \in \mathcal{H}_u} \left ( \mathbb{E} \left [ \hat{S}_m(t) \right ] \mathbb{E} \left [ \hat{Q}_m(t) \right ] \right ).
\end{aligned}
\end{equation}

The following lemma gives a lower bound on term $\sum_{m \in \mathcal{H}_u} \mathbb{E} \left [ \hat{S}_m(t) \hat{Q}_m(t) \right ]$. For proof of the following lemma, refer to Lemma $B.18$ in \cite{xie2016schedulingPhD}.

\begin{lemma}
$$\sum_{m \in \mathcal{H}_u} \mathbb{E} \left [ \hat{S}_m(t) \hat{Q}_m(t) \right ] \geq \sum_{m \in \mathcal{H}_u} \alpha \mathbb{E} \left[ \hat{Q}_m(t) \right ] - C_1,$$
where $C_1$ is a constant.
\end{lemma}

As we are studying the system in steady state, $\mathbb{E} \left [ \hat{Q}_m(t + 1) \right ] = \mathbb{E} \left [ \hat{Q}_m(t) \right ],$ $ \ $ $ \forall m $ $ \in \mathcal{H}_u$, so

$$\mathbb{E} \left [ \hat{A}_m(t) - \hat{S}_m(t) \right ] = \mathbb{E} \left [ \hat{Q}_m(t + 1) - \hat{Q}_m(t) \right ] = 0,$$

which results in $\mathbb{E} \left [ \hat{S}_m(t) \right ] = \mathbb{E} \left [ \hat{A}_m(t) \right ]$, so

\begin{equation*}
\begin{aligned}
& \mathbb{E} \left [ \sum_{m \in \mathcal{H}_u} \hat{S}_m(t) \right ] \mathbb{E} \left [ \sum_{n \in \mathcal{H}_u} \hat{Q}_n(t) \right ] - \sum_{m \in \mathcal{H}_u} \left ( \mathbb{E} \left [ \hat{S}_m(t) \right ] \mathbb{E} \left [ \hat{Q}_m(t) \right ] \right ) \\
& = \mathbb{E} \left [ \sum_{m \in \mathcal{H}_u} \hat{A}_m(t) \right ] \mathbb{E} \left [ \sum_{n \in \mathcal{H}_u} \hat{Q}_n(t) \right ]  - \sum_{m \in \mathcal{H}_u} \left ( \mathbb{E} \left [ \hat{A}_m(t) \right ] \mathbb{E} \left [ \hat{Q}_m(t) \right ] \right ) \\
& = \mathbb{E} \left [ \sum_{m \in \mathcal{H}_u} \left ( A_m^l(t) + A_m^k(t) \right) \right ] \mathbb{E} \left [ \sum_{n \in \mathcal{H}_u} \hat{Q}_n(t) \right ]  \\
& \ \ \ \ \ \ \ \ \ \ \ \ \ \ \ \ \ \ \ \ \ \ \ \ \ \ \ \ \ \ \ \ \ \ \ \ \ \ \ \ \ - \sum_{m \in \mathcal{H}_u} \left ( \mathbb{E} \left [ A_m^l(t) + A_m^k(t) \right ] \mathbb{E} \left [ \hat{Q}_m(t) \right ] \right ) \\
& \overset{(a)}{\geq} \mathbb{E} \left [ \sum_{m \in \mathcal{H}_u} \left ( A_m^l(t) + A_m^k(t) \right) \right ] \mathbb{E} \left [ \sum_{n \in \mathcal{H}_u} \hat{Q}_n(t) \right ]  - \sum_{m \in \mathcal{H}_u} \left ( \alpha \rho_h^* \mathbb{E} \left [ \hat{Q}_m(t) \right ] \right ),
\end{aligned}
\end{equation*}
where $(a)$ follows from the following lemma. Refer to Lemma $B.17$ in \cite{xie2016schedulingPhD} for the proof of the following lemma.

\begin{lemma}
$$\forall m \in \mathcal{H}_o \cup \mathcal{H}_u, \ \exists \ 0 \leq \rho_h < 1, \text{ where $\rho_h$ does not depend on $\epsilon$, } s.t.$$ $$ \mathbb{E} \left [ \frac{A_m^l}{\alpha} \right ] \leq \rho_h.$$
\end{lemma}

Then we have the following:

\begin{equation*}
\begin{aligned}
& \mathbb{E} \left [ \langle {\bf c}_h, \hat{{\bf Q}}(t) \rangle \langle {\bf c}_h, \hat{{\bf S}}(t) - \hat{{\bf F}}(t) \rangle \right ] \\
& \geq \frac{1}{M_{\mathcal{H}_o}} \Bigg \{ \sum_{m \in \mathcal{H}_u} \alpha \mathbb{E} \left [ \hat{Q}_m(t) \right ] - C_1 + \mathbb{E} \left[ \sum_{m \in \mathcal{H}_u} \left ( A_m^l(t) + A_m^k(t) \right ) \right ]  \mathbb{E} \left[ \sum_{n \in \mathcal{H}_u} \hat{Q}_n(t) \right ]  \\
& \ \ \ \ \ \ \ \ \ \ \ \ \ \ \ \ \ \ \ \ \ \ \ \ \ \ - \sum_{m \in \mathcal{H}_u} \left ( \alpha \rho_h^* \mathbb{E} \left [ \hat{Q}_m(t) \right ] \right ) - \mathbb{E} \left [ \sum_{m \in \mathcal{H}_u} \hat{Q}_m(t) \right ] \left ( \sum_{\bar{L} \in \mathcal{L}_{\mathcal{H}_u}^*} \lambda_{\bar{L}} \right ) \Bigg \}
\end{aligned}
\end{equation*}
\begin{equation*}
\begin{aligned}
& = \frac{1}{M_{\mathcal{H}_o}} \left \{ \alpha (1 - \rho_h^*) + \mathbb{E} \left [ \sum_{m \in \mathcal{H}_u} \left ( A_m^l(t) + A_m^k(t) \right ) \right ] - \sum_{\bar{L} \in \mathcal{L}_{\mathcal{H}_u}^*} \lambda_{\bar{L}} \right \} \mathbb{E} \left [ \sum_{m \in \mathcal{H}_u} \hat{Q}_m(t) \right ] \\
& \ \ \ \ \ \ \ \ \ \ \ \ \ \ \ \ \ \ \ \ \ \ \ \ \ \ \ \ \ \ \ \ \ \ \ \ \ \ \ \ \ \ \ \ \ \ \ \ \ \ \ \ \ \ \ \ \ \ \ \ \ \ \ \ \ \ \ \ \ \ \ \ \ \ \ \ \ \ \ \ \ \ \ \ \ \ \ \ \ \ \ \ - \frac{C_1}{M_{\mathcal{H}_o}}.
\end{aligned}
\end{equation*}

The following can be driven from proof of Lemma \ref{lemma27} (or Lemma $B.23$ in \cite{xie2016schedulingPhD}):

\begin{equation}
\begin{aligned}
& \sum_{\bar{L} \in \mathcal{L}_{\mathcal{H}_u}^*} \lambda_{\bar{L}} - \mathbb{E} \left [ \sum_{m \in \mathcal{H}_u} \left ( A_m^l(t) + A_m^k(t) \right ) \right ] \\
& = \mathbb{E} \left [ A_{\mathcal{H}_u \mathcal{H}_o}^l + A_{\mathcal{H}_u \mathcal{B}_o}^l + A_{\mathcal{H}_u \mathcal{H}_o}^k + A_{\mathcal{H}_o \mathcal{H}_o}^k + A_{\mathcal{H}_u \mathcal{H}_u}^k + A_{\mathcal{H}_u \mathcal{B}_o}^r + A_{\mathcal{H}_u \mathcal{H}_o}^r + A_{\mathcal{H}_u \mathcal{H}_u}^r \right ] \\
& \leq C \epsilon,
\end{aligned}
\end{equation}
where $C$ is a constant that is only a function of $\alpha, \beta,$ and $\gamma$. Furthermore, the definition of the ideal arrival process yields the following:
$$\sum_{m \in \mathcal{H}_u} \hat{F}_m(t) = \sum_{\bar{L} \in \mathcal{L}_{\mathcal{H}_u}^*} \lambda_{\bar{L}} \geq \sum_{m \in \mathcal{H}_u} \hat{A}_m(t).$$
Then we have the following:

\begin{equation}
\begin{aligned}
& \mathbb{E} \left [ \langle {\bf c}_h , \hat{{\bf Q}}(t) \rangle \langle {\bf c}_h , \hat{{\bf S}}(t) - \hat{{\bf F}}(t) \right ] \\
& \geq \frac{1}{M_{\mathcal{H}_o}} \left [ \alpha (1 - \rho_h^*) - C \epsilon \right ] \mathbb{E} \left [ \sum_{m \in \mathcal{H}_u} \hat{Q}_m(t) \right ] - \frac{C_1}{M_{\mathcal{H}_o}}.
\label{41}
\end{aligned}
\end{equation}

The number of arriving tasks and services are bounded, so

\begin{equation}
\mathbb{E} \left [ \langle {\bf c}_h , \hat{{\bf F}}(t) - \hat{{\bf S}}(t) \rangle^2 \right ] \leq C_2,
\label{42}
\end{equation}

\begin{equation}
\mathbb{E} \left [ \langle {\bf c}_h , \hat{{\bf A}}(t) - \hat{{\bf F}}(t) \rangle^2 \right ] \leq C_3,
\label{43}
\end{equation}
where $C_2 > 0$ and $C_3 > 0$ are constants not depending on $\epsilon$. Then we have the following for the term in equation (\ref{39}):
\begin{equation}
\begin{aligned}
& \mathbb{E} \left [ \langle {\bf c}_h , \hat{{\bf Q}}(t) + \hat{{\bf F}}(t) - \hat{{\bf S}}(t) \rangle \langle {\bf c}_h , \hat{{\bf A}}(t) - \hat{{\bf F}}(t) \rangle \right ] \\
& = \mathbb{E} \left [ \langle {\bf c}_h , \hat{{\bf Q}}(t) \rangle \langle {\bf c}_h , \hat{{\bf A}}(t) - \hat{{\bf F}}(t) \rangle \right ] + \mathbb{E} \left [ \langle {\bf c}_h , \hat{{\bf F}}(t) - \hat{{\bf S}}(t) \rangle \langle {\bf c}_h , \hat{{\bf A}}(t) - \hat{{\bf F}}(t) \rangle \right ] \\
& \overset{(a)}{\leq} \mathbb{E} \left [ \langle {\bf c}_h , \hat{{\bf F}}(t) - \hat{{\bf S}}(t) \rangle \langle {\bf c}_h , \hat{{\bf A}}(t) - \hat{{\bf F}}(t) \rangle \right ] \\
& \overset{(b)}{\leq} C_4,
\label{44}
\end{aligned}
\end{equation}
where $(a)$ is true as $\langle {\bf c}_h , \hat{{\bf A}}(t) - \hat{{\bf F}}(t) \rangle \leq 0$, and $(b)$ is true as the number of task arrival is bounded, and $C_4$ is a constant.

From equations (\ref{39}), (\ref{41}), (\ref{42}), (\ref{43}), and (\ref{44}), the following is derived:
$$\frac{2}{M_{\mathcal{H}_o}} \left [ \alpha (1 - \rho_h^*) - C \epsilon \right ] \mathbb{E} \left [ \sum_{m \in \mathcal{H}_u} \hat{Q}_m(t) \right ] \leq \frac{2C_1}{M_{\mathcal{H}_o}} + C_2 + C_3 + 2C_4,$$
so for any $0 < \epsilon < \frac{\alpha (1 - \rho_h^*)}{C}$, the following is true:

$$\mathbb{E} \left [ \sum_{m \in \mathcal{H}_u} \hat{Q}_m(t) \right ] \leq \frac{C_5}{\alpha (1 - \rho_h^*) - C \epsilon},$$
where $C_5 = C_1 + (C_2 + C_3 + 2C_4) \frac{M_{\mathcal{H}_o}}{2},$ so

$$\lim_{\epsilon \downarrow 0} \mathbb{E} \left [ \sum_{m \in \mathcal{H}_u} \left ( Q_m^{l (\epsilon)}(t) + Q_m^{k (\epsilon)}(t) \right ) \right ] \leq \frac{C_5}{\alpha (1 - \rho_h^*)},$$
that is equivalent to the following:
$$\lim_{\epsilon \downarrow 0} \epsilon \mathbb{E} \left [ \sum_{m \in \mathcal{H}_u} \left ( Q_m^{l (\epsilon)}(t) + Q_m^{k (\epsilon)}(t) \right ) \right ] = 0.$$
Similarly, we can prove the following:

$$\lim_{\epsilon \downarrow 0} \epsilon \mathbb{E} \left [ \sum_{m \in \mathcal{H}_o} Q_m^{l (\epsilon)}(t) \right ] = 0.$$

\subsection{Proof of Theorem \ref{state-space-collapse}}
\label{proofstate-space-collapse}

The following lemma is given for ideal load decomposition for the case $\mathcal{O} \neq$ $ \oldemptyset$. For proof of this lemma, refer to Lemma $B.19$ in \cite{xie2016schedulingPhD}.

\begin{lemma}
$$\exists \lambda_0 > 0 \text{ not depending on } \epsilon, \text{ such that:}$$
\begin{enumerate}
\item Defining
$$\omega_m = \sum_{\bar{L} : m \in \bar{L}} \frac{\lambda_{\bar{L}, m}^*}{\alpha} + \sum_{\bar{L} : m \in \bar{L}_k} \frac{\lambda_{\bar{L}, m}^*}{\beta} + \sum_{\bar{L} : m \in \bar{L}_r} \frac{\lambda_{\bar{L}, m}^*}{\gamma},$$
we have the following:
$$
\omega_m =
\begin{cases}
	1 - \gamma \epsilon_0, & \forall m \in \mathcal{H}_u \\
	1 - \beta \epsilon_0, & \forall m \in \mathcal{H}_o \\
	1 - \alpha \epsilon_0, & \forall m \in \mathcal{B}_o \\
\end{cases},
$$
where $\epsilon_0 = \frac{\epsilon}{|| \hat{{\bf c}} ||^2}$.

\item Denote the set of task types that are only local to $\mathcal{B}_o$ by $\mathcal{L}_{\mathcal{B}_o}$. Then,
\begin{equation*}
\begin{aligned}
& \forall \bar{L} \in \mathcal{L}_{\mathcal{B}_o}, \text{ and } \forall m \in \{ i \in \mathcal{M} | i \in \bar{L}, \text{ or } i \in \mathcal{H}_u, \text{ or } i \in \bar{L}_k \cap \mathcal{H}_o \}, \\
& \exists \kappa > 0, \text{ independent of } \epsilon, \text{ such that: } \lambda_{\bar{L}, m}^* = \sum_{n \in \bar{L}} \lambda_{\bar{L}, n, m}^* \geq \kappa.
\end{aligned}
\end{equation*}
\end{enumerate}

\label{lemma23}
\end{lemma}

For the evenly loaded scenario, the following three lemmas are used. For the proof of these lemmas, refer to Lemmas $B.20$, $B.21$, and $B.22$ in \cite{xie2016schedulingPhD}.

\begin{lemma}
Using the Balanced-Pandas algorithm, we have the following:
$$\mathbb{E} \left [ \langle {\bf W}(t), {\bf A}(t) \rangle - \langle {\bf W}, {\mathbold{\omega}} \rangle | Z(t) \right ] \leq - \lambda_{min} || {\bf W}_{\perp}(t) ||, \ \ \ \forall t \geq 0,$$
where $\lambda_{min} > 0$ is a constant not depending on $\epsilon$.
\label{lemma24}
\end{lemma}

\begin{lemma}
Using the Balanced-Pandas algorithm, we have the following:
$$\mathbb{E} \left [ \langle {\bf W}(t), {\mathbold{\omega}} \rangle - \langle {\bf W}(t), {\bf S}(t) \rangle | Z(t) \right ] = - \frac{\epsilon}{|| \hat{{\bf c}} ||} \langle {\bf c} , {\bf W} \rangle, \ \ \ \forall t \geq 0.$$
\label{lemma25}
\end{lemma}

\begin{lemma}
$$\mathbb{E} \left [ \langle {\bf c}, {\bf W}(t) \rangle \langle {\bf c}, {\bf A}(t) - {\bf S}(t) \rangle | Z(t) \right ] \geq - \frac{\epsilon}{|| \hat{{\bf c}} ||} \langle {\bf c} , {\bf W} \rangle, \ \ \ \forall t \geq 0.$$
\label{lemma26}
\end{lemma}
In order to prove Theorem \ref{state-space-collapse}, consider the following Lyapunov function:

$$F(Z) = || {\bf W}_{\perp} ||.$$
The drift of this Lyapunov function is given as below:

$$ \Delta F(Z) \leq \frac{1}{2 || {\bf W}_{\perp} ||} \left ( \Delta V(Z) - \Delta V_{||}(Z) \right ),$$
where $\Delta V(Z)$ and $\Delta V_{||}(Z)$ are the drifts for Lyapunov functions $V(Z) = || {\bf W} ||^2$ and $V_{||}(Z) = || {\bf W}_{||} ||^2$, respectively. Then, we have the following:
\begin{equation*}
\begin{aligned}
& \mathbb{E} [\Delta V(Z(t)) - \Delta V_{||}(Z(t)) | Z(t)] \\
& \leq 2 \mathbb{E} [ \langle {\bf W}(t), {\bf A}(t) - {\bf S}(t) \rangle - \langle {\bf c} , {\bf W}(t) \rangle \langle {\bf c}, {\bf A}(t) - {\bf S}(t) \rangle | Z(t)] + C_1.
\end{aligned}
\end{equation*}

Using Lemmas \ref{lemma24}, \ref{lemma25}, and \ref{lemma26}, we get the following upper bound on $\mathbb{E} [ \Delta F(Z(t)) | Z(t) ]:$

$$\mathbb{E} [ \Delta F(Z(t)) | Z(t) ] \leq - \lambda_0 + \frac{C}{|| {\bf W}_{\perp}(t) ||},$$
where $\lambda_0 > 0$ and $C > 0$ are constants not depending on $\epsilon.$ This last inequality satisfies the negative drift condition, so there exists finite series of constants $\{ C_r' \}_{r \in \mathbb{N}}$ such that $\mathbb{E} \left [ || {\bf W}_{\perp}^{(\epsilon)}(t) ||^r \right ] \leq C_r'$ for any $\epsilon \in (0, M \alpha).$

\subsection{Proof of Theorem \ref{lower-bound}}
\label{prooflower-bound}

The lower bound on $\mathbb{E} \left [ \Phi^{(\epsilon)}(t) \right ]$ can be driven by constructing a system with a single server and a single queue with arrival process $\left \{ \sum_{\bar{L} \in \mathcal{L}_{\mathcal{B}_o}} A_{\bar{L}}^{(\epsilon)}(t), \ t \geq 0 \right \}$ and service process $\Big \{ b^{(\epsilon)}(t) = \sum_{i \in \mathcal{B}_o} X_i(t) + \sum_{j \in \mathcal{H}_o} Y_j(t) + \sum_{n \in \mathcal{H}_u} V_n(t), \ $ $t $ $\geq$ $0 \Big \}$. Denote the queue length of the constructed system by $\Psi^{(\epsilon)}(t)$. By the definitions of $X_i, Y_j,$ and $V_n$, $\mathbb{E} \left [ \sum_{i \in \mathcal{B}_o} X_i(t) \right ]$, $\mathbb{E} \left [ \sum_{j \in \mathcal{H}_o} Y_j(t) \right ]$, and $\mathbb{E} \left [ \sum_{n \in \mathcal{H}_u} V_n(t) \right ]$ are the maximum amount of local, rack-local, and remote services that can be given to $\sum_{\bar{L} \in \mathcal{L}_{\mathcal{B}_o}} A_{\bar{L}}^{(\epsilon)}(t)$. Hence, it is obvious that in steady state $\Psi^{(\epsilon)}(t)$ is stochastically smaller than or equal to $\Phi^{(\epsilon)}(t)$. Then the lower bound on $\mathbb{E} \left [ \Phi^{(\epsilon)}(t) \right ]$ is derived by using Lemma 4 in \cite{eryilmaz2012asymptotically}.

\subsection{Proof of Theorem \ref{upper-bound}}
\label{proofupper-bound}

The ideal scheduling, service, and arrival processes are defined as follows: \\
\textbf{Ideal Scheduling Decision Process} $\boldsymbol{ \eta'} (t)$: Under ideal scheduling, a beneficiary server in an over-loaded rack is only giving service to its local tasks queued in its local sub-queue, and an idle helper server in an overloaded rack that has no local tasks in its local sub-queue is only scheduled to give service to its rack-local tasks queued at its rack-local sub-queue. In other works,
\begin{equation*}
\begin{aligned}
& \forall m \in \mathcal{B}_o, \eta_m'(t) = 0, \\
& \forall m \in \mathcal{H}_o, \eta_m'(t) = \eta_m(t) \text{ if } \eta_m(t) = 0, \text{ and } \eta_m'(t) = 1 \text{ if } f_m(t^-) = -1, Q_m^l(t) = 0 \\
& \forall m \in \mathcal{H}_u, \eta_m'(t) = \eta_m(t).
\end{aligned}
\end{equation*}
\textbf{Ideal Service Process ${\bf D}(t)$:}
$$\forall m \in \mathcal{B}_o, \ D_m^l(t) = X_m^l(t), \ D_m^k(t) = 0, \ D_m^r(t) = 0,$$
where $X_m^l(t) \sim Bern(\alpha)$, and each process $X_m^l(t)$ is i.i.d. and is coupled with $S_m(t)$ as follows: If $\eta_m(t) = 0,$ $X_m^l(t) = S_m^l(t)$; if $\eta_m(t) = 1$, $X_m^l(t) = 0$ when $S_m^k(t) = 0$, and $X_m^l(t) \sim Bern(\frac{\alpha}{\beta})$ when $S_m^k(t) = 1$; if $\eta_m(t) = 2$, $X_m^l(t) = 0$ when $S_m^r(t) = 0$, and $X_m^l(t) \sim Bern(\frac{\alpha}{\gamma})$ when $S_m^k(t) = 1$. Furthermore,
$$\forall m \in \mathcal{H}_o, \ D_m^l(t) = S_m^l(t), \ D_m^k(t) = Y_m^k(t), \ D_m^r(t) = 0,$$
where $Y_m^k(t) \sim Bern(\beta I_{\{ \eta_m(t) \neq 0 \}})$ and each process $Y_m^k(t)$ is i.i.d. Finally, $\forall m \in \mathcal{H}_u$, ${\bf D}_m(t) = {\bf S}_m(t)$,
$$D_m^l(t) = S_m^l(t), \ D_m^k(t) = S_m^k(t), \ D_m^r(t) = S_m^r(t).$$
\textbf{Ideal Arrival Process} ${\bf F}(t)$:  Ideally, any task type that has a local server in the set $\mathcal{H}_u$ should receive service locally. In other words, $\forall \bar{L} \in \mathcal{L}_{\mathcal{H}_u}^*$, task of type $\bar{L}$ is routed to one of its local servers in the set $\mathcal{H}_u$. Hence, unwanted arrivals $\sum_{m : m \notin \bar{L}, m \notin \mathcal{H}_u} A_{\bar{L}, m}$ should be reassigned evenly among their local servers in $\mathcal{H}_u$. Similarly, $\forall \bar{L} \in \mathcal{L}_{\mathcal{H}_o}$, the task of type $\bar{L}$ should ideally be assigned to its local servers in $\mathcal{H}_o$, that is unwanted arrivals $\sum_{m : m \notin \bar{L}, m \notin \mathcal{H}_o} A_{\bar{L}, m}$ should be reassigned evenly among their local servers in $\mathcal{H}_o$. On the other hand, $\forall \bar{L} \in \mathcal{L}_{\mathcal{B}_o}$, task of type $\bar{L}$ should either receive service locally from a server in $\mathcal{B}_o$, or rack-locally from a server in $\mathcal{H}_o$, or remotely from a server in $\mathcal{H}_u$. Hence, we reassign tasks so that the above conditions hold in the ideal case. Then, the dynamics of $\tilde{{\bf Q}}$ can be written as follows:
$$\tilde{{\bf Q}}(t + 1) = \tilde{{\bf Q}}(t) + \tilde{{\bf F}}(t) - \tilde{{\bf D}}(t) + \tilde{{\bf V}}(t),$$
where $\tilde{{\bf V}}(t) = \tilde{{\bf A}}(t) - \tilde{{\bf F}}(t) + \tilde{{\bf D}}(t) - \tilde{{\bf S}}(t) + \tilde{{\bf U}}(t).$ Note that in steady state, we have the following:
\begin{equation}
\begin{aligned}
& 2 \mathbb{E} \left [ \langle {\bf c}, \tilde{{\bf Q}}(t) \rangle \langle {\bf c}, \tilde{{\bf D}}(t) - \tilde{{\bf F}}(t) \rangle \right ] \\
& = \mathbb{E} \left [ \langle {\bf c}, \tilde{{\bf F}}(t) - \tilde{{\bf D}}(t) \rangle^2 \right ] + \mathbb{E} \left [ \langle {\bf c} , \tilde{{\bf V}}(t) \rangle^2 \right ] \\
& \ \ \ \ \ \ \ \ \ \ \ \ \ \ \ \ \ \ \ \ \ \ \ \ \ \ \ \ \ \ \ \ \ \ \ \ \ \ \ \ \ \ \ \ + 2 \mathbb{E} \left [ \langle {\bf c}, \tilde{{\bf Q}}(t) + \tilde{{\bf F}}(t) - \tilde{{\bf D}}(t) \rangle \langle {\bf c}, \tilde{{\bf V}}(t) \rangle \right ].
\end{aligned}
\label{48}
\end{equation}
On the other hand,
\begin{equation}
\begin{aligned}
\Phi^{(\epsilon)}(t) \leq & \sum_{m \in \mathcal{H}_u} \gamma \frac{Q_m^r}{\gamma} +  \sum_{m \in \mathcal{H}_o} \beta \left ( \frac{Q_m^k}{\beta} + \frac{Q_m^r}{\gamma} \right ) + \sum_{m \in \mathcal{B}_o} \alpha \left ( \frac{Q_m^l}{\alpha} + \frac{Q_m^k}{\beta} + \frac{Q_m^r}{\gamma} \right ) \\
= & || \tilde{{\bf c}} || \langle {\bf c}, \tilde{{\bf Q}} \rangle,
\label{49}
\end{aligned}
\end{equation}
so in order to find an upper bound on $\mathbb{E} \left [ \Phi^{(\epsilon)}(t) \right ]$, we need to find an upper bound on $\mathbb{E} \left [ \langle {\bf c}, \tilde{{\bf Q}}(t) \rangle \right ]$. To this aim, We start by analyzing different terms in equation (\ref{48}). For simplicity, we omit the superscripts $^{(\epsilon)}$ in the following equations temporarily. The definition of ideal arrival process yields the following:
$$\langle \hat{{\bf c}}, \tilde{{\bf F}}(t) \rangle = \sum_{m \in \mathcal{B}_o} \alpha \cdot \frac{F_m^l(t)}{\alpha} + \sum_{m \in \mathcal{H}_o} \beta \cdot \frac{F_m^k(t)}{\beta} + \sum_{m \in \mathcal{H}_u} \gamma \cdot \frac{F_m^r(t)}{\gamma}  = \sum_{\bar{L} \in \mathcal{L}_{\mathcal{B}_o}} A_{\bar{L}}(t). $$
Therefore,
$$\mathbb{E} \left [ \langle \hat{{\bf c}}, \tilde{{\bf F}}(t) \rangle \right ] = \sum_{\bar{L} \in \mathcal{L}_{\mathcal{B}_o}} \lambda_{\bar{L}},$$
$$Var \left [ \langle \hat{{\bf c}}, \tilde{{\bf F}}(t) \rangle \right ] = \left ( \sigma^{(\epsilon)} \right )^2.$$
The definition of ideal service process yields the following:
$$\langle \hat{{\bf c}}, \tilde{{\bf D}}(t) \rangle = \sum_{m \in \mathcal{B}_o} \alpha \cdot \frac{D_m^l(t)}{\alpha} + \sum_{m \in \mathcal{H}_o} \beta \cdot \frac{D_m^k(t)}{\beta} + \sum_{m \in \mathcal{H}_u} \gamma \cdot \frac{D_m^r(t)}{\gamma}. $$
For a server $m$, define $\rho_m^l$ as the proportion of time in steady state the server spends on giving local service to the tasks queued in its local sub-queue. Then we have the following:
\begin{equation*}
\begin{aligned}
& \mathbb{E} \left [ \langle \hat{{\bf c}}, \tilde{{\bf D}}(t) \rangle \right ] = \alpha M_{\mathcal{B}_o} + \sum_{m \in \mathcal{H}_o} \beta \left ( 1 - \rho_m^l \right ) + \sum_{m \in \mathcal{H}_u} \gamma \left ( 1 - \rho_m^l \right ), \\
& Var \left [ \langle \hat{{\bf c}}, \tilde{{\bf D}}(t) \rangle \right ] = \alpha (1 - \alpha) M_{\mathcal{B}_o} + \sum_{m \in \mathcal{H}_o} \beta \left ( 1 - \rho_m^l \right ) \left [ 1 - \beta \left ( 1 - \rho_m^l \right ) \right ] \\
& \ \ \ \ \ \ \ \ \ \ \ \ \ \ \ \ \ \ \ \ \ \ \ \ \ \ \ \ \ \ \ \ \ \ \ \ \ \ \ \ \ \ + \sum_{m \in \mathcal{H}_u} \gamma \left ( 1 - \rho_m^l \right ) \left [ 1 - \gamma \left ( 1 - \rho_m^l \right ) \right ] \\
& \ \ \ \ \ \ \ \ \ \ \ \ \ \ \ \ \ \ \ \ = \left ( \nu^{(\epsilon)} \right )^2.
\end{aligned}
\end{equation*}
Then,
$$\mathbb{E} \left [ \langle \hat{{\bf c}}, \tilde{{\bf D}}(t) \rangle \right ] - \mathbb{E} \left [ \langle \hat{{\bf c}}, \tilde{{\bf F}}(t) \rangle \right ] = \epsilon + \sum_{m \in \mathcal{H}_o} \beta \left ( \rho_m^{l(\epsilon)} - \rho_m^l \right ) + \sum_{m \in \mathcal{H}_u} \gamma \left ( \rho_m^{l(\epsilon)} - \rho_m^l \right )$$ $$= \epsilon + \delta,$$
where $\delta = \sum_{m \in \mathcal{H}_o} \beta \left ( \rho_m^{l(\epsilon)} - \rho_m^l \right ) + \sum_{m \in \mathcal{H}_u} \gamma \left ( \rho_m^{l(\epsilon)} - \rho_m^l \right ) \geq 0$, and $\delta \rightarrow 0 $ as $\epsilon \rightarrow 0$.
Hence, we have the following for the left-hand side term in equation (\ref{48}):
\begin{equation}
\begin{aligned}
& \mathbb{E} \left [ \langle {\bf c}, \tilde{{\bf Q}}(t) \rangle \langle {\bf c}, \tilde{{\bf D}}(t) - \tilde{{\bf F}}(t) \rangle \right ] \\
& = \frac{1}{|| \hat{{\bf c}} ||} \mathbb{E} \left [ \langle {\bf c}, \tilde{{\bf Q}}(t) \rangle \left ( \langle \hat{{\bf c}}, \tilde{{\bf D}}(t) \rangle - \langle \hat{{\bf c}}, \tilde{{\bf F}}(t) \rangle \right ) \right ] \\
& = \frac{\epsilon + \delta}{|| \hat{{\bf c}} ||} \mathbb{E} \left [ \langle {\bf c}, \tilde{{\bf Q}}(t) \rangle \right ].
\end{aligned}
\label{50}
\end{equation}

The first term on the right-hand side of equation (\ref{48}) can be simplified as follows:
\begin{equation}
\begin{aligned}
& \mathbb{E} \left [ \langle {\bf c}, \tilde{{\bf F}}(t) - \tilde{{\bf D}}(t) \rangle^2 \right ] \\
& = \frac{1}{|| \hat{{\bf c}} ||^2} \left \{ Var \left [ \langle \hat{{\bf c}}, \tilde{{\bf D}}(t) \rangle \right ] + Var \left [ \langle \hat{{\bf c}}, \tilde{{\bf F}}(t) \rangle \right ] + \left ( \mathbb{E} \left [ \langle {\bf c}, \tilde{{\bf F}}(t) - \tilde{{\bf D}}(t) \rangle \right ] \right )^2 \right \} \\
& = \frac{1}{|| \hat{{\bf c}} ||^2} \left \{ \left ( \sigma^{(\epsilon)} \right )^2 + \left ( \nu^{(\epsilon)} \right )^2 + (\epsilon + \delta)^2 \right \}.
\label{51}
\end{aligned}
\end{equation}
The second term on the right-hand side of equation (\ref{48}) is upper bounded as the following lemma suggests.

\begin{lemma}
$$\mathbb{E} \left [ \langle {\bf c}, \tilde{{\bf V}}(t) \rangle^2 \right ] \leq C \epsilon,$$
where $C$ is a constant that does not depend on $\epsilon$.
\label{lemma27}
\end{lemma}

In order to find an upper bound on the third term on the right-hand side of equation (\ref{48}), we do the following. The system is in steady state, so

$$\mathbb{E} \left [ \langle {\bf c}, \tilde{{\bf F}}(t) - \tilde{{\bf D}}(t) + \tilde{{\bf V}}(t) \rangle \right ] = \mathbb{E} \left [ \langle {\bf c}, \tilde{{\bf Q}}(t+1) - \tilde{{\bf Q}}(t) \rangle \right ] = 0,$$
so
$$\mathbb{E} \left [ \langle {\bf c}, \tilde{{\bf V}}(t) \rangle \right ] = \mathbb{E} \left [ \langle {\bf c}, \tilde{{\bf F}}(t) - \tilde{{\bf D}}(t) \rangle \right ] = \frac{\epsilon}{M \alpha},$$
then
\begin{equation*}
\begin{aligned}
& \mathbb{E} \left [ \langle {\bf c}, \tilde{{\bf F}}(t) - \tilde{{\bf D}}(t) \rangle \langle {\bf c}, \tilde{{\bf V}}(t) \rangle \right ] \\
& \leq \mathbb{E} \left [ \langle {\bf c}, \tilde{{\bf F}}(t) \rangle \langle {\bf c}, \tilde{{\bf V}}(t) \rangle \right ] \\
& \leq \frac{C_A}{\sqrt{M} \alpha} \mathbb{E} \left [ \langle {\bf c}, \tilde{{\bf V}}(t) \rangle \right ] \\
& = \frac{C_A}{M \sqrt{M} \alpha^2} \epsilon,
\end{aligned}
\end{equation*}
so we have the following upper bound on the third term on the right-hand side of equation (\ref{48}):
\begin{equation*}
\begin{aligned}
& \mathbb{E} \left [ \langle {\bf c}, \tilde{{\bf Q}}(t) + \tilde{{\bf F}}(t) - \tilde{{\bf D}}(t) \rangle \langle {\bf c}, \tilde{{\bf V}}(t) \rangle \right ] \\
& = \mathbb{E} \left [ \langle {\bf c}, \tilde{{\bf Q}}(t) \rangle \langle {\bf c}, \tilde{{\bf V}}(t) \rangle \right ] + \mathbb{E} \left [ \langle {\bf c}, \tilde{{\bf F}}(t) - \tilde{{\bf D}}(t) \rangle \langle {\bf c}, \tilde{{\bf V}}(t) \rangle \right ] \\
& \leq \mathbb{E} \left [ \langle {\bf c}, \tilde{{\bf Q}}(t) \rangle \langle {\bf c}, \tilde{{\bf V}}(t) \rangle \right ] + \frac{C_A}{M \sqrt{M} \alpha^2} \epsilon.
\end{aligned}
\end{equation*}
We then simplify the term $\langle {\bf c}, \tilde{{\bf Q}}(t) \rangle \langle {\bf c}, \tilde{{\bf V}}(t) \rangle$ as follows:
\begin{equation}
\begin{aligned}
& \langle {\bf c}, \tilde{{\bf Q}}(t) \rangle \langle {\bf c}, \tilde{{\bf V}}(t) \rangle \\
& = \langle \tilde{{\bf Q}}(t), \tilde{{\bf V}}(t) \rangle - \langle \tilde{{\bf Q}}_{\perp}(t), \tilde{{\bf V}}_{\perp}(t) \rangle \\
& = \langle \tilde{{\bf Q}}(t), \tilde{{\bf D}}(t) - \tilde{{\bf S}}(t) \rangle + \langle \tilde{{\bf Q}}(t), \tilde{{\bf A}}(t) - \tilde{{\bf F}}(t) \rangle + \langle \tilde{{\bf Q}}(t), \tilde{{\bf V}}(t)\rangle - \langle \tilde{{\bf Q}}_{\perp}(t), \tilde{{\bf V}}_{\perp}(t) \rangle.
\label{53}
\end{aligned}
\end{equation}
The following two lemmas give a bound for the first two terms in equation (\ref{53}).

\begin{lemma}
$$\mathbb{E} \left [ \langle \tilde{{\bf Q}}(t), \tilde{{\bf D}}(t) - \tilde{{\bf S}}(t) \rangle \right ] = 0.$$
\label{lemma28}
\end{lemma}

\begin{lemma}
$$\mathbb{E} \left [ \langle \tilde{{\bf Q}}(t), \tilde{{\bf A}}(t) - \tilde{{\bf F}}(t) \rangle \right ] = o(\epsilon).$$
\label{lemma29}
\end{lemma}
For the proof of Lemmas \ref{lemma28} and \ref{lemma29} refer to Lemmas $B.24$ and $B.25$ in \cite{xie2016schedulingPhD}.

By Lemma \ref{lemma123}, the third term in equation (\ref{53}) is equal to zero. In order to find an upper bound for the last term in equation (\ref{53}), we first find an upper bound on $\mathbb{E}\left [ || \tilde{{V}}(t) ||^2 \right ]$. Using lemma \ref{lemma27}, we have the following:
$$\mathbb{E}\left [ || \tilde{{V}}(t) ||^2 \right ] \leq R \epsilon,$$
where $R$ is a constant not depending on $\epsilon$. Then we use Cauchy-Schwartz inequality and the result on state space collapse to find the following bound:
$$\mathbb{E} \left [ - \langle \tilde{{\bf Q}}_{\perp}(t), \tilde{{\bf V}}_{\perp}(t) \right ] \leq \sqrt{\mathbb{E} \left [ || \tilde{{\bf Q}}_{\perp}(t) ||^2 \right ] \mathbb{E} \left [ || \tilde{{\bf V}}_{\perp}(t) ||^2 \right ]} \leq \sqrt{C_2'R\epsilon}.$$
Hence, we have the following bound on the last term on the right-hand side of equation (\ref{48}):
\begin{equation}
\mathbb{E} \left [ \langle {\bf c}, \tilde{{\bf Q}}(t) + \tilde{{\bf F}}(t) - \tilde{{\bf D}}(t) \rangle \langle {\bf c}, \tilde{{\bf V}}(t) \rangle \right ] \leq \frac{C_A}{M \sqrt{M} \alpha^2} \epsilon + \sqrt{C_2' R \epsilon} + o(\epsilon).
\label{54}
\end{equation}

Using Lemma \ref{lemma27}, equations (\ref{48}), (\ref{50}), (\ref{51}), and (\ref{54}) in equation (\ref{49}) and bringing the superscript $^{(\epsilon)}$ back in the equations, we have the following:
\begin{equation*}
\begin{aligned}
& 2 \frac{\epsilon + \delta}{|| \hat{{\bf c}} ||} \mathbb{E} \left [ \langle {\bf c} , \tilde{{\bf Q}}(t) \rangle \right ] \\
& \leq \frac{1}{|| \hat{{\bf c}} ||^2} \left ( \left ( \sigma^{(\epsilon)} \right )^2 + \left ( \nu^{(\epsilon)} \right )^2 + (\epsilon + \delta)^2 \right ) + C \epsilon + \frac{2C_A}{M \sqrt{M} \alpha^2} \epsilon + 2 \sqrt{C_2' R \epsilon} + 2 o(\epsilon),
\end{aligned}
\end{equation*}
since $\delta \geq 0$,
\begin{equation*}
\begin{aligned}
& 2 \frac{\epsilon}{|| \hat{{\bf c}} ||} \mathbb{E} \left [ \langle {\bf c} , \tilde{{\bf Q}}(t) \rangle \right ] \\
& \leq \frac{1}{|| \hat{{\bf c}} ||^2} \left ( \left ( \sigma^{(\epsilon)} \right )^2 + \left ( \nu^{(\epsilon)} \right )^2 + (\epsilon + \delta)^2 \right ) + C \epsilon + \frac{2C_A}{M \sqrt{M} \alpha^2} \epsilon + 2 \sqrt{C_2' R \epsilon} + 2 o(\epsilon),
\end{aligned}
\end{equation*}
so,
\begin{equation*}
\begin{aligned}
& || \hat{{c}} || \mathbb{E} \left [ \langle {\bf c}, \tilde{{\bf Q}}(t) \rangle \right ] \\
& \leq \frac{\left ( \sigma^{(\epsilon)} \right )^2 + \left ( \nu^{(\epsilon)} \right )^2 + (\epsilon + \delta)^2}{2 \epsilon} + \left ( \frac{C}{2} + \frac{C_A}{M \sqrt{M} \alpha^2} \right ) || \hat{{\bf c}} ||^2 + || \hat{{\bf c}} ||^2 \sqrt{\frac{C_2' R}{\epsilon}} + o(1).
\end{aligned}
\end{equation*}
Note that,
\begin{equation*}
\begin{aligned}
& \mathbb{E} \left [ \Phi^{(\epsilon)}(t) \right ] \\
& = \mathbb{E} \bigg [ \sum_{m \in \mathcal{B}_o} \left ( Q_m^{l(\epsilon)}(t) + Q_m^{k(\epsilon)}(t) + Q_m^{r(\epsilon)}(t) \right ) + \sum_{m \in \mathcal{H}_o} \left ( Q_m^{k(\epsilon)}(t) + Q_m^{r(\epsilon)}(t) \right ) \\
& \ \ \ \ \ \ \ \ \ \ \ \ \ \ \ \ \ \ \ \ \ \ \ \ \ \ \ \ \ \ \ \ \ \ \ \ \ \ \ \ \ \ \ \ \ \ \ \ \ \ \ \ \ \ \ \ \ \ \ \ \ \ \ \ \ \ \ \ \ \ \ \ \ + \sum_{m \in \mathcal{H}_u} Q_m^{r(\epsilon)}(t) \bigg ] \\
& \leq \mathbb{E} \left [ \sum_{m \in \mathcal{B}_o} \alpha \left ( \frac{Q_m^l}{\alpha} + \frac{Q_m^k}{\beta} + \frac{Q_m^r}{\gamma} \right ) + \sum_{m \in \mathcal{H}_o} \beta \left ( \frac{Q_m^k}{\beta} + \frac{Q_m^r}{\gamma} \right ) + \sum_{m \in \mathcal{H}_u} \gamma \frac{Q_m^r}{\gamma} \right ] \\
& = \mathbb{E} \left [ || \tilde{{\bf c}} || \langle {\bf c}, \tilde{{\bf Q}} \rangle \right ].
\end{aligned}
\end{equation*}
Hence,

$$\mathbb{E} \left [ \Phi^{(\epsilon)}(t) \right ] \leq \frac{\left ( \sigma^{(\epsilon)} \right )^2 + \left ( \nu^{(\epsilon)} \right )^2 + (\epsilon + \delta)^2}{2 \epsilon} + B^{(\epsilon)},$$
where $B^{(\epsilon)} = \left ( \frac{C}{2} + \frac{C_A}{M \sqrt{M} \alpha^2} \right ) || \hat{{\bf c}} ||^2 + || \hat{{\bf c}} ||^2 \sqrt{\frac{C_2' R}{\epsilon}} + o(1)$, i.e., $B^{(\epsilon)} = o(\frac{1}{\epsilon})$. This proves Theorem \ref{upper-bound} as $\epsilon \rightarrow 0$.

\end{appendices}